\documentclass[prx,aps,twocolumn,showpacs,preprintnumbers,amsmath,amssymb,notitlepage,10pt]{revtex4-1}
\usepackage{graphicx}
\usepackage[table]{xcolor}
\DeclareMathOperator{\sgn}{sgn}
\DeclareMathOperator{\Tr}{Tr}
\renewcommand{\Im}{\mathfrak{Im}}
\renewcommand{\Re}{\mathfrak{Re}}
\DeclareGraphicsExtensions{.pdf,.PDF,.eps,.png,.PNG,}

\begin{document}

\title{ Topological phase transitions, invariants and enriched bulk-edge correspondences in fermionic gapless systems with extended Fermi surface }
\author{ Fadi Sun$^{1,2}$ and Jinwu Ye$^{1,2,3}$   }
\affiliation{ 
$^{1}$ The School of Science, Great Bay University, Dongguan 523000, Guangdong, China \\
$^{2}$ Institute for Theoretical Sciences, Westlake University, Hangzhou 310024, Zhejiang, China  \\
$^{3}$  Department of Physics and Astronomy, Mississippi State University, MS 39762, USA     }
\date{\today }
\begin{abstract}
 Topological phases and topological phase transitions (TPT) are among the most fantastic phenomena in Nature.
 Here we show that injecting a current  may lead to new topological phases, especially new gapless topological metallic phases with extended Fermi surfaces (FSs)
 through novel class of TPTs in the bulk or the boundary.
 Specifically, we study the quantum anomalous Hall (QAH) system in a square lattice
 under various forms of injecting currents. In addition to the previously  known Chern insulator ( which will be called even Chern insulator  here ),
 band insulator  and band metal (BM), we find three new topological phases we name as: the gapped odd Chern insulator (Odd CI),
 the gapless odd Chern metal (Odd CM) and even Chern metal (Even CM).
The Chern number may not be effective anymore in characterizing the topological gapless phases with extended FS.
It is the Hall conductance which acts as the new topological invariant in such gapless systems.
Its jump is a universal integer or non-integer across the even CM/BM or odd CM/BM TPT respectively where
there is also a corresponding TPT in the Longitudinal (L-) edge modes.  The Odd/even CM to BM transition is a novel class of TPT
without any non-analyticity in the ground state energy density.
This  presents the first example of a TPT which is not a quantum phase transition (QPT).
The original bulk-edge correspondence is enriched into bulk/Longitudinal (L-)/Transverse (T-) edge correspondence.
The L- edge reconstruction may happen earlier, later or at the same time as the bulk TPT respectively in the even CI/odd CI/odd CM sequence
with the edge dynamic exponent $ z_L=3 $, in the even CI/even CM/odd CM sequence with $ z_L=2 $ or a direct even CI/odd CM with a flat edge.
The disappearance of the T- edge always happen at the same time as the bulk TPT with a universal edge critical  behaviour.
 We  classify all the possible bulk and edge TPTs and also evaluate the thermodynamic quantities  such as the density of states, specific heat, compressibility and the Wilson ratio in all the phases and also their quantum scaling forms near all these TPTs.
 The methods may be applied to explore new topological phases of other Hamiltonians in any
 lattice in any dimension in any forms of injecting current.
 Doing various experiments by injecting different sorts of currents
 may become an effective way to drive various topological phases to new topological gapped or gapless metallic phases through novel bulk or edge TPTs.
\end{abstract}


\maketitle
\section{Introduction}
The Anomalous Hall Effect(AHE) due to the Berry phase of itinerant electrons in a quantum Ferromagnet in real space \cite{AHE1} or
in momentum space \cite{AHE2,Haldane2004} has been under intense investigations since more than 20 years ago.
It involves the spin-orbit coupling which couples the orbital motion of electrons to the spin polarization\cite{AHE3}.
The AHE in a metallic Ferromagnet is in general, not quantized, so can take any value.
However, it can be quantized in an insulator, called Quantum Anomalous Hall (QAH) effect. The original QAH was proposed \cite{haldane} even earlier in a honeycomb lattice model soon after the discovery of integer and fractional quantum Hall effects (FQHE).
Since the first experimental realization of the quantum anomalous Hall (QAH) effect in Cr doped Bi(Sb)$_2$Te$_3$ thin films \cite{QAHthe,QAHexp},
it has also been observed in many other materials such as both Cr doped and V doped (Bi,Sb)$_2$Te$_3$ films \cite{QAHexp2}.
More recently, it was also discovered in the twisted bilayer graphene \cite{TBGAHE}.
It was realized with cold atoms  with the fermions $^{40}$K in  \cite{haldaneexp}.
The bosonic version of QAH was also implemented with spinor bosons $^{87}$Rb  in \cite{2dsocbec}.
The connections between the non-interacting fermionic QAH
and the interacting bosonic version of QAH and various quantum or topological phase transitions in interacting bosonic systems were explored in \cite{QAHboson}.

 It turns out that the FQHE and QAH are just two early members of the vast number of topological quantum matter \cite{kane,zhang,tenfold,wenrev}.
 The topological quantum matter is one of the main themes in condensed matter physics,
 lattice gauge theories and topological quantum field theories \cite{kane,zhang,tenfold,wenrev}.
 It not only contain new physical concepts, rich and profound phenomena with deep mathematical structures, but may also have some potential
 industry applications such as dissipationess  transmissions, quantum communications and  topological quantum computing.

 There have been flurries of classifications of both gapped and gapless topological phases \cite{kane,zhang,tenfold,wenrev}.
 In the known non-interacting gapless case, the gapless source comes from the Dirac fermions, Weyl fermions and various line/ring nodes in the bulk.
 We call this class of bulk gapless system with
 Fermi points or lines, linear dispersion $ \omega \sim k $ and vanishing density of states (DOS) $ D(\omega) \sim \omega $
 as having the dynamic exponent $ z=1 $.
 There are still little works on  how to characterizing a bulk topological gapless system with extended Fermi surface (FS), neither much work on
 topological phase transitions (TPT) between various topological phases.
 Furthermore, how to define topological invariants to characterize a bulk gapless system with extended FS and what are the  associated edge modes
 remain outstanding problems. Because these extended FS originates from point-like Fermi points with quadratic band touching
 $ \omega \sim k^2 $ and constant DOS $ D(\omega) \sim const. $  at a TPT,
 we call such class of gapless systems as having the bulk dynamic exponent $ z=2 $ which are the main focus of the present work.

\begin{figure}[tbhp]
\centering
\includegraphics[width=.6 \linewidth]{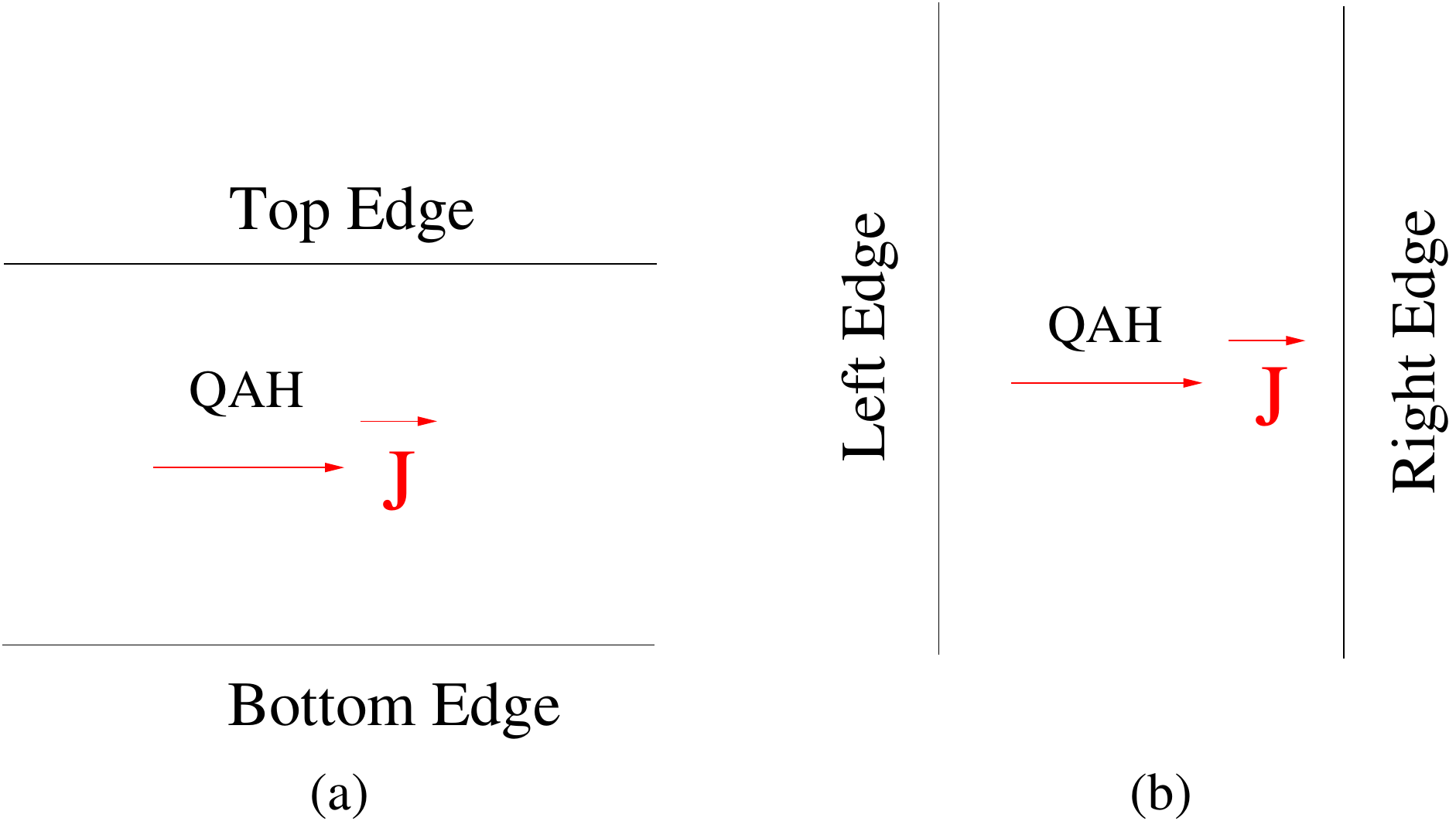}
\caption{
 New topological phases and bulk or edge  TPTs are generated by
 (a) Injecting a longitudinal current into the QAH sample in a strip geometry.
 (b) Injecting a transverse current into the QAH sample in a strip geometry.}
\label{frames}
\end{figure}

\begin{figure}[tbhp]
\centering
\includegraphics[width=.6 \linewidth]{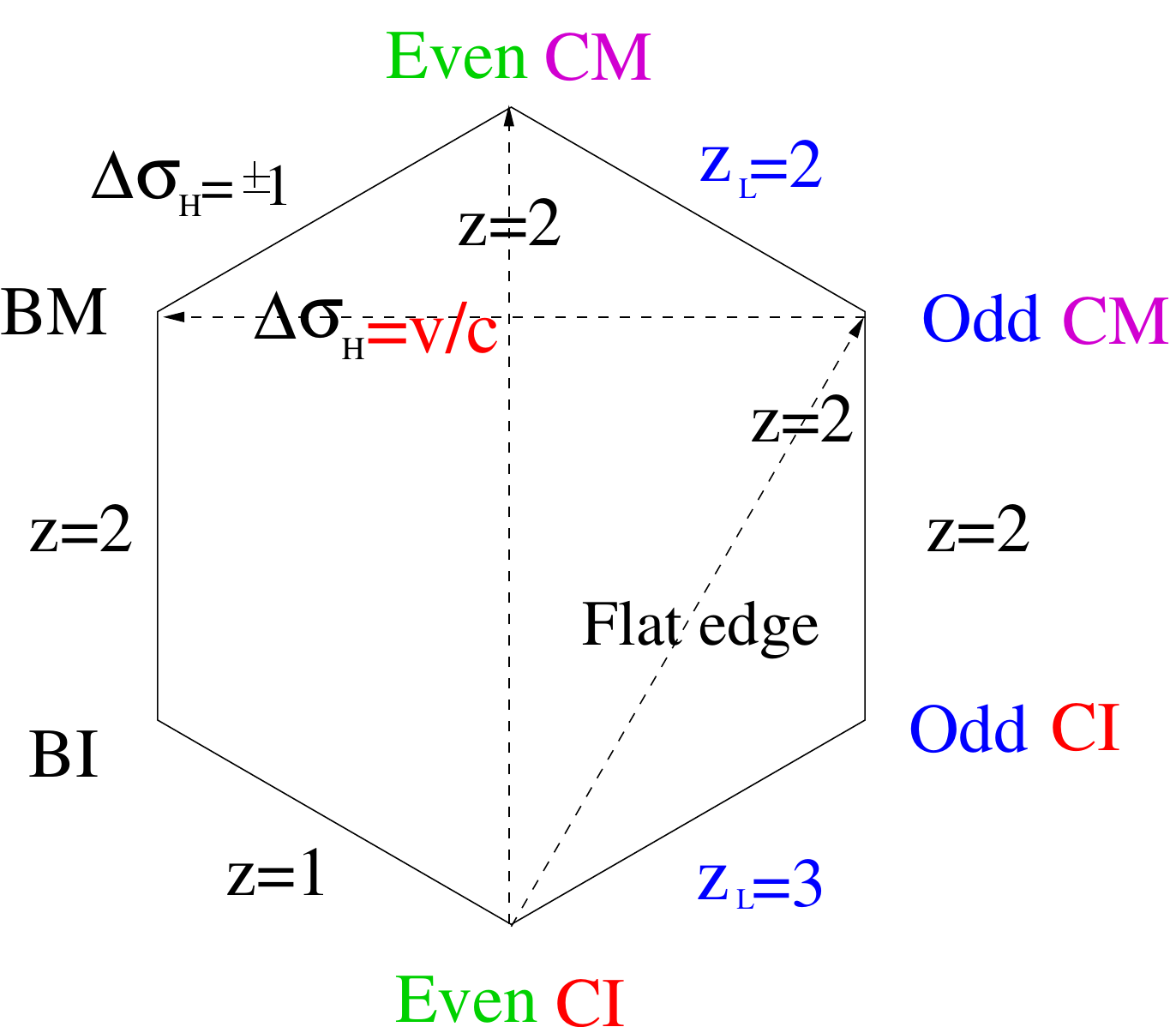}
\caption{ The six-fold hexagon phase diagram of the 6 quantum/topological phases:
 3 insulators: BI: band insulator, Even CI: even Chern insulator, Odd CI: odd Chern insulator;
 3 metals: BM: band metal, Even CM: even Chern metal,  Odd CM: odd Chern metal are sitting in the 6 vertices of the hexagon.
 $ z=1, 2 $ are the dynamic exponent of the bulk QPT,  $ z_L=2, 3 $ are the dynamic exponent of the longitudinal edge TPT.
 The bulk TPT from the even or odd  CM to the BM is infinite order with the unit
 jump in the Hall conductance $ \Delta \sigma_H= \pm 1 $ or non-integer $ \Delta \sigma_H= \pm v/c < 1 $ respectively.
 It is $ \Delta \sigma_H $ which plays the role of topological invariants in the 3 metal phases.
 It maybe the first example of a TPT which is not a QPT.
 The TPT/QPT from the even CI to the odd CM usually goes in two steps:
 even CI/odd CI/odd CM (the longitudinal-edge TPT happens earlier than the bulk)
 or even CI/even CM/odd CM (the L- edge TPT happens later than the bulk)
 with the odd CI or the even CM as the intermediate phase  respectively\cite{SSS},
 but it could also go directly (the L-edge TPT happens at the same time as the bulk QPT)
 with $ z=2 $ and completely flat L- edge shown in Fig.\ref{fig:phaseLattice}.
 The direct TPT/QPT may be viewed as a fine tuning  or a multi-critical point.
 The Transverse (T-) edge mode disappears always at the same time as the bulk TPT with its velocity vanishing as $ \sqrt{v^2-c^2} $.
 All the metallic phases have no T-edge mode.
 Due to the absence of $ C- $ symmetry, the even CM to the even CM transition is absent.
 The Universal Hall conductance jump from the odd CM to its T-reversal odd CM partner is twice of that from the odd CM to the BM.
 Some BM may also have non-vanishing Chern number and associated floating edge modes above the bulk modes. But practically, it is the same phase as the BM
 except it has a larger AHE. If no lines connecting two phases, then it means there is no direct transition between the two except through
 some fine tuned multi-critical point.}
\label{sixfolds}
\end{figure}

 There are two complementary approaches to address these outstanding problems.
 One way is to use SPT or SET to classify by various mathematical tools such as Co-homology, K-theory (for non-interacting fermions), Co-bordisms and tensor categories or its higher order versions \cite{tenfold,wenrev}.
 Another is to start from a concrete parent Hamiltonian hosting various topological phases, then
 explore its topological phases and TPTs under various deformations breaking different symmetries \cite{haldane,dimer1,dimer2,dimer3,Kit1,Kit2,tenfold}.
 Here we take the second approach, but limit to non-interacting fermionic systems\cite{haldane}.
 We focus on the simplest and widely experimentally realized topological phase:
 the quantum anomalous Hall (QAH) phase \cite{haldane,QAHthe,QAHexp,QAHexp2,haldaneexp,QAHboson} and also establish its connection to
 the un-quantized AHE.
 Specifically, we start from a QAH Hamiltonian which hosts some known gapped topological phases such as Chern insulator and band insulator
 and see how a Parity (P-) breaking  injecting current Fig.\ref{frames}a,b or a P- persevering chemical potential
 or energy dispersion drives the  parent QAH  Hamiltonian into new topological gapped or gapless phases through
 new classes of topological phase transitions (TPT).
 We also define the new "topological bulk invariants" of these $ z=2 $ topological gapless phase,
 investigate the associated new edge state structure, explore enriched bulk-edge correspondence
 and new longitudinal/transverse edge-edge correspondence. During this establishment,
 we propose a classification of the QAH insulators and AHE metals in Fig.\ref{sixfolds}.

This work contains 3 parts. In part I,
under an injecting current (Fig.\ref{frames}a,b), we evaluate the Hall conductivity at both zero and finite temperatures.
Then we study the non-analytical behaviours of the ground state energy
and map out the global phase diagram in the Zeeman field/injecting current plane Fig.\ref{fig:phaseLattice}.
 The 2 old phase turn into 4 phases: the old (even) Chern insulator with the Chern number ${\rm Ch}_{-}= \pm 1 $ and
 quantized Hall conductivity $ \sigma_{xy}= \pm 1 \times e^2/h $ when $ c< v $,
 a new $ z=2 $ Odd Chern metal (OCM) phase with ${\rm Ch}_{-}= \pm 1 $, extended Fermi surfaces (FS) and un-quantized
 universal Hall conductivity $ |\sigma_{xy}|=v/c < 1 $ (in the unit of $ e^2/ h $)  when $ c > v $.
 The emergence of the particle and hole FS in the Odd Chern metal phase is responsible for the
 reduced Hall conductivity  which is still independent of $ h/t $ and  many other microscopic details of the Hamiltonian.
 The old band insulator with  Ch$_{-}=0, \sigma_{xy}= 0 $ and  a new band metal phase with Ch$_{-}=0, \sigma_{xy}= 0 $
 with extended FSs. The TPT from the Chern insulator to the band insulator is a 3rd order with the dynamic exponent $ z=1 $,
 while both the Chern insulator to Odd Chern metal,
 and the band insulator to band metal are second order ones with $ z=2 $.
 Strikingly, the TPT from the Odd Chern metal to the band metal is novel: it has
 no non-analyticity to infinite order in the ground state energy, but the Chern number of the band jumps   $ \Delta {\rm Ch}_{-}= \pm 1 $
 and the Hall conductivity has a universal non-integer jump $ \Delta \sigma_{xy} =v/c $. This presents the first example of
 a TPT from both bulk and associated edge properties, but not a QPT in conventional wisdom \cite{sachdev,tqpt,weyl}.
 We also evaluate various thermodynamic quantities  such as the density of states, specific heat, compressibility and Wilson ratio
 at a finite $ T $ in all the 4 phases and also their quantum scaling forms near all these bulk TPTs.

Then we study the edge states in a strip geometry  when the injecting current is either parallel (longitudinal) or transverse to the edge,
from both the lattice model and from the effective low energy theory.
During the bulk TPT from the even Chern insulator to the Odd Chern metal,
the edge in a parallel injection also undergoes very unusual edge TPT:
Inside the Chern insulator $ c < v $ before the TPT,  the two edge modes
on the two opposite sides of the sample (Fig.\ref{frames}c) flow along the opposite directions $ v_L >0, v_R < 0 $,
but one side flow  slower than the other side $ |v_R| < |v_L| $. Only the edge mode contributes to $ \sigma_H= \pm 1 $.
At the TPT $ c=v $,  one edge mode becomes completely flat  with zero velocity $ v_R=0 $ which can be viewed as a fine tuning to
a multi-critical point (Fig.\ref{sixfolds}).
Inside the odd Chern metal  after the TPT  $ c > v $, the two opposite sides of the sample start to flow along
the {\em same} direction $ v_L >0, v_R > 0 $,
but one side flows much slower than the other side $ |v_R| \ll |v_L| $. Both bulk and edge contribute to the transport.
However, in a transverse boost, the edge mode velocity behaves as $ \sqrt{v^2 - c^2 } $ before the TPT $ c < v $,
but  were squeezed out after the TPT $ c > v $, so no T-edge modes in the Odd Chern metal anymore.
Only the bulk contributes to the transport. So the $ z=2 $ Odd Chern metal has the exotic edge modes with  $ v_Lv_R > 0 $ in a parallel injection (Fig.\ref{frames}a),
but not in a transverse injection (Fig.\ref{frames}b).
We may call this new phenomenon a new longitudinal ( L-) /transverse( T-) edge correspondence under the current injection.
As alerted in the last paragraph, the novel bulk TPT from Odd Chern metal to band metal can also be viewed from the edge in a parallel boost:
the former has one edge mode with $ v_L v_R > 0 $, the latter none.
Obviously, the Odd Chern metal with $ z=2 $ is clearly different from the previously known topological semi-metals
such as Dirac metal or Weyl semi-metal with $ z=1 $ \cite{tenfold}.

It maybe necessary to stress that one need to distinguish two mathematical quantities
which have different physical meanings inside different phases:
the Chern number $ Ch_{-} $ versus the Hall conductivity $ \sigma_{xy} $.
They are the same in the Chern insulator, but different in the Odd Chern metal:
the Chern number is defined for a band only, therefore independent of the filling of the band \cite{bosonicQAH}. Even so,
it still has a clear physical meaning when the boost is parallel to the edge in a strip geometry even in a Odd Chern metal:
it stands for the contribution from the edge states in both Chern insulator and Odd Chern metal.
Inside the Chern insulator $ c < v $, this is the only contribution to
the $ \sigma_{xy} $, but inside the Odd Chern metal $ c > v $, due to the gapless extended FS in the bulk,
the bulk states also contribute, so one have the decomposition $ \sigma_{xy}= 1 + ( v/c-1 )=v/c < 1 $
where the first term comes from the
edge state which is quantized, the second term is from the bulk FS which is un-quantized.
So when the boost is parallel to the edge, there is an enriched bulk-edge correspondence inside the topological Odd Chern metal:
the Chern number defined for the bulk band gives the quantized edge contribution, independent of the fillings of the band. But
the Hall conductivity $ \sigma_{xy} $ receives the total contribution from the edge + the bulk, depends on the fillings, so not quantized.
However, when the boost is perpendicular to the edge, the Chern number defined for a band only has mathematical sense, but no physical meaning
in the Odd Chern metal phase. In this case, there is no edge state anymore, so no contributions from the edge,
$ \sigma_{xy}= 0 +  v/c $ completely comes from the bulk.

In Part II which is on the gauge-invariant current injection case,
one need to consider the combined effects of the two currents,
the first is a NN  $ n=1 $ Non-Abelian gauge-invariant current term,
the second is a NNN $ n=2 $ (higher order) current term. The first term can be treated exactly  in a transformed basis  by
combining it with the NN hopping and SOC term in the original QAH Hamiltonian. It leads to a very counter-intuitive effect:
an band insulator near the TPT from the Chern insulator to the band insulator in the lab frame turns into a Chern insulator,
but not the other way around.
  In the transformed basis, if treating the NNN  $ n=2 $ current term  as an independent
one just like a NNN injecting current, the results achieved on NN $ n=1 $ injecting current
in the Part-I can be applied here with some notable differences:
(1) The Global phase diagram changes to Fig.\ref{fig:phaseLattice2} where one can see that
a new topologically gapped phase we named odd Chern insulator phase intervening between the
the even CI and the odd CM in Fig.\ref{fig:phaseLattice}. It has the same bulk properties
as the even Chern insulator, but with different edge properties. Its L- edge modes satisfy the exotic
relation $ v_L v_R > 0 $ similar to the odd CM, its T- edge modes satisfy the conventional relation
$ v_L v_R < 0 $ just as in the even Chern insulator.
So the direct TPT from the Even CI to the odd CM in Fig.\ref{fig:phaseLattice} splits into two in Fig.\ref{fig:phaseLattice2}:
In the L- edge, the edge mode undergoes
its own edge TPT from the even Chern insulator to the odd Chern insulator with an L- edge dynamic exponent $ z_L=3 $
before the bulk TPT from the odd CI to the odd CM.
(2) The TPT from the odd CI to the odd CM does not happen at a constant $ c=v $, but depends on the Zeeman field $ h/t $ in a lobe shape.
    The Odd Chern metal's Hall conductivity is not just given by $ v/c $, but also depends on the Zeeman field $ h/t $.
    The band metal's Hall conductivity is not zero anymore, but also depends on the Zeeman field $ h/t $.
(3) Remarkably, the Hall conductivity jump from the odd Chern metal to the band metal remains the same universal number as the $ n=1 $ case.
    So does the jump from the odd Chern metal to its T-reversal odd Chern number partner which is twice as that from the OCM to the BM.
    This could be a universal salient feature of the odd Chern metal (Fig.\ref{sixfolds}) !
    This fact establish the Hall conductivity jump as the new "topological invariants" characterizing the gapless topological phases with extended FS.
(4) Due to the $ n=2 $ NNN feature, the Doppler shift in the 4 nodes become the same sign.
    This is contrast to the $ n=1 $ NN injecting current case where the 4 nodes have 2+ and 2- Doppler shifts.

As a byproduct, taking some results from \cite{moving},
for the QAH or AHE due to the artificially generated  SOC which is a non-relativistic effect,
we find that a Galileo boost $ v $ on a lattice leads to the $ n=1 $ NN gauge-invariant current  and $ n=2 $ NNN current.
So the results achieved in Sec.V and Sec.VI can also be applied to a moving sample.
So if an insulator is band or Chern type may depend on if the observer is
moving relative to the lattice.
Then after absorbing the  $ n=1 $ NN gauge-invariant current into the QAH Hamiltonian by a unitary transformation,
the NNN current term in the transformed basis changes sign after some critical boost velocity $ v_c \sim 1 cm /s $ solely determined  by the Wannier functions.
So does the Doppler shift near the four nodes.  All these new features are subject to the scattering measurements in the moving frame in Fig.\ref{detector}.

 In terms of the SPT language, despite the original QAH Hamiltonian breaks the time-reversal symmetry explicitly,
 it still  has a charge (C-) conjugation symmetry and also a parity (P-) symmetry.
 An injecting current or a moving sample breaks P-symmetry, but keeps the C-symmetry.
 Because it is a non-interacting system, the C-symmetry is never broken during the evolution,
 so it is the C-symmetry protected topological phases and TPTs driven by the $ n=1 $ and $ n=2 $ current.

 In part III, we study the P-preserving deformation such as an energy dispersion.
 It leads to a even Chern metal phase which has the same bulk properties as the odd Chern metal.
 But the Universal Hall conductance jump from the even CM to the band metal is an integer number.
 It  also has a dramatically different L-edge mode properties than the odd CM:
 the L-edge mode satisfies the conventional $ v_L v_R < 0 $ instead of the exotic $ v_L v_R > 0 $.
 A real material contains both P-breaking and P-persevering components and  is  examined in Sec. VIII.
 We find that as the parameter changes, the generic AHE  will be either in even-like Chern metal  or odd-like Chern metal:
 there is a edge reconstruction between the two with a L- edge exponent $ z_l=2 $.
 We propose a complete classification of AHE metals leading to un-quantized QAH effect as
 the Band metal (BM), odd Chern metal and even Chern metal, while the gapped phases leading to quantized QAH effect as BI, CI
 and odd CI (Fig.\ref{sixfolds}).
 The BM is nothing but the previously well studied one contributing to the un-quantized AHE \cite{AHE2,Haldane2004}.
 While the itinerant metal contributing the AHE due to the Berry phase acquired by electrons moving
 in the non-coplanar spin texture in the real space in a Ferromagnet does not fall into this non-interacting classification \cite{AHE1}.

 Experimentalists got used to apply magnetic field, electric field, or strain, pressure,
 neutron scattering,  muon spin rotation, etc. Here, we show that injecting various forms of currents
 may be an effective way to bring out a lot of information on the topological phases, also drive them to new phases through novel TPTs.
 Alternatively, for SOC which is a non-relativistic effect,
 putting the sample in a strip shape to move in a trail, then perform various scattering experiments such as neutron,
 X-ray scattering or ARPES may also be helpful for artificially generated QAH systems.

The rest of the paper is organized as follows:
we will first study the QAH under a P-breaking injecting current ( Fig.\ref{frames}a,b ).
We will study the bulk properties of both systems
via lattice theory and continuum effective theory in the thermodynamic limit,
then investigate the corresponding edge properties in  a strip geometry in both longitudinal
and transverse edge via also both lattice theory and the continuum effective theory.
Then in the first two appendices, we investigate the QAH under a P-preserving chemical potential or energy dispersion
by the similar approaches.
The P-breaking and P-preserving  Hamiltonian are two different kinds of deformations leading to
different bulk  phases and topological phase transitions, also different edge properties.
A real material contains both and will be examined in Sec.VIII.
In Sec.IX, we summarize "Topological invariants" and the enriched bulk/L-edge/T-edge correspondence in gapless fermionic systems with extended Fermi surface.
The experimental detections are analyzed in Sec.X.

\section{The bulk properties: the microscopic lattice theory }

The quantum anomalous Hall model on a square lattice takes the form
\begin{align}
	&H_\text{QAH} =
	-\!\sum_i[c_i^\dagger (t\sigma_z-it_s\sigma_x) c_{i+x}
		\!+\!c_i^\dagger (t\sigma_z-it_s\sigma_y) c_{i+y}
		\nonumber   \\
	&\quad +h.c.]
	-h\sum_i(n_{i\uparrow}-n_{i\downarrow})
	+ U \sum_i n^2_i -\mu\sum n_i\>.
\label{QAH0}
\end{align}
Without loss of generality, we assume that $t>0$ and $t_s>0$. In this work, we focus on the non-interacting limit $ U=0 $,
but the chemical potential can be zero or non-zero.

\begin{figure}
    \centering
    \includegraphics[width=0.95\linewidth]{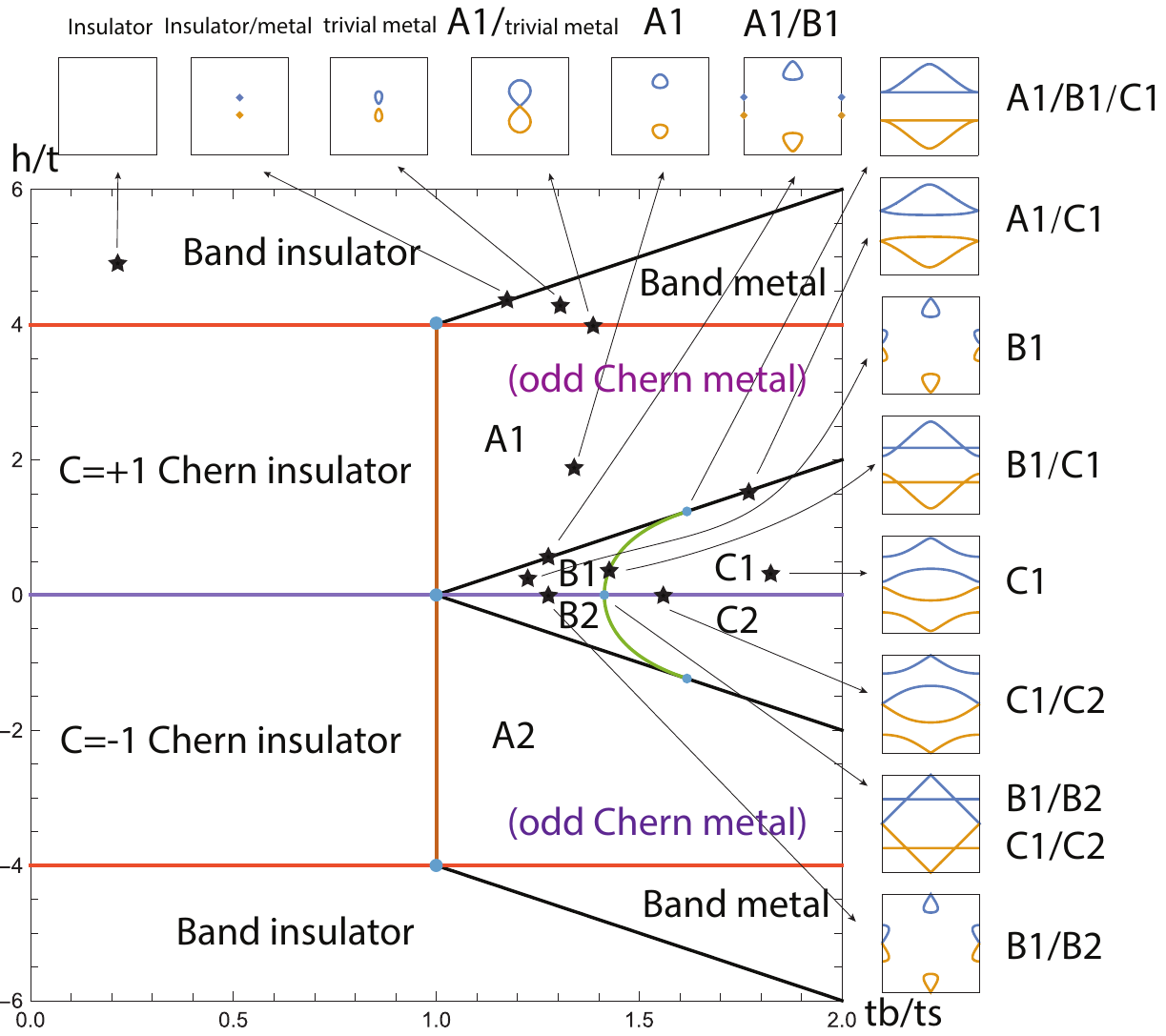}
    \caption{The global Phase diagram of the Lattice Hamiltonian Eq.\ref{eq:Boosted_H} with a fixed $t_s/t=1$.
    The phase diagram does depend on the ratio $t_s/t$. For simplicity, we take $t_s/t=1$ in this figure only.
    Then  the black phase boundary is a straight line. But $ t_s/t $ is taken as any value in the continuum effective actions.
    The topological metal phases are named as A1,A2,B1,B2,C1,C2. Green means the Particle (P-) FS, the yellow Hole (H-) FS.
    The inserts plot the Fermi surface of different metal phases within the 1st Brillouin Zone,
    where we use the notation A1/B1 for the phase boundary between A1 and B1, etc.
    The first box stands for the band  insulator where there is no FS, the second is the metal-insulator transition
    where there are two Fermi points (one P-, one H-) emerging  with the quadratic band touchings
    (so the dynamic exponent $ z=2 $), they evolve into two (one P-, one H-) Fermi pockets
    in the band metal. Then they touch quadratically at $ (0,0) $ in the A1-CM/BM  QCP with {\em not} any non-analyticity.
    A1 corresponds to Fig.9d, A1/B1 to Fig.13d.
    The P- FS expands to A1/C1, the bottom part of the FS becomes straight in A1/B1/C1. B1 corresponds to Fig.13e.
    The collision of the two P- FS from B1 leads to B1/C1. B1/B2 differs from B1 by the conic touching
    of P- and H- Fermi pocket at $(0,\pi) $ and $ ( \pi, 0) $ with {\em not} any non-analyticity.
    Then the split  between the P- and H- FS from C1/C2 leads to C1.
    See also Fig.\ref{fig:Dirac0}, \ref{fig:Dirac0_FS} and Fig.\ref{fig:Dirac12}, \ref{fig:Dirac12_FS}
    near $ h/t =\pm 4 $ and $ h/t = 0 $ respectively in the continuum calculation. See also Fig.23 for higher order $ n=2 $ boost. }
\label{fig:phaseLattice}
\end{figure}

  Under an injecting current (for its motivation from a Galileo transformation, see  appendix E), one obtains the injected Hamiltonian:
\begin{align}
	H_\text{inj}
	=H_{QAH}+\sum_{i}(it_{b,x} c_i^\dagger c_{i+x}+it_{b,y} c_i^\dagger c_{i+y}+h.c.)
\label{eq:Boosted_H}
\end{align}

The non-injected Hamiltonian Eq.\ref{QAH0} has the Charge conjugation C-symmetry and
the parity P-symmetry $ k_x \to -k_x, k_y \to -k_y; \sigma_x \to - \sigma_x,  \sigma_y \to - \sigma_y $.
but no Time Reversal T-symmetry.
The injected Hamiltonian Eq.\ref{eq:Boosted_H} breaks the P-symmetry, still respects the C-symmetry:
$ \sigma_x \mathcal{K}$,($\mathcal{K} $ denotes the complex conjugate),
\begin{align}
 (\sigma_x \mathcal{K})H(\mathbf{k})(\sigma_x \mathcal{K})=-H(-\mathbf{k})
\label{Csymmetry}
\end{align}
The C-symmetry guarantees a relation between upper band and lower band $E_+(\mathbf{k})=-E_-(-\mathbf{k})$,
and the Berry curvature $\Omega_+(\mathbf{k})=-\Omega_-(-\mathbf{k})$.
We also have $\Omega_{\pm}(\mathbf{k})=\Omega_{\pm}(-\mathbf{k})$.
Note that the QAH belongs to the Class A, but still with the particle number conserved.
Because this C-symmetry does not conserve the particle number,
so it can not be understood  as the existence of an anticomutating symmetry operators,
otherwise the system would belong to the Class D instead of the Class A \cite{tenfold}.

For simplicity, we study the injection along the $y$ direction,
thus $t_{b,x}=0$ and $t_{b,y}=t_b$. It can be easily generalized to any injection direction.
In momentum space, the Hamiltonian becomes
\begin{align}
	H_\text{inj}
	=\sum_{k}c_k^\dagger\{&\!-[h+2t(\cos k_x+\cos k_y)]\sigma_z
    +2t_s\sin k_x\sigma_x
    \nonumber\\
	&\!
    +2t_s\sin k_y\sigma_y
	-2t_b\sin k_y\sigma_0\}c_k
\label{eq:Boosted_Hk}
\end{align}
The Diagonalization of Eq.\eqref{eq:Boosted_Hk} leads to the two bands
\begin{align}
	&E_\pm(\mathbf{k})
	=-2t_b \sin k_y\nonumber\\
    &\pm\sqrt{[h+2t(\cos k_x\!+\!\cos k_y)]^2\!+\!4t_s^2(\sin^2k_x\!+\!\sin^2k_y)}
\label{pmbands}
\end{align}

Since $E_+(\mathbf{k})\geq E_-(\mathbf{k})$ always holds for a fixed $\mathbf{k}$,
we will call the $E_+$ the upper band and the $E_-$ the lower band.
When $t_b$ is sufficiently small,
it is in a insulating phase;
When $t_b/t_s$ is sufficiently large,
it is in a metallic phase,
with hole surface is given by $E_-(\mathbf{k})=0$
and electronic surface  given by $E_+(\mathbf{k})=0$.

The critical $t_b$ are determined by the minimization problem
$\min_k E_+(\mathbf{k};t_b)=0$. In the full range of $t_b/t_s$, the Fermi surfaces (FS ) can be rather complicated, see Fig.\ref{fig:phaseLattice}.

When $4t>h>0$, it is in the Chern insulator phase. At the critical $t_{b}/t_s= 1 $ and
the Fermi points are located at $(\pi,\arccos(\frac{2t-h}{2t}))$, it moves into the A1 Odd Chern metal phase,
then as $ t_b $ increases further to $t_{b}=\sqrt{t_s^2+h(h+4t)/4}$, the previous Fermi points grow-up into a Fermi surface
with the emergence of the other Fermi point at $(0,\arccos[\frac{-t(2t+h)}{2(t^2+t_b^2-t_s^2)}])$ ( See A1/B1 ),
it moves into the B1 Odd Chern metal phase.
When $t>t_b$ increases further, these two Fermi surfaces can collide at B1/C1 to move into the  C1 phase.

When $h>4t$, it is in the band  insulator phase. At the critical $t_b=\sqrt{t_s^2+h(h-4t)/4}$,
the Fermi points are located at $(0,\arccos(\frac{2t}{h-2t}))$, it moves into the band metal phase.

    The TPTs in Fig.\ref{fig:phaseLattice} can be classified into 4 classes:
    (1) The linear band touching due to the Dirac points are 3rd TPT with $ z=1 $.
    In the $ t_b=0 $, it is a Dirac point with emergent Lorentz invariance. The $ t_b \neq 0 $ drives it into a boosted Dirac point.
    (2) The emergency of the P- or H- Fermi point in insulator/metal, A1/B1 are quadratic band touching
    2nd order TPT with the dynamic exponent $ z=2 $.
    (3) Band metal/A1, B1/B2,C1/C2, even the M point B1/B2 (C1/C2) can be understood as the
    the conic band touching between the P- and the H- FS. They do {\em not} have any non-analyticity in the ground state energy.
    (4) The P-/P- FS ( equivalently H-/H- FS ) collision  in A1/C1, B1/C1 are TPT with universal sub-leading scalings \cite{weyl,class3}.

\begin{widetext}
\begin{center}
\begin{table}[!tbhp]
    \setlength{\tabcolsep}{10pt}
        \caption{ The classification of bulk QPT or TPT.}
        \begin{tabular}{c|c|c|c|c}
        \hline\hline
        4 classes TPTs    &  CI-BI   & CI-OCM  &  OCM-BM  & P-FS/P-FS collision    \\  \hline
        Dynamic exponent     &   $ z=1 $    &    $z=2 $     &    No    &  saddle point cone    \\
        Order                &   3rd     &      2nd     &   Infinite order    &   3rd, 5th..       \\
        Scaling                &   Yes     &      Yes     &   No   &   sub-leading      \\
       $R_W=T\kappa_u/C_v$   &   $ \frac{2\ln2}{9\zeta(3)} $    &  $ \frac{3}{\pi^2}  $
        &   $ \frac{3}{\pi^2}$   &   sub-leading      \\
        \hline\hline
        \end{tabular}
\end{table}
\end{center}
\end{widetext}
  where we only list one representative of  CI-BI, CI-CI in the second column, CI-OCM, BI-BM, OCM-OCM in the 3rd column,
  OCM-BM, OCM-OCM in the 4th column, H-FS/H-FS in the 5th column. The Odd CM can also be replaced by the Even CM except
  the ECM-ECM transition does not exist as demonstrated in Sec.VII and VIII ( See Fig.\ref{evenoddphase} and Fig.\ref{even2phase} ).
  The Wilson ratio (WR) was evaluated in the lattice in Sec.II-C-3 and in the continuum effective theories in Sec.III-A-3c and III-B-3c.
  The WR for all the gapped phases ( CI and BI ) $R_W=(T/\tilde{\Delta})^2$ ( See Eq.\ref{tildeD} ) is not listed in the Table.
  In fact, $R_W=\frac{3}{\pi^2} $ also holds inside the all the gapless phases ( CM and BM ), because the OCM to the BM
  transition has no analyticity anyway.
  We did not list the edge reconstruction transition from the  CI to odd CI in Fig.23 and even CM to odd CM in Fig.\ref{evenoddphase} with the  longitudinal dynamic exponent $ z_L=3 $  and $ z_L=2 $ respectively. For the bulk or edge properties of these phases, see Table II.

In fact, as elucidated  in Fig.\ref{fig:phaseLattice}, if one look at
$ B1 \to B1/C1 \to C1 $,  the B1/C1 QCP is reached by the collision of the two P- FS (or equivalently, the two H-FS)
from B1, so this class of TPT can be similarly investigated by the method developed in \cite{weyl}, universal {\em subleading} scalings
can be derived. One can also similarly study the TPT from $ A1 \to A1/C1 \to C1 $.
In the following, we are mainly interested in the experimentally most relevant case $t_b/t_s$ is not too large,
so focus on A1,A2,B1,B2 phases and class-1 and class-2 TPTs, but do not discuss C1 and C2 phase and the class-3 TPT in any details.
The A1,A2,B1,B2 phases can be distinguished by $\sigma_{xy}$ and Fermi surfaces (FS) topology:
The A1 phase has $0<\sigma_{xy}<1$ and the FS is just one part,
the B1 phase has $0<\sigma_{xy}<1$ and the FS consists of two disconnected parts.
A2(B2) have the same FSs as A1(B1), but with opposite sign of $\sigma_{xy}$.

\subsection{ The quantum Hall response at zero and finite temperature: Topological phase transitions (TPT). }


In order to calculate Berry connections $\mathbf{A}$ and Berry curvatures $\Omega$,
it is convenient to rewrite the boosted QAH Hamiltonian Eq.\eqref{eq:Boosted_H} in terms of $ ( d_0, \vec{d} ) $ vectors:
\begin{align}
    &H(k)=d_0(k)\sigma_0+d_x(k)\sigma_x+d_y(k)\sigma_y+d_z(k)\sigma_z, \nonumber  \\
    &d_0(k)=-2t_b\sin k_y,\>\mathbf{d}=(d_x,d_y,d_z)  \nonumber   \\
    &d_x(k)=2t_s\sin k_x,\>
    d_y(k)=2t_s\sin k_y,\>
                         \nonumber   \\
    & d_z(k)=-[h+2t(\cos k_x+\cos k_y)]
\label{dvector}
\end{align}

 Then the Berry Connections and Berry curvatures can be evaluated as:
\begin{align}
    A_{\pm,i}(\mathbf{k})
        &\!=\!i\langle \pm,\mathbf{k}|\partial_{k_i}| \pm,\mathbf{k}\rangle
        =\frac{(d_y\partial_{k_i} d_x-d_x\partial_{k_i} d_y)}{2|\mathbf{d}|(|\mathbf{d}|\pm d_z)}  \nonumber  \\
    \Omega_{\pm,xy}(\mathbf{k})
        &\!=\!\partial_{k_x}\! A_{\pm,y}\!-\!\partial_{k_y}\! A_{\pm,x}
        \!=\!\mp\frac{1}{2|\mathbf{d}|^3}\mathbf{d}\!\cdot\! \partial_x \mathbf{d}\!\times\! \partial_y \mathbf{d}
\label{Aomega}
\end{align}
Since the $t_b$ term is proportional to the unit matrix $ \sigma^0 $, so it
does not affect the eigenvectors, then
the Berry connections and Berry curvatures are exactly the same as $t_b=0$ case.
Note that $\sigma_{xy}$ is related to $\Omega_{\pm,xy}$
and $\sigma_{yx}=-\sigma_{xy}$ is related to $\Omega_{\pm,yx}$.
For later calculations on Hall response, we will only consider $\sigma_{xy}$ and drop subscript $xy$ in $\Omega$.

\subsubsection{Zero temperature Hall conductance and TPT}

As long as $h\neq 0,\pm 4t$, the upper band and the lower band are well separated,
thus one can calculate the Chern number of the lower band via integrating the Berry curvature $\Omega_-(\mathbf{k})$
over the entire Brillouin zone (BZ) which is a torus $\mathbb{T}^2$:
\begin{align}
	{\rm Ch}_-
    \!=\!\frac{1}{2\pi}\int_{\mathbb{T}^2} d^2\mathbf{k}\> \Omega_-(\mathbf{k})
	=\begin{cases}
		0, & |h/t|>4;\\
		+1,& 4>h/|t|>0;\\
		-1,& 0>h/|t|>-4;\\
	\end{cases}
\label{eq:Ch}
\end{align}

In the insulating phase, only the lower band is full occupied,
thus the Chern number is the zero temperature Hall conductance in unit $e^2/h$,
that is $\sigma_{H}={\rm Ch}_-$.
In the metallic phase, both band are partially filled,
thus the zero temperature Hall conductance reduces to $\sigma_{H}={\rm Ch}_- \times |t_s/t_b|$.

 In any case, the zero temperature Hall conductance can be expressed as
\begin{align}
    \sigma_{H} & =\frac{1}{2\pi}\int_{\mathbb{T}^2} d^2\mathbf{k}\>
    \sum_{s=\pm}\Omega_s(\mathbf{k})\Theta(-E_s(\mathbf{k}))
     \nonumber   \\
	& ={\rm Ch}_-\times \min(1,|t_s/t_b|)\>.
\label{eq:sigma_xy}
\end{align}

\begin{figure}[!tbhp]
    \centering
    \includegraphics[width=\linewidth]{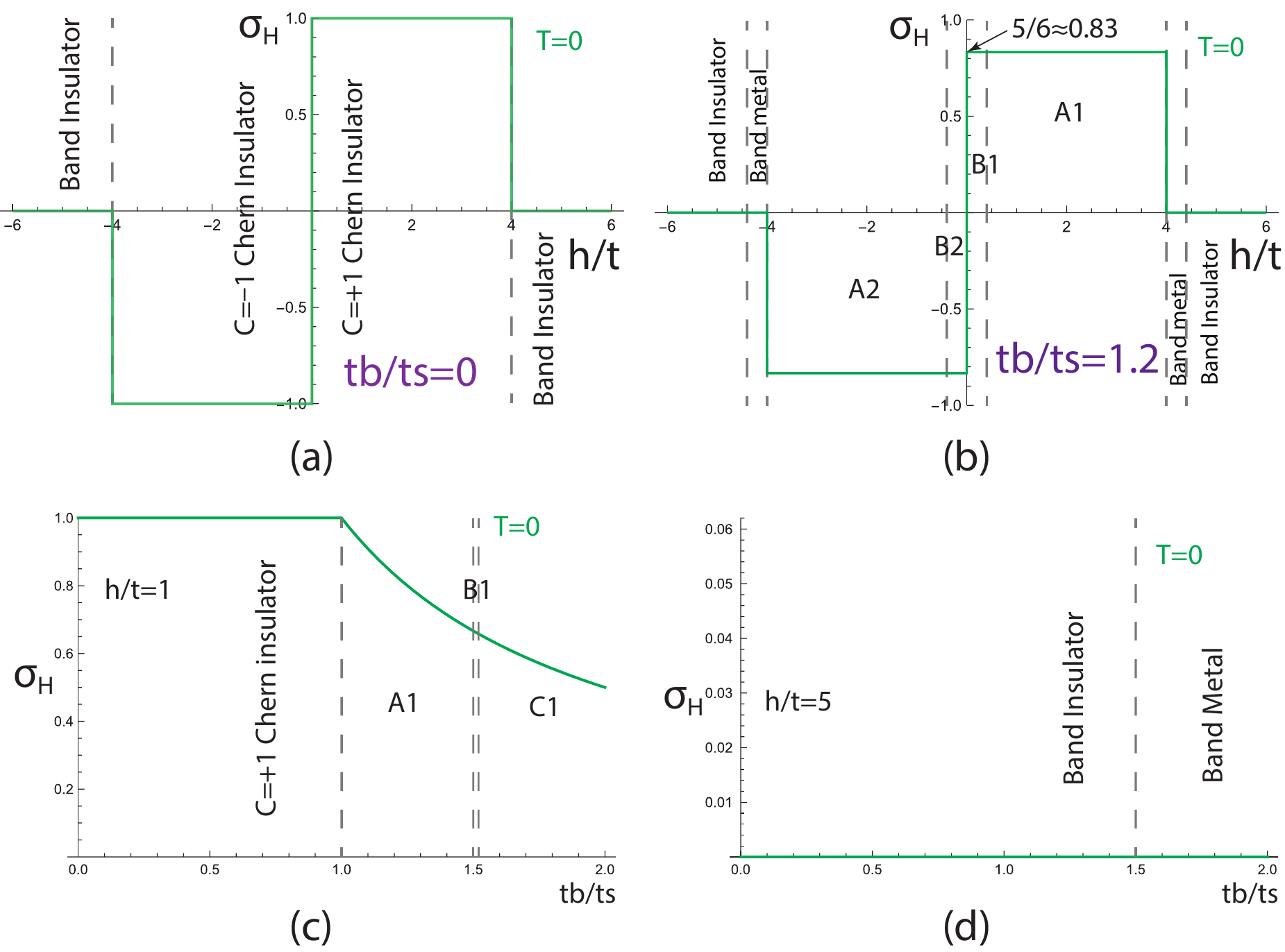}
    \caption{The zero temperature $ T=0 $ Hall conductance $\sigma_{H}$  as a function of $h/t$
    for fixed $t_b/t_s=0,1.2$ in (a) Across the 3rd order TPT with $ z=1 $ from  BI/CI, CI/CI, CI/BI,
	the $\sigma_{H}(T=0)$ has an integer jump $ \Delta \sigma_H=-1, 2, 1 $ respectively,
    (b)  As shown in Fig.\ref{fig:E_Lattice2}, there is no non-analytical behaviours ( infinite order ) for
	the BM/A2, B1/B2, A1/BM TPTs,
	but $\sigma_{H}(T=0)$ still has a universal non-integer jump $ \Delta \sigma_H= -t_s/t_b=-1/1.2=-5/6, 2 \times 5/6, 5/6 $
    respectively, independent of any other microscopic details.
	When across the 2nd order TPT with $ z=2 $ from the BI/BM, A2/B2, and A1/B1,
    $\sigma_{H}(T=0)$ has no changes.
    In a generic case, the BM should contribute to a un-quantized AHE ( See Fig.24 and Table 1 ),
    but it vanishes in this particular $ n=1 $ case due to some fine tuning.
    As a function of $t_b/t_s$ for fixed $h/t=1,5$ in (c) a 2nd order QPT with $ z=2 $ and
    (d) $ \sigma_H=0 $ in both the BI and BM.
	We expect the class-3 TPT of B1/C1 and A1/C1 \cite{weyl} has no changes
    in $\sigma_{H}(T=0)$ either. See also Fig.\ref{fig:sigma_C1} and Fig.\ref{fig:sigma_C2}
    near $ h/t =\pm 4 $ and $ h/t = 0 $ respectively in the continuum calculations.  }
    \label{fig:Hall_T0}
\end{figure}

As shown in Fig.\ref{fig:Hall_T0}, the Hall conductivity show plateau structure
during the scanning of $h$: it is zero for band  insulator and
band metal,
show a plateau with value $-t_s/t_b$ for A2 phase and B2 phase,
show a plateau with value $t_s/t_b$ for  A1 phase and B1 phase,
becomes zero again for band  insulator and band metal.

We conclude that the zero temperature Hall conductance in the metallic phase
is reduced  by a factor $|t_s/t_b|<1$ relative to its quantized value in the Chern insulator phase ( Fig.\ref{fig:Hall_T0} ).
In both the insulting phase and metallic phase,
they are not that sensitive to the microscope details.
As to be shown in the following sections, they can all be reproduced
via the  analytical evaluations of relevant integrals in the continuum theory.

\subsubsection{Finite temperature Hall conductance}

\begin{figure}[!tbhp]
    \centering
    \includegraphics[width=\linewidth]{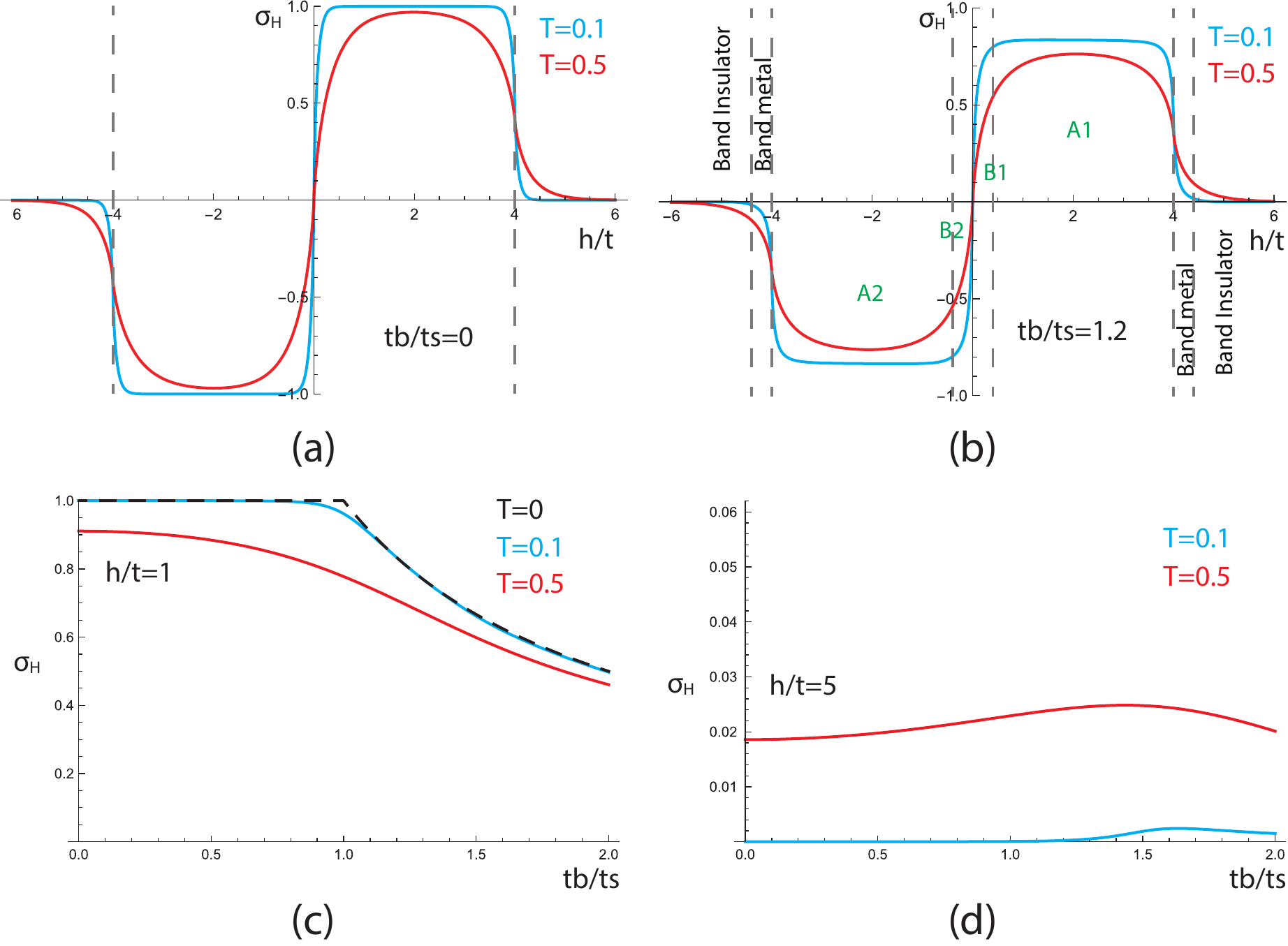}
    \caption{The finite temperature $ T > 0 $  Hall conductance $\sigma_{H}$ as a function of $h$ for fixed $t_b/t_s=0,1.2$
    in (a) and (b) and as function of $t_b/t_s$ for fixed $h/t=1,5$ in (c) and (d),
	where we also choose $t=t_s=1$, thus $T=0.1$ means $k_BT/t=0.1$.  }
    \label{fig:Hall_T1}
\end{figure}

At finite temperature, one only need to replace the step function in Eq.\ref{eq:sigma_xy}  by the Fermi distribution function
\begin{align}
    \sigma_{H}(T)
    &=\frac{1}{2\pi}\int_{\mathbb{T}^2} d^2\mathbf{k}\,
    \sum_{s=\pm}\Omega_s(\mathbf{k})f(E_s(\mathbf{k}))
                  \nonumber   \\
    &=\frac{1}{2\pi}\int_{\mathbb{T}^2} d^2\mathbf{k}\,
    \Omega_-(\mathbf{k})[f(E_-(\mathbf{k}))-f(E_+(\mathbf{k}))]   \nonumber  \\
    &={\rm Ch}_- + \frac{1}{\pi} \int_{\mathbb{T}^2} d^2\mathbf{k}\Omega_+(\mathbf{k})f(E_+(\mathbf{k}))
\label{finiteTsigmaH}
\end{align}
where $f(E)=1/[\exp(E/T)+1]$.
The finite $ T $ Hall conductances as varying parameters of $ h/t $ or $ t_b/t_s $ are plotted in Fig.\ref{fig:Hall_T1}

\subsection{ The ground-state energy and Quantum phase transitions (QPT) }

We are interested in the non-analytical behaviours in the ground-state energy density on the lattice
which can be numerically calculated via
\begin{align}
	E_\text{GS}^\text{(lat)}
	\!\!=\!\!\int_{\mathbb{T}^2} \!\!\frac{d^2\mathbf{k}}{(2\pi)^2}
	[E_{+}(\mathbf{k})\Theta(-E_{+}(\mathbf{k}))
    +E_{-}(\mathbf{k})\Theta(-E_{-}(\mathbf{k}))]
\label{eq:E_lattice}
\end{align}
The quantum phase transitions can be driven either by tuning by the Zeeman field $h/t$ or the boost $t_b/t$.

\subsubsection{QPTs driven by the boost $t_b/t $}

\begin{figure}[!tbhp]
    \centering
    \includegraphics[width=\linewidth]{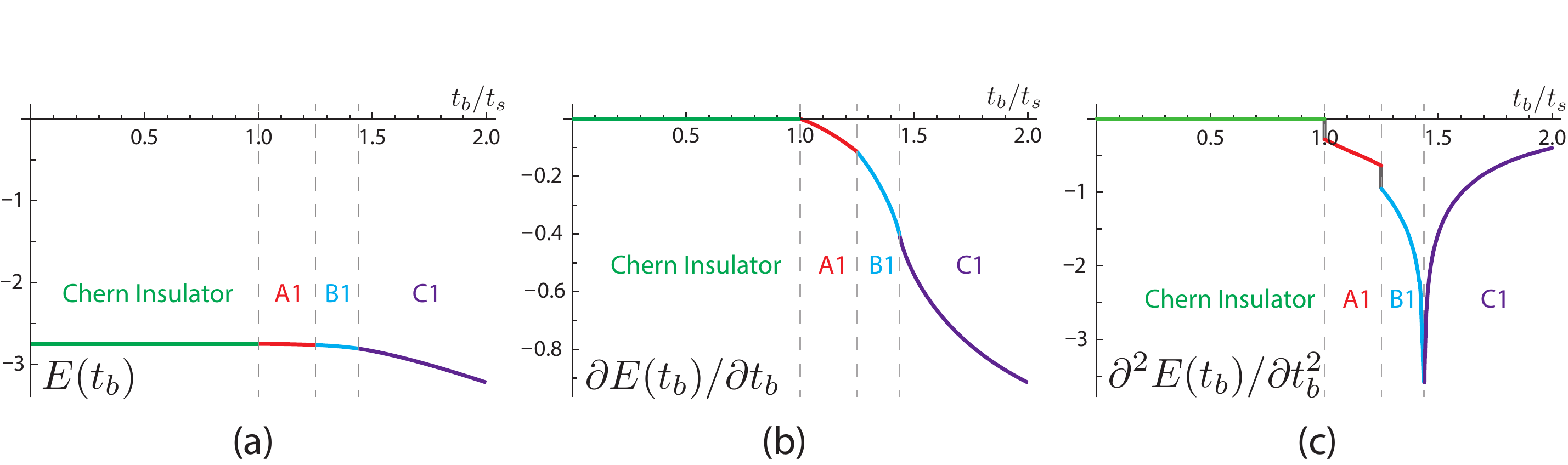}
    \caption{The ground-state energy density $E_\text{GS}^\text{(lat)}$ as the function of $t_b/t$,
	with fixed $t=t_s=1$ and $h=0.5$ in Fig.\ref{fig:phaseLattice}.
	From (a),(b),(c) are $E_\text{GS}^\text{(lat)}$'s zeroth-/first-/second-order derivative with respect to $t_b$.
	The vertical dashed lines correspond to the critical $t_b$, which separate
	the Chern insulator, A1,B1 and C1.
	An obvious second order QPT discontinuity is shown at $t_b=1.0,1.25$ both with $ z=2 $,
    the cusp at $ t_b =1.44 $ shows a third order one in the class-3 \cite{weyl} respectively. See also table I.  }
\label{latticetb1}
\end{figure}

\begin{figure}[!tbhp]
    \centering
    \includegraphics[width=\linewidth]{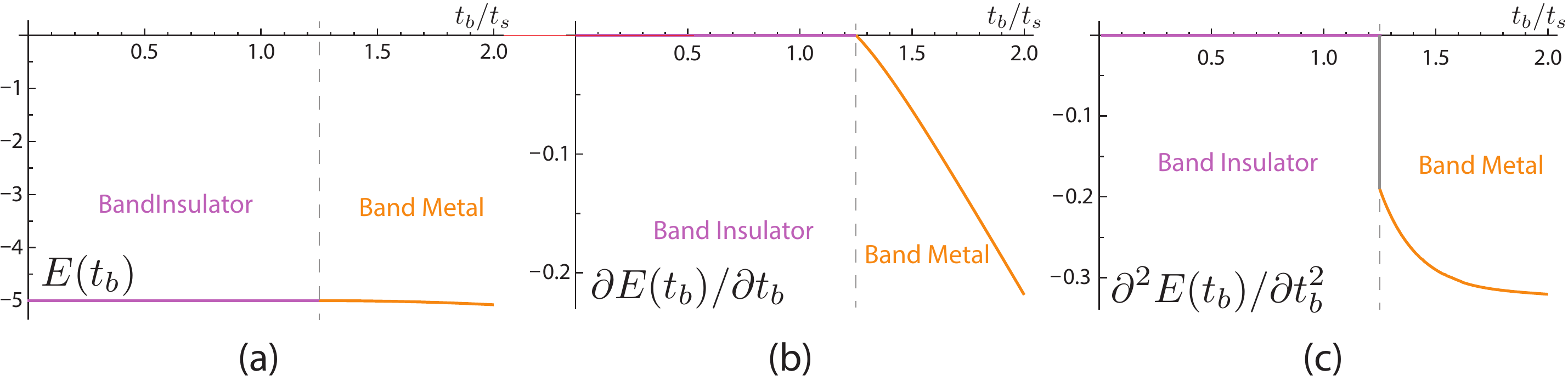}
    \caption{The same as Fig.\ref{latticetb1}, but with $h=4.5$ in Fig.\ref{fig:phaseLattice}.
	The only vertical dashed line corresponds to the critical $t_b$ which separates
	the band  insulator from the band metal. An obvious second order QPT discontinuity with $ z=2 $ is shown at $t_b=1.25$.  }
\label{latticetb2}
\end{figure}

In the insulating phase, the lower band is full occupied, the ground-state energy density is
\begin{align}
	E_\text{GS}^\text{(lat)}
	 \!=\!\!\int_{\mathbb{T}^2} \!\frac{d^2\mathbf{k}}{(2\pi)^2}E_{-}(\mathbf{k};t_b)
	 \!=\!\!\int_{\mathbb{T}^2} \!\frac{d^2\mathbf{k}}{(2\pi)^2}E_{-}(\mathbf{k};t_b\!=\!0)
\label{groundinsulator}
\end{align}
 where the last equality is due to that the $t_b$-dependent part in Eq.\ref{pmbands} is odd in $ \vec{k} $, so
 vanishes after integration over entire Brillouin zone (BZ).
Thus the $E_\text{GS}^\text{(lat)}$ is independent of $t_b$ in the insulating phase.
However, in the metallic phase, due to the partial filling of the upper and lower bands, the $E_\text{GS}^\text{(lat)}$ does  depend on  $t_b$.
Thus near the insulating-metallic transition driven by $t_b$ at some critical value $t_{b,c}$,
the ground-state energy density  $E_\text{GS,I}^\text{(lat)}=const$ in the insulating side $(t_b<t_{b,c})$,
$E_\text{GS,M}^\text{(lat)}\propto (t_b-t_{b,c})^\alpha$ in the metallic side $(t_b>t_{b,c})$.
The index $\alpha>0$ indicates the order of the phase transition is $\lceil\alpha\rceil$.
For example,
if $\alpha\leq 1$, then is the first order transition with a cusp at $ t_b=t_{b,c} $;
if $1<\alpha\leq 2$, then is the second order QPT shown in Fig.\ref{latticetb1},\ref{latticetb2}.

Because any phase transitions are independent of how they are approached or scanned, so scanning $ t_b/t $ or $ h/t $
should research consistent results. That is indeed the case as shown in the following section.

\subsubsection{QPTs driven by $h/t$}

\begin{figure}[tbhp]
    \centering
    \includegraphics[width=\linewidth]{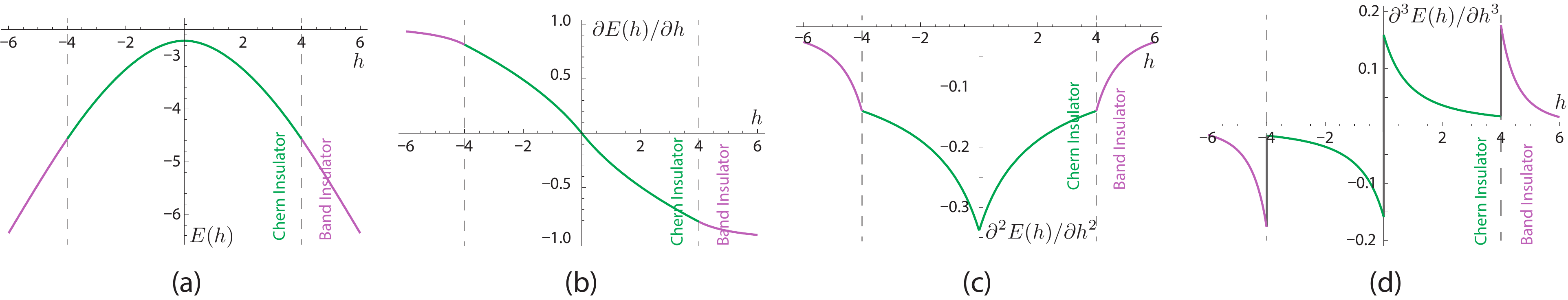}
    \caption{The ground-state energy density $E_\text{GS}$ as function of $h$,
	with fixed $t=t_s=1$, $t_b=0$.
	From left to right are $E_\text{GS}$'s 0th-/1st-/2nd-/3rd-order derivative with respect to $h$.
	The vertical dashed lines correspond to the critical $h$, which separate
	band  insulator, Chern insulator (-1), Chern insulator  (+1), band  insulator.
	An obvious 3rd order QPT discontinuity with $ z=1 $ is shown at $h/t=-4,0,+4$.  }
    \label{fig:E_Lattice1}
\end{figure}

We first study the TPT between band  insulator and Chern Insulator.
Because $ E_{GS} $ is $t_b$ independent in the insulating phases, so the TPTs are the same as the no-boost $t_b=0$ case.
In Fig.\ref{fig:E_Lattice1} 
we numerically evaluate the ground-state energy Eq.\eqref{groundinsulator} as a function of
scanning $ -6 < h/t < 6 $ with fixed $t_b=0$ which is identical to the  $t_b/t < 1 $ case.
An obvious third order discontinuity is shown between
the band  insulator and the Chern insulator.

\begin{figure}[tbhp]
    \centering
    \includegraphics[width=\linewidth]{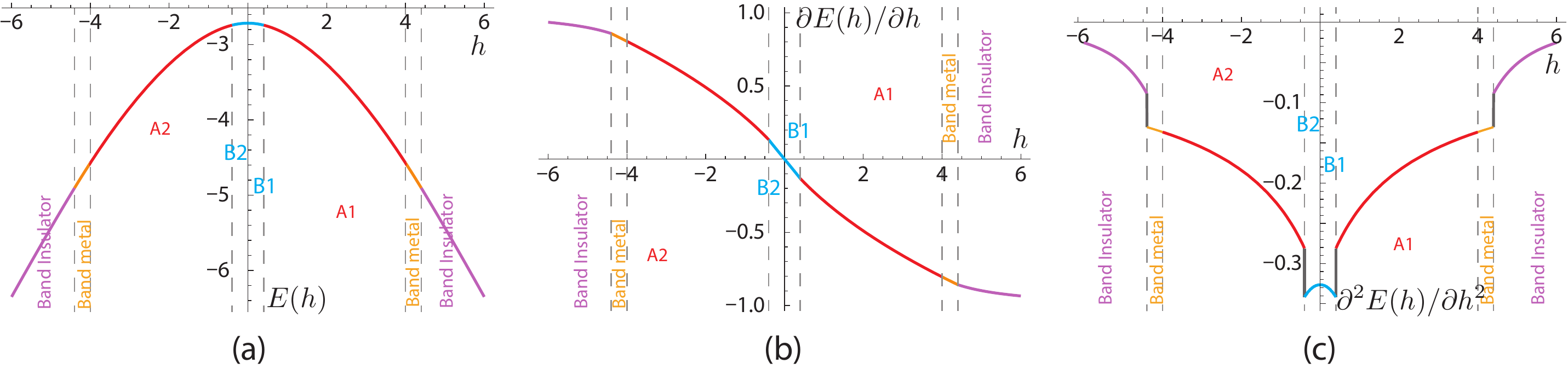}
    \caption{The ground-state energy density $E_\text{GS}$ as function of $h$,
	with fixed $t=t_s=1$, $t_b=1.2$.
	From left to right are $E_\text{GS}$'s 0th-/1st-/2nd-/3rd-order derivative with respect to $h$.
	The vertical dashed lines correspond to the critical $h$, which separate
	band  insulator, tribal metal, A2, B2, B1, A1, band metal, band  insulator.
	A clear 2nd order QPT discontinuity with $ z=2 $ is obtained between  band  insulator/band metal
	A2/B2, A1/B1 phase. However, no non-analytical behaviours are found for the
	BM/A2, B2/B1, A1/BM transitions, even in the third-order derivatives and beyond.
    The underlying physical mechanisms for these infinite order TPT are explored from the continuum effective theory in Sec.III-A and III-B.    }
    \label{fig:E_Lattice2}
\end{figure}

Now we study TPTs between metals where  $ E_{GS} $ becomes $t_b$ independent.
Scanning $ -6 < h/t < 6 $ with fixed $t_b>t_s$, we meet consecutively
band  insulator, band  metal, A2 Odd Chern metal, B2 Odd Chern metal,
B1, A1, band  metal, band  insulator.
In the Fig.\ref{fig:E_Lattice2}, a clear second order discontinuity appears between  band  insulator/band metal
A2/B2 phase, and A1/B1 phase. However, no any order discontinuity is found
The band  metal/A2, B2/B1, A1/band  metal transitions, so they
could be just infinite-order TPT. But they can still be distinguished by the Hall conductance shown in Fig.\ref{fig:Hall_T0}.

\subsection{ Thermodynamic Quantities}

 We will first discuss the density of states ( DOS ), then use it to compute several experimentally measurable quantities.

\subsubsection{The density of states (DOS) }

From the Hamiltonian \eqref{dvector}, one can find the Matsubara Green's function
\begin{align}
    G(\mathbf{k},i\omega_n)&=[i\omega_n-H(\mathbf{k})]^{-1}
                    \nonumber   \\
    &=\frac{P_+(\mathbf{k})}{i\omega_n-E_+(\mathbf{k})}
    +\frac{P_-(\mathbf{k})}{i\omega_n-E_-(\mathbf{k})}
\end{align}
where $P_\pm=\frac{1}{2}[1\pm\mathbf{d}\cdot\boldsymbol{\sigma}/|\mathbf{d}|]$ are
the projection operators onto the $s=\pm$ upper/lower bands
and $P_+ + P_{-}=1$.

The total DOS $ D(\omega)=D_{+}(\omega) + D_{-}(\omega) $ is
\begin{align}
    D(\omega)& =-\frac{1}{\pi}\int\frac{d^2\mathbf{k}}{(2\pi)^2} \Im\Tr[G_R(\mathbf{k},\omega)]
    \nonumber   \\
    & =\int\frac{d^2\mathbf{k}}{(2\pi)^2}[\delta(\omega-E_{+}(\mathbf{k}))+\delta(\omega-E_{-}(\mathbf{k}))]
\end{align}

It should be convenient to introduce the DOS for each band
\begin{align}
    D_{\pm}(\omega)=\int \frac{d^2\mathbf{k}}{4\pi^2}\delta(\omega-E_{\pm}(\mathbf{k})),
\end{align}

\begin{figure}[tbhp]
    \centering
    \includegraphics[width=\linewidth]{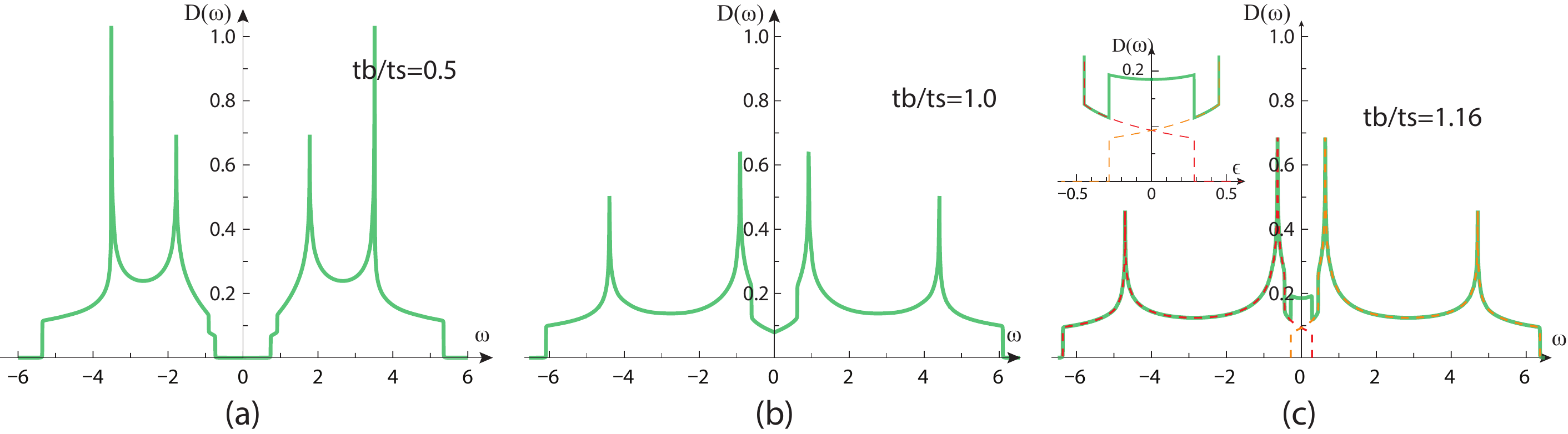}
    \caption{ At fixed $t=t_s=1$ and $h=1$, the density of states (DOS) of Eq.\eqref{eq:Boosted_Hk} at different $ t_b/t_s $ values:
    (a) $t_b/t_s=0.5$ in Chern insulator,
    (b) $t_b/t_s=1.0$ on the QCP between Chern Insulator and A1 Odd Chern metal with $ z=2 $, (c) $t_b/t_s=1.16$ inside the A1 Odd Chern metal.
     The cusps are Van-Hove singularities.
    Note that (c) inset is showing the total density of states is nearly, but not exactly flat near $\omega=0$.
    Compare to Fig.\ref{fig:DOS_C1} and Fig.\ref{fig:DOS_C2}  near $ h/t =\pm 4 $ and $ h/t = 0 $ respectively in the continuum calculation. }
\label{fig:DOS_L2}
\end{figure}

The DOS on a lattice contain some van-Hove singularities when $\omega$ far away from $0$,
which makes numerical calculation on DOS time consuming. The lattice DOS is plotted in Fig.\ref{fig:DOS_L2}.
For $\omega\sim 0$, in a metallic phase where $ t_b/t > 1 $, the DOS is nearly, but not exactly a constant.

\subsubsection{The specific heat and  compressibility}

 Here, we will make use of the DOS to evaluate the two conserved quantities, then compute the Wilson ratio.
  The Helmholtz free energy density
\begin{align}
    &F(T)
    =-T\int\frac{d^2\mathbf{k}}{(2\pi)^2}
	\ln[2(1+\cosh(E_+(\mathbf{k})/T))]
    \nonumber  \\
    &\!=\!E_\text{GS}
    \!+\!\!\int\!\!\frac{d^2\mathbf{k}}{(2\pi)^2}
    (|E_+\!(\mathbf{k})| \!-\! T\ln[2(1\!+\!\cosh\!\frac{E_+\!(\mathbf{k})}{T})]
    )
\end{align}
where the zero temperature part of $F(T)$ is nothing but
the ground-state energy density $E_\text{GS}$  in Eq.\ref{eq:E_lattice}.

 The specific heat (at a constant volume) $ C_v(T)=-T\frac{\partial^2 F}{\partial T^2} $:
\begin{align}
	&C_v(T)  =\int\frac{d^2\mathbf{k}}{(2\pi)^2}
            \frac{[E_+(\mathbf{k})]^2}
            {T^2[1+\cosh(E_+(\mathbf{k})/T)]}    \nonumber    \\
           & =\!\!\int\!\! \frac{d\omega \> \omega^2 D_+(\omega)}{T^2[1+\cosh(\omega/T)]}
            =\!\!\int\!\! \frac{d\omega \> \omega^2 D(\omega)}{2T^2[1+\cosh(\omega/T)]}
\end{align}

  The (isothermal) uniform compressibility  $ \kappa_u(T)
    =\sum_{s=\pm}\int \frac{d^2k}{(2\pi)^2}\frac{\partial f(E_{s,k})}{\partial E_{s,k}} $ is
\begin{align}
    &\kappa_u(T)
     =\int \frac{d^2k}{(2\pi)^2}\frac{[E_+(\mathbf{k})]^2}{T^2[1+\cosh(E_+(\mathbf{k})/T)]}   \nonumber   \\
    &=\!\int\! \frac{d\omega \> D(\omega)}{2T[1+\cosh(\omega/T)]}     
    =\!\int\! \frac{d\omega \> D_+(\omega)}{T[1+\cosh(\omega/T)]}
\end{align}

Below we discuss their low temperature behaviours.

In the gapped phases, i.e. Chern insulator or band  insulator,
we denote the gap as
\begin{align}
\tilde{\Delta}=\min_k E_+(k)>0,
\label{tildeD}
\end{align}
then $D(\omega)=0$ for entire $-\tilde{\Delta}<\omega<\tilde{\Delta}$
and $D(\tilde{\Delta})=D_2>0$ at the gap edge ( Fig.\ref{fig:DOS_L2}a ),
thus $C_v(T)=2D_2\tilde{\Delta}^2 T^{-1}e^{-\tilde{\Delta}/T}$. Similarly, $\kappa_u(T)=2D_2 Te^{-\tilde{\Delta}/T}$.
Both are  exponentially suppressed in $ T $.

In the gapless phases, i.e. A or B Odd Chern metal phase or band metal phase,
we have $\tilde{\Delta}=\min_k E_+(k)<0$ and $D(0)=D_0>0$ ( Fig.\ref{fig:DOS_L2}c ),
thus $C_v(T)=\frac{\pi^2}{3} D_0 T$, linear in $ T $. Similarly, $\kappa_u(T)=D_0+\cdots$ where
the sub-leading $ T $ dependence $ \cdots $ can be best evaluated in the continuum theory to be evaluated in Sec.III.

For the 3rd order QCPs with $h=0,\pm 4t$ and $t_b/t < 1 $,
we have $\tilde{\Delta}=\min_k E_+(k)=0$ and $D(\omega)=D_1\omega$ for small $\omega$ just like a Dirac fermion,
thus $C_v(T)=9\zeta(3)D_1 T^2$ where $\zeta(3)\approx1.2021$ is the Riemann Zeta function:
\begin{align}
 \zeta(s)= \sum^{\infty}_{n=1} 1/n^s = \frac{1}{\Gamma(s)} \int^{\infty}_{0} \frac{ x^{s-1} }{ e^x-1} dx
\end{align}
Similarly, $\kappa_u(T)=2\ln2 D_1 T$ where $\ln(2)\approx0.6931$.

For the other QCPs which are 2nd order, i.e. $  \tilde{\Delta}=0$ and $t_b/t_s> 1$, or at the boundary between $A1$ and $B1$,
we always have non-zero $D(0)=D_0>0$( Fig.\ref{fig:DOS_L2}b ) , thus $C_v(T)=\frac{\pi^2}{3}D_0T $. Similarly, $\kappa_u(T)=D_0+\cdots$.
where again the sub-leading $ T $ dependence $ \cdots $ can also be best evaluated in
the continuum theory to be evaluated in Sec.III.

\subsubsection{The Wilson ratio}

The Wilson ratio is defined as the ratio of the two conserved quantities $R_W=T\kappa_u/C_v$ which
has the following low temperature behaviours.

In the gapped phases, i.e. Chern insulator phase or band  insulator phase, $R_W=(T/\tilde{\Delta})^2$ where
$ \tilde{\Delta} $ is given in Eq.\ref{tildeD}.

In the gapless phases, i.e. A or B Odd Chern metal phase or band metal phase, $R_W=\frac{3}{\pi^2}\approx0.3040$.

For the 3rd order QCP with $ z=1 $ near $h=0,\pm 4t$ and $t_b/t_s < 1 $, $R_W=\frac{2\ln2}{9\zeta(3)}\approx0.1281$.

For the other QCP which are 2nd order with $ z=2 $,  $R_W=\frac{3}{\pi^2}\approx0.3040$.

These results will also be confirmed by the analytic calculations from the continuum theory in Sec.III-A-3(c) near $ h/t= \pm 4 $
and III-B-3(c) near $ h/t =0 $ respectively. They are also listed in the last line of the Table-I.

So we conclude the Wilson ratio can be used to distinguish all the gapped, gapless and QCPs.
However, one need $ Ch_{-} $ or $ \sigma_H $, especially the longitudinal/transverse edge to be discussed
in Sec.V and VI to distinguish the topology.

\subsection{ QPT versus TPT }

  Quantum phase transition (QPT) is characterized by the change of the ground state energy as shown in Sec.II-B.
  One diagnose the QPT by the non-analytical behaviours of ground state energy density: namely,
by taking consecutive derivatives on the ground state energy density at $ T=0 $
with respect to the tuning parameter such as the injection/boost or the Zeeman field until hitting the singularity \cite{sachdev,tqpt,weyl}.
The number of derivatives needed to reach the singularity gives the 1st, 2nd, or higher order QPT.
The DOS and the dynamic exponent $ z $ can also be extracted.
Then at a finite $ T $ near the QPT, various physical quantities such as the specific heat, compressibility and Wilson ratio
satisfy the corresponding scaling functions or sub-leading scalings \cite{weyl}.

  While Topological phase transition (TPT) is characterized by the change of topological invariants
  such as the Chern number $ Ch_{-} $ and quantum Hall conductance  as shown in Sec.II-A.
  For the non-interacting Fermi system \cite{tqpt,weyl,topoSF}, there is also the corresponding changes
  in the Fermi surface topology as shown in Fig.\ref{fig:phaseLattice}.

  QPT, especially in interacting bosonic or quantum spin system may not necessarily be a TPT.
  For example, the QPT from the BI-BM   in Fig.\ref{fig:Dirac0} is a pure 2nd order QPT with $ z=2 $ which is not a TPT.
  Of course, the well known SF-Mott QPT, AFM to VBS, etc are not TPT \cite{sachdev}.
  However, the QPT from CI to OCM is also a 2nd order one with $ z=2 $, but despite there is no change in the Chern number $ Ch_{-} $,
  there is also a corresponding changes in  both longitudinal and transverse edge modes. So it is also a TPT.
  However, in general, a TPT must be also a QPT.
  For example, the TPT from the CI to BI is a TPT where the Chern number $ Ch_{-} $ and the Hall conductance $ \sigma_H $ changes by  $ \pm 1 $.
  Of course, there is also an corresponding changes in the edge modes due to the conventional bulk-edge correspondence.
  At the same time, it is also a 3rd order QPT with $ z=1 $.
  Of course, the well-known TPT from FQH to insulator transition \cite{moving} is also a QPT with $ z=1 $.
  However, for the very first time, we discover an counter-example to this general believe: the TPT from the OCM to the BM is not a QPT !
  This maybe the very first example of a TPT which is NOT a QPT.
  Similar classifications also apply to Fig.\ref{fig:Dirac12}. See also Table I and Table II.
  See also Sec.VIII-C for more concrete discussions.

\section{ The bulk effective theory in the continuum limit }

In the momentum space, Eq.\ref{eq:Boosted_H} becomes:
\begin{align}
	H(\mathbf{k})&=-[h+2t(\cos k_x+\cos k_y)]\sigma_z
	 +2t_s\sin k_x\sigma_x
                                \nonumber  \\
	& +2t_s\sin k_y\sigma_y -2t_b\sin k_y\sigma_0
\end{align}
When $h\sim 4t$, there are low-energy excitations near $ \mathbf{K}_3=(\pi,\pi)$, it reduces to
\begin{align}
	H_3(\mathbf{K}_3+\mathbf{k})& =-[h-4t+t(k_x^2+k_y^2)]\sigma_z-2t_sk_x\sigma_x
                      \nonumber  \\
 &-2t_sk_y\sigma_y+2t_bk_y\sigma_0
\end{align}
When $h\sim 0$, there are low-energy excitations near both $\mathbf{K}_1=(\pi,0)$ and $\mathbf{K}_2=(0,\pi)$, it reduces to
\begin{align}
    H_1(\mathbf{K}_1+\mathbf{k}) & =-[h+t(k_x^2-k_y^2)]\sigma_z -2t_sk_x\sigma_x     \nonumber  \\
    &+2t_sk_y\sigma_y-2t_bk_y\sigma_0
    \nonumber  \\
	H_2(\mathbf{K}_2+\mathbf{k}) & =-[h-t(k_x^2-k_y^2)]\sigma_z +2t_sk_x\sigma_x
        \nonumber  \\
   &-2t_sk_y\sigma_y+2t_bk_y\sigma_0
\end{align}
When $h\sim -4t$, there are low-energy excitations  near $ \mathbf{K}_0=(0,0)$, it reduces to
\begin{align}
	H_0(\mathbf{k})&=-[h+4t-t(k_x^2+k_y^2)]\sigma_z  +2t_sk_x\sigma_x+2t_sk_x\sigma_x
     \nonumber  \\
    &-2t_bk_y\sigma_0
\end{align}


The low-energy physics near all the 4 Dirac points can be written in a generic form
\begin{align}
    H(\mathbf{k})=& (\Delta+\alpha_x k_x^2+\alpha_y k_y^2)\sigma_z
        \!+v_xk_x\sigma_x\!+v_yk_y\sigma_y
        \!-ck_y\sigma_0\>
\label{eq:Dirac}
\end{align}
where the velocities $v_{x,y}$ and ``mass'' $\alpha_{x,y}$ must be non-zero.
In this work, all the above equations correspond to the isotropic case $|v_x|=|v_y|$ and $|\alpha_x|=|\alpha_y|$.
As stressed below Eq.\ref{eq:Boosted_H},
both the C-symmetry (charge symmetry) and P-symmetry (parity symmetry) exist at $c=0$,
and P-symmetry is broken at $c\neq 0$, but the C-symmetry still holds at $c\neq 0$.

\begin{figure}[tbhp]
    \centering
    \includegraphics[width=0.8\linewidth]{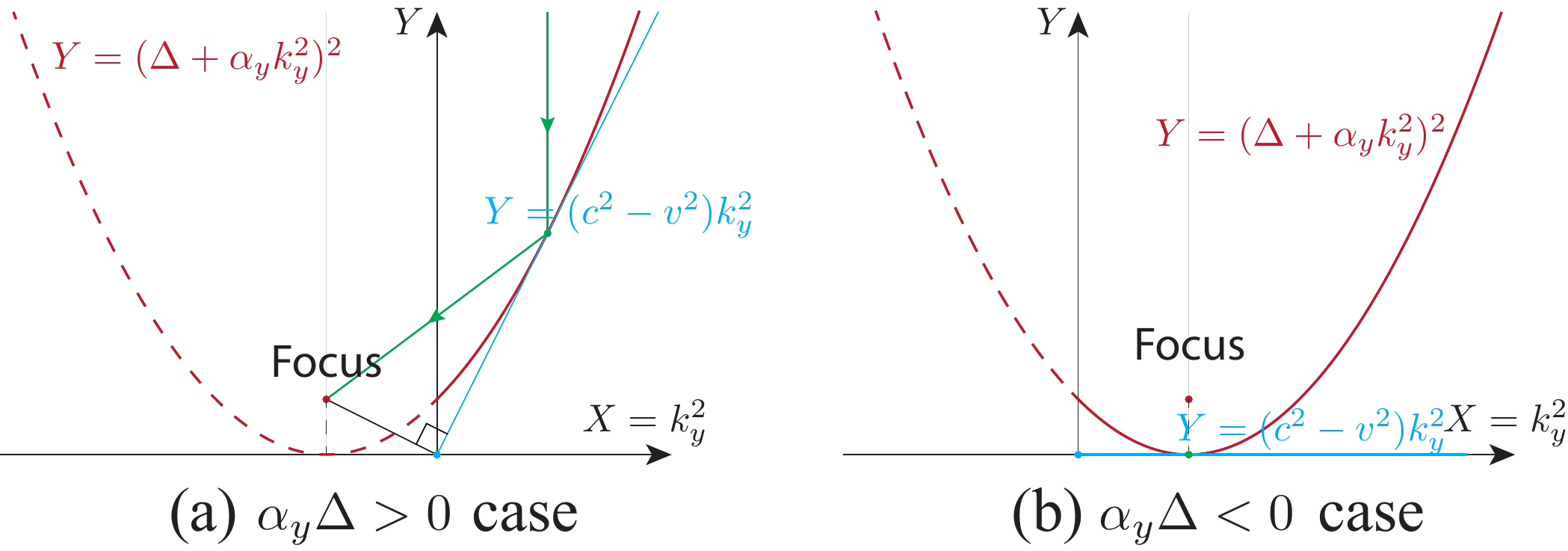}
    \caption{Geometric determination of the
    critical velocity  (a) $\alpha\Delta>0$ and (b) $\alpha\Delta<0$.
    The condition $0=\min_\mathbf{k} \epsilon_+(\mathbf{k})$ is equivalent to
    $0=\min_k[(\Delta+\alpha_y k_y^2)^2-(c^2-v^2) k_y^2]$.
    The  symmetry axis of the parabola is $k_y^2=-\Delta/\alpha$.
    %
    In both cases, the two Fermi points are located at momentum $ \pm \sqrt{ |\Delta/\alpha| } $ shown in Fig.\ref{fig:Dirac0_FS}.
    It may be contrasted the Geometric determination of the critical velocities
    inside a superfluid with or without a roton minimum \cite{moving}.  }
    \label{fig:parabola}
\end{figure}

Diagonalization of the effective Hamiltonian leads to the energy dispersion
\begin{align}
	\epsilon_\pm(\mathbf{k})
	\!=\!\pm\sqrt{(\Delta\!+\!\alpha_x k_x^2\!+\!\alpha_y k_y^2)^2\!+\!v_x^2k_x^2\!+\!v_y^2k_y^2}
    \!-\!c k_y
\end{align}
The half-filling condition and C-symmetry ensure the Fermi energy is always zero.
By examining the minima of upper band $\min_{\mathbf{k}} \epsilon_+(\mathbf{k})$,
there exists a critical velocity $c_0$
and $c<c_0$ the $\epsilon_+$ is empty and the system is in an insulating phase;
and $c>c_0$ the $\epsilon_+$ is partially filled and the system is in a metallic phase;
When $\alpha_y\Delta>0$, the critical velocity is $c_0=\sqrt{\smash[b]{v_y^2+4\alpha_y\Delta}}$;
when $\alpha_y\Delta<0$, the critical velocity is $c_0=|v_y|$.
See Fig.\ref{fig:parabola} for a geometric interpretation of the critical velocity in the two cases.

   When re-write the continuum Hamiltonian Eq.\ref{eq:Dirac} in the form of Eq.\ref{dvector}:
\begin{align}
    &H(k)=d_0(k)\sigma_0+d_x (k)\sigma_x+d_y (k)\sigma_y+d_z (k)\sigma_z,  \nonumber  \\
    &d_0(k)=-c k_y,\>\mathbf{d}=(d_x,d_y,d_z)    \nonumber  \\
    &d_x(k)=v_x k_x,\>   d_y(k)=v_y k_y,\>
                               \nonumber  \\
    &d_z(k)=\Delta+\alpha_x k_x^2+\alpha_y k_y^2
\end{align}
and then the Berry Connections and Berry curvatures stay the same form as in Eq.\ref{Aomega}:
\begin{align}
    A_{\pm,i}(\mathbf{k})
        & =\frac{(d_y\partial_{k_i} d_x-d_x\partial_{k_i} d_y)}{2|\mathbf{d}|(|\mathbf{d}|\pm d_z)}   \nonumber  \\
    \Omega_{\pm,xy}(\mathbf{k})
        & =\mp\frac{1}{2|\mathbf{d}|^3}\mathbf{d}\cdot \partial_x \mathbf{d}\times \partial_y \mathbf{d}
\end{align}
  which takes the identical form as Eq.\ref{Aomega}.
  But it is integrated over $ R^2 $ here, while it is over a Torus $ \mathbb{T}^2 $ there.

  It can be written in an explicit form
\begin{align}
    A_{\pm}(\mathbf{k})
        &=\frac{v_xv_y(k_y,-k_x)}{2\sqrt{A^2+ \bar{k}^2 }
        [\sqrt{ A^2+\bar{k}^2 }\pm A ]}   \nonumber\\
    \Omega_{\pm,xy}(\mathbf{k})
        &=\mp\frac{v_xv_y(\Delta-\alpha_x k_x^2-\alpha_y k_y^2)}
        {2[A^2+\bar{k}^2]^{3/2}}
\end{align}
 where $ A= \Delta+\alpha_x k_x^2+\alpha_y k_y^2, \bar{k}^2=v_x^2k_x^2+v_y^2k_y^2 $.

Since the $c$ boost term does not affect the eigenvectors,
the Berry connections and Berry curvatures are exactly the same as $c=0$ case.

Because the upper band and lower band are always separated either directly or indirectly,
thus a Chern number of lower band, independent of its filling,  can always be evaluated
via the integral ${\rm Ch}_-=\frac{1}{2\pi}\int_{\mathbb{R}^2} \Omega_-(\mathbf{k})d^2\mathbf{k}$,
which gives
\begin{align}
	\mathrm{Ch}_-\!=\!\sgn(v_xv_y)
    [2\sgn(\Delta)\!-\!\sgn(\alpha_x)\!-\!\sgn(\alpha_y)]/4,
\label{eq:C}
\end{align}
where $\sgn()$ denotes the sign function \cite{footnote00}.

The band Chern number, which is independent of the filling,
is closely related to the zero temperature Hall conductance which does depend on the filling.
The linear response theory gives the intrinsic Hall conductance
\cite{Qi2006} in the unit $e^2/h$ as
\begin{align}
\sigma_{H}=\frac{1}{2\pi}\int_{\mathbb{R}^2}
\sum_{s=\pm}\Omega_s(\mathbf{k})\Theta(-\epsilon_{s}(\mathbf{k}))d^2\mathbf{k}
\end{align}
where the unit step function $\Theta(\epsilon)=1$
when $\epsilon>0$ and $=0$, when $\epsilon\leq0$. it just replaces
the $ \mathbb{T}^2$  in a lattice Eq.\ref{eq:sigma_xy} by $ R^2 $ in the continuum.

Due to the C-symmetry, the intrinsic Hall conductance can be separated into two parts
\begin{align}
    \sigma_{H}& =\frac{1}{2\pi}\int_{\mathbb{R}^2} \Omega_-(\mathbf{k})d^2\mathbf{k}
                +\frac{1}{\pi}\int_{\mathbb{R}^2} \Omega_+(\mathbf{k})\Theta(-\epsilon_{+}(\mathbf{k}))d^2\mathbf{k}
                \nonumber  \\
               & ={\rm Ch}_-+\nu_b
\label{eq:Ch+nu}
\end{align}
whose interpretation in terms of the edge states in a strip geometry will be given in Sec.IV-A-1.

When $\max\epsilon_{-}(\mathbf{k})<0<\min\epsilon_{+}(\mathbf{k})$,
$\nu_b=0$ and $\sigma_{H}={\rm Ch}_-$.
However, when $\min\epsilon_{+}(\mathbf{k})<0<\max\epsilon_{-}(\mathbf{k})$, the
evaluation of the second integral in Eq.\eqref{eq:Ch+nu} gives
\begin{align}
    \nu_b=\sgn({\rm Ch}_-)\frac{\sgn(\alpha_y\Delta)-1}{2}\frac{c-|v_y|}{c}
\label{nub}
\end{align}
which indicates that $\nu_b$ and ${\rm Ch}_-$ always have opposite sign. So
$|\sigma_{H}|$ is always smaller than $|{\rm Ch}_-|$.

Using Eq.\eqref{eq:C}, one can  reproduce the correct Chern number
calculated from the lattice Hamiltonian Eq.\ref{dvector}.

\vspace{2mm}
I) When $h\sim 4t$, $\Delta=4t-h$, $\alpha_x=\alpha_y=-t$,$v_x=v_y=-2t_s$, thus ${\rm Ch}_-=[\sgn(4t-h)+\sgn(t)]/2$;

If $t>0$, ${\rm Ch}_-=+1$ when $h<4t$ (aka $h/|t|<4$), 0 otherwise;

If $t<0$, ${\rm Ch}_-=-1$ when $h>4t$ (aka $h/|t|>-4$), 0 otherwise.

\vspace{2mm}
II) When $h\sim 0$, $\Delta_1=\Delta_2=-h$, $\alpha_{1x}=-\alpha_{1y}=-\alpha_{2x}=\alpha_{2y}=t$,
$v_{1x}=-v_{1y}=-v_{2x}=v_{2y}=2t_s$,
thus ${\rm Ch}_-={\rm Ch}_{1-}+{\rm Ch}_{2-}=-\sgn(-h)=\sgn(h)$;

\vspace{2mm}
III) When $h\sim -4t$, $\Delta=-4t-h$, $\alpha_x=\alpha_y=t$, $v_x=v_y=2t_s$,
thus ${\rm Ch}_-=[\sgn(-4t-h)-\sgn(t)]/2$;

If $t>0$, ${\rm Ch}_-=-1$ when $h>-4t$ (aka $h/|t|>-4$), 0 otherwise;

If $t<0$, ${\rm Ch}_-=+1$ when $h<-4t$ (aka $h/|t|<4$), 0 otherwise.

\vspace{2mm}
These I),II),III) are indeed consistent with Eq.\eqref{eq:Ch} achieved on a lattice.
Notice that ${\rm Ch}_-$, if it is not zero, only depends  on $\sgn(h)$,
but independent of $\sgn(t)$ and $\sgn(t_s)$.

The $\alpha_x\alpha_y>0$ case corresponding to $|h/t|\sim 4$ transition,
and $\alpha_x\alpha_y<0$ case corresponding to $|h/t|\sim 0$ transition.
Since the sign of $\alpha_x\alpha_y$ makes the topological property dramatically different,
in the following, we will discuss $\alpha_x\alpha_y>0$ and $\alpha_x\alpha_y<0$ separately.

\subsection{ The bulk topological transitions near $|h/t|= 4$ (the $\alpha_x\alpha_y>0$ case) }

Without loss of generality, we consider $h/t\sim -4$ and near $ \mathbf{K}_0=(0,0)$ case,
which belongs to the $\alpha_x\alpha_y>0$ case,
\begin{align}
    H_0(\mathbf{k})=[\Delta+\alpha (k_x^2+k_y^2)]\sigma_z
        +vk_x\sigma_x+vk_y\sigma_y-ck_y\sigma_0
\label{eq:Dirac0}
\end{align}
where $\Delta=-(h+4t)$, $\alpha=t$, $v=2t_s$, $c=2t_b$.
Diagonalization of Eq.\eqref{eq:Dirac0} leads to two bands
\begin{align}
	\epsilon_{\pm}(\mathbf{k})=\pm\sqrt{v^2k^2+(\Delta+\alpha k^2)^2}-ck_y
\label{Dirac0eng}
\end{align}

\begin{figure}[tbhp]
    \centering
    \includegraphics[width=0.6\linewidth]{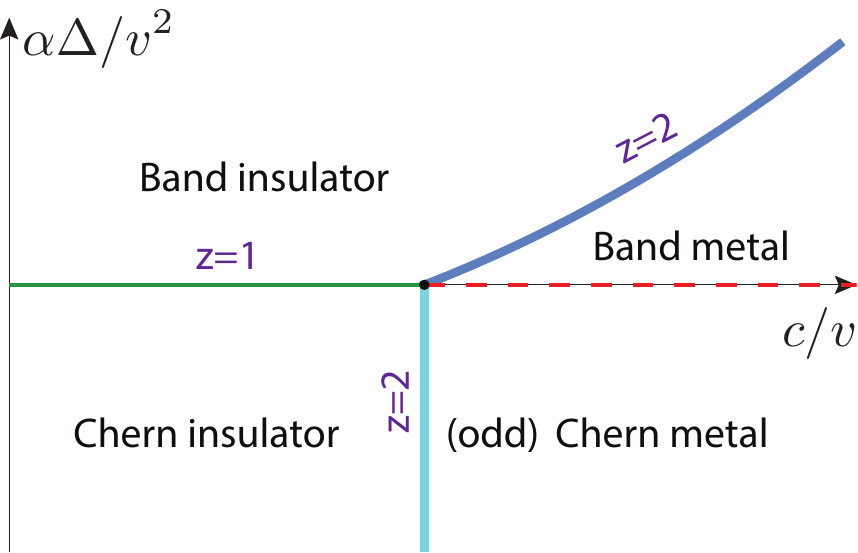}
    \caption{The phase diagram of Eq.\eqref{eq:Dirac0}.
    It is the part near $ h/t \sim \pm 4 $ in the global phase diagram in Fig.\ref{fig:phaseLattice} where the Odd Chern metal is the A1 phase.
	Thick/thin/dashed line are 2nd/3rd/infinite order Topological phase transitions (TPTs) respectively.
    The Chern insulator with $ Ch_-=1, \sigma_H=1 $ to the insulator with $ Ch_-=0, \sigma_H =0 $
    is 3rd order TPT due to the linear band touching of the Dirac fermion with $ z=1 $.
    The Dirac fermion at $ c/v=0 $ has the emergent Lorentz invariance which is broken by any $ 0 < c/v < 1 $.
    The Chern insulator to the Odd Chern metal with $ Ch_-=1, \sigma_H < 1 $ and $ z=2 $
    transition can be read from Fig.\ref{fig:Dirac0_FS}. The band  insulator to the band metal transition  with $ z=2 $
    can be read from the first 3 boxes in Fig.\ref{fig:phaseLattice}.
    The dashed line from the Odd Chern metal to  the band metal with  $ Ch_-=0, \sigma_H =0 $  due to the conic band touching
    between the P- and the H- FS is infinite order and can be read from the 3rd to the 5th boxes in Fig.\ref{fig:phaseLattice}.   }
\label{fig:Dirac0}
\end{figure}

As shown in Fig.\ref{fig:parabola}, when $\alpha\Delta>0$,
the critical velocity $c_0=\sqrt{v^2+4\alpha\Delta}$.
When $\alpha\Delta<0$, the critical velocity $c_0=|v|$.
The phase diagram of Eq.\eqref{eq:Dirac0} is given in Fig.\ref{fig:Dirac0}.

At a fixed $ \Delta $, when $c\to c_0$, energy bands overlap.
There is an electron pocket near $\mathbf{k}=(0,\sqrt{|\Delta/\alpha|})$ with $ z=2 $
in the $k_y>0$ regime
and the corresponding  hole pocket near $\mathbf{k}=(0,-\sqrt{|\Delta/\alpha|})$ with $ z=2 $
in the $k_y<0$ regime.
When $|c-c_0|\ll v$, the gap in Eq.\ref{tildeD}  vanishes linearly as $c\to c_0$ with
\begin{align}
\tilde{\Delta}=\min_k\epsilon_{+}(k)=\sqrt{|\Delta/\alpha|}(c-c_0)+\cdots
\label{gap1}
\end{align}

At a fixed $c<v$, when $\Delta\to 0$, then the upper band and lower band conic touch at $k=0$ with $ z=1 $.
When $|\Delta|\ll v^2/\alpha$, the gap vanishes also linearly as $\Delta\to 0$ with
\begin{align}
\tilde{\Delta}=\min_k\epsilon_{+}(k)=\sqrt{1-c^2/v^2}|\Delta|+\cdots
\label{gap2}
\end{align}
Both Eq.\ref{gap1} and Eq.\ref{gap2} will be used in evaluating various thermodynamic quantities in Sec.III-A-3.

\begin{figure}[tbhp]
    \centering
    \includegraphics[width=\linewidth]{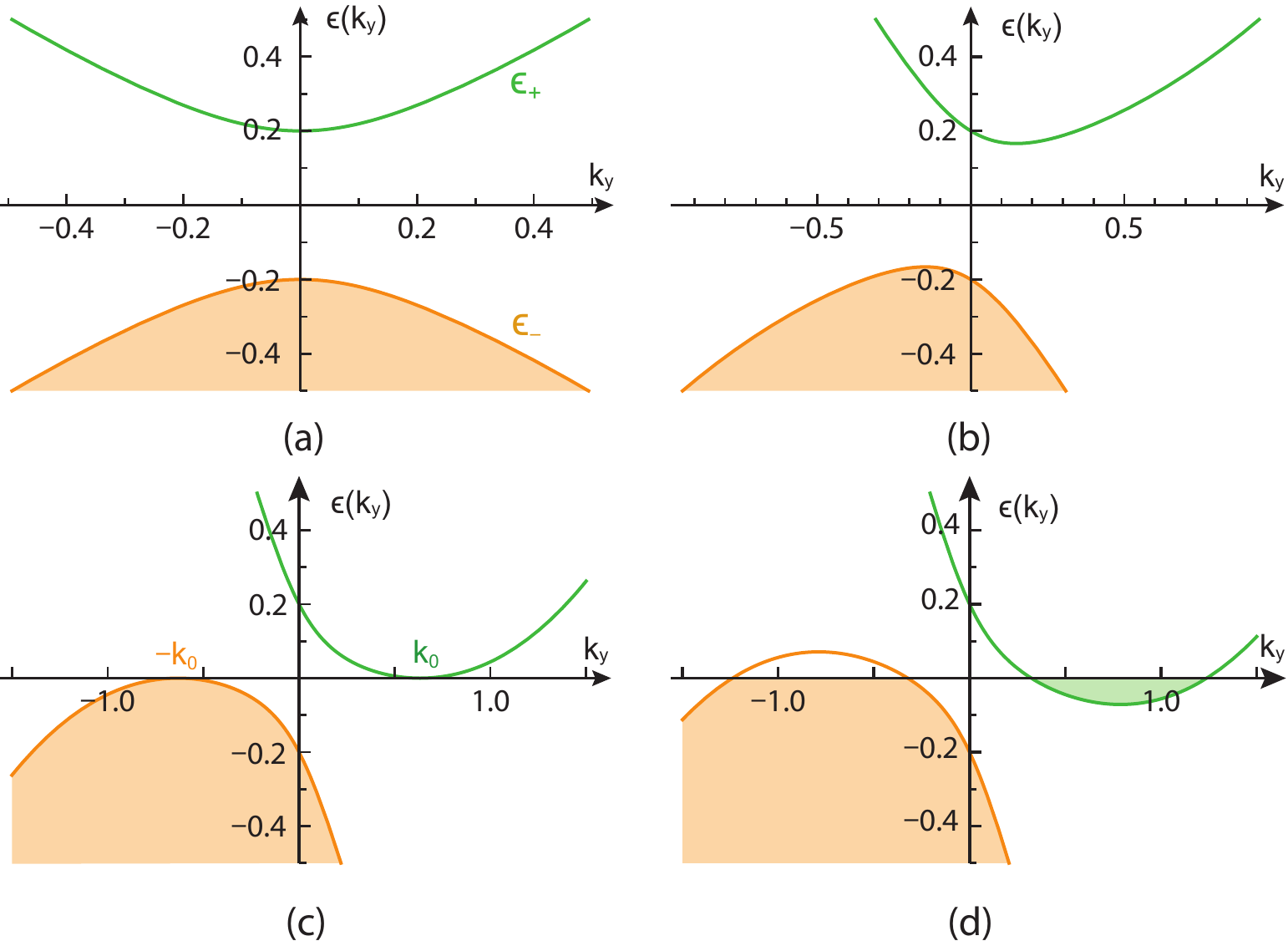}
    \caption{ The dispersion $\epsilon_{\pm}(k)$ as function of $k_y$
    with fixed $k_x=0$ and $v=1$, $\Delta=1/5$, $\alpha=-1/2$, $c=0,0.5,1.0,1.1$.
    (a) it has a direct gap at $k_y=0$,
    (b) it has an indirect gap at $k_y\neq 0$,
    (c) it becomes gapless and shows Fermi points at $k_y=\pm k_0$ with the quadratic band touching and $ z=2 $,
    (d) it becomes gapless and show finite Fermi pockets. It is the A1 phase in Fig.\ref{fig:phaseLattice}.   }
    \label{fig:Dirac0_FS}
\end{figure}

\subsubsection{ Hall conductance at zero and finite temperatures }

We will also evaluate the Hall conductance at zero and finite T respectively.

{\sl (a). Zero temperature Hall conductance}

When $c<c_0$, the $ T=0 $  intrinsic Hall conductance in unit $e^2/h$
can be evaluated as $\sigma_{H}=\frac{1}{2\pi}\int_{\mathbb{R}^2} \Omega_-(\mathbf{k})d^2\mathbf{k}$
\begin{align}
	\sigma_{H} & ={\rm Ch}_-=[\sgn(\Delta)-\sgn(\alpha)]/2
                \nonumber  \\
     & =\sgn(\Delta)[1-\sgn(\alpha\Delta)]/2
\end{align}
which can be obtained by just setting $ \sgn(\alpha_x) = \sgn(\alpha_y) $ in  Eq.\ref{eq:C}.
It suggests that $\sigma_{H}=-1$ for $h<-4t$ and $\sigma_{xy}=0$ for $h>-4t$
If choosing $t>0$, this is consistent with calculation on lattice scale.

When $c>c_0$, there is an electron pocket near $\mathbf{k}=(0,\sqrt{|\Delta/\alpha|})$
in the $k_y>0$ regime,
and a hole pocket near $\mathbf{k}=(0,-\sqrt{|\Delta/\alpha|})$
in the $k_y<0$ regime ( Fig.\ref{fig:Dirac0_FS} ).
Both $+/-$ band contribute to the Hall conductance as shown in Eq.\ref{eq:Ch+nu}.
\begin{align}
    \sigma_{H}=\frac{1}{2\pi}\int_{\mathbb{R}^2} \Omega_-(\mathbf{k})d^2\mathbf{k}
                +\frac{1}{\pi}\int_{\mathbb{R}^2} \Omega_+(\mathbf{k})\Theta(-\epsilon_{+}(\mathbf{k}))d^2\mathbf{k}
\label{eq:Ch+nu1}
\end{align}

\begin{figure}[tbhp]
    \centering
    \includegraphics[width=\linewidth]{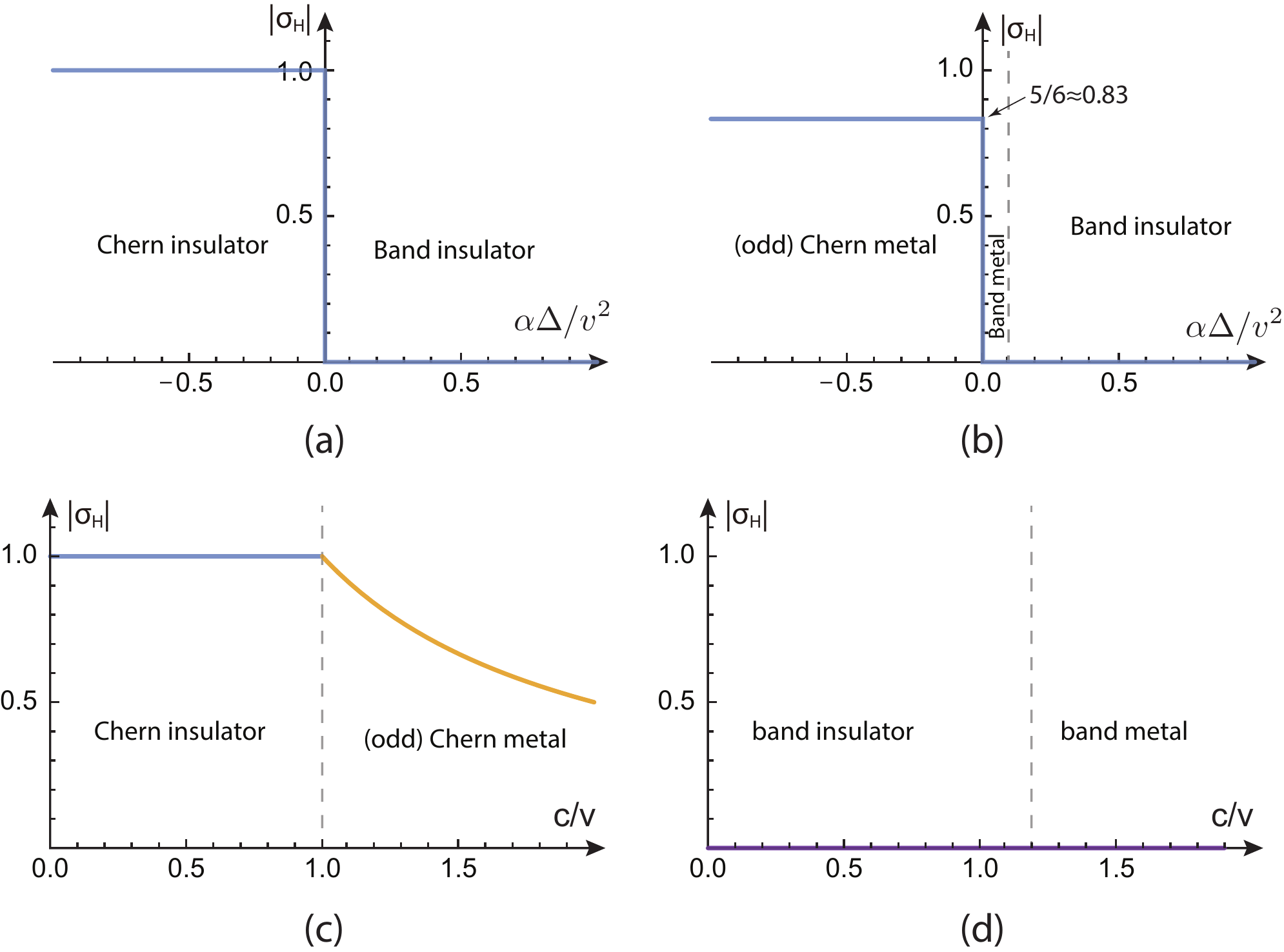}
    \caption{ The Hall conductance of Eq.\eqref{eq:Dirac0} at different $\alpha\Delta/v^2$ or $c/v$ values:
    Varying $\alpha\Delta/v^2$ in (a) and (b):  (a) fixing $c/v=0.5$,
    (b) fixing $c/v=1.2$,
    There is a jump from Odd Chern metal to the band metal with the magnitude $  v/c = 5/6 $.
    However, as indicated in Fig.\ref{fig:Dirac0}, it is infinite order.
    Varying  $c/v$ in (c) and (d): (c) fixing $\alpha\Delta/v^2=-0.1$,
    (d) fixing $\alpha\Delta/v^2=+0.1$.
    Again, as already alerted in Fig.\ref{fig:Hall_T0}d, in a generic case,
    the BM should contributes to a un-quantized AHE ( See Fig.24 and Table 1 ).
    But it vanishes in this particular $ n=1 $ case due to the fine tuning.
    They match all the TPTs achieved on the lattice shown in Fig.\ref{fig:Hall_T0}, so the captions of Fig.\ref{fig:Hall_T0}
    on the universality classes of the TPTs also apply here. }
\label{fig:sigma_C1}
\end{figure}

Because the electron FS is a simple closed loop,
Eq.\ref{eq:Ch+nu1} can be expressed as the Berry phase
for an adiabatic path around the FS \cite{Haldane2004}
\begin{align}
    \sigma_{H}&={\rm Ch}_-+\frac{1}{\pi}\oint d\mathbf{k}_F \cdot A_+(\mathbf{k}_F)
    ={\rm Ch}_-+\frac{\phi_{+}}{\pi}\>.
\label{eq:phi}
\end{align}
Evaluation of the integral Eq.\ref{eq:Ch+nu1} or Eq.\eqref{eq:phi}
leads to the same result
\begin{align}
    \sigma_{H}& ={\rm Ch}_-[1+\frac{c-v}{c}\Theta(c-v)\Theta(-\alpha\Delta)]
       \nonumber  \\
    &=\sgn(\Delta)\times \min(1,v/c) \quad \text{if $\alpha\Delta<0$}
\end{align}
   otherwise $0$. The Hall conductance developed a cusp at the critical velocity $c=v$ as shown in  Fig.\ref{fig:sigma_C1}.

{\sl (b). Finite temperature Hall conductance}

When $c<c_0$,
the intrinsic Hall conductance \cite{Qi2006} in unit $e^2/h$
can be evaluated as
\begin{align}
    \sigma_{H}(T)=\frac{1}{2\pi}\sum_{s=\pm}
        \int_{\mathbb{R}^2} d^2\mathbf{k}
        \Omega_s(\mathbf{k})f(\epsilon_s(k))
\end{align}
  which is just replacing $ T^2 $ in Eq.\ref{finiteTsigmaH}  by $ R^2 $.

Due to the C-symmetry, the finite temperature Hall conductance can be expressed as
\begin{align}
    \sigma_{H}(T)
        &={\rm Ch}_-+\frac{1}{\pi}\int_{\mathbb{R}^2} d^2\mathbf{k}\>
            \Omega_+(\mathbf{k})f(\epsilon_+(\mathbf{k}))
               \nonumber  \\
        & ={\rm Ch}_-+\nu(T)
\label{finiteTsigmaHh4}
\end{align}
which is also just replacing $ T^2 $ in Eq.\ref{finiteTsigmaH}  by $ R^2 $.

The $\nu(T=0)$ and ${\rm Ch}_-$ has been evaluated in Eq.\ref{eq:C} and Eq.\ref{nub} respectively.
At low temperature $T$ where $T$ is the lowest energy scale,
one can get a low temperature expansion of $\nu(T)$.
By keeping the leading low $T$ dependence, assuming $|\alpha\Delta/v^2|\ll 1/2$, we have

1) When $\alpha\Delta>0$ in the band insulating phase,
$\nu(T)-\nu(T=0)\propto -\sgn(\Delta)Te^{-|\tilde{\Delta}|/T};~~~-\sgn(\Delta)T^2;~~~-\sgn(\Delta)T^2 $
at  $c<c_0,~c=c_0,~c>c_0$ respectively.

2) When $\alpha\Delta < 0$ in the Chern insulating phase, $\nu(T)-\nu(T=0)\propto -\sgn(\Delta)Te^{-|\tilde{\Delta}|/T};~~~-\sgn(\Delta)T;~~~-\sgn(\Delta)T^2 $
at  $c<c_0,~c=c_0,~c>c_0$ respectively.


\subsubsection{ Ground-state energy density and topological phase transitions }

When $c<c_0$, $\min\epsilon_+\geq0\geq\max\epsilon_-$,
the lower band is full occupied and the upper band is complete empty,
thus the ground-state energy density is
$E_\text{GS}=(4\pi^2)^{-1}\int_{\mathbb{R}^2}d^2\mathbf{k} \> \epsilon_-(\mathbf{k})$.
The integral is divergent in $ R^2 $, we need a cut-off $\Lambda$ for $k$.
Due to its odd property, the $ck_y$ part in Eq.\ref{Dirac0eng} drops off, one obtain
\begin{align}
    E_0=-\frac{|\Delta|^3}{6\pi v^2}+E_\Lambda
\label{ELambda1}
\end{align}
where $E_\Lambda$ is $\Lambda$, $\alpha\Delta$ and $v$ dependent function
which is differentiable up to 3 degrees of differentiation, but
may contain higher order non-analyticity.

When $c>c_0$, $\min\epsilon_+<0<\max\epsilon_-$,
both the lower band and the upper band are only partially occupied,
thus the ground-state energy density becomes
$E_\text{GS}=(4\pi^2)^{-1}\sum_{s=\pm}\int_{\mathbb{R}^2} d^2\mathbf{k} \>
\epsilon_{s}(\mathbf{k})\Theta(-\epsilon_{s}(\mathbf{k}))$.
Due to the C-symmetry, it can be expressed as
$E_\text{GS}=E_0+E_1\Theta(c-c_0)$, where the $c$-dependent part is
$E_1=(2\pi^2)^{-1}\int_{\mathbb{R}^2} d^2\mathbf{k} \>
\epsilon_{+}(\mathbf{k})\Theta(-\epsilon_{+}(\mathbf{k}))$.
The integral is convergent, we obtain
\begin{align}
    E_1  =-\frac{1}{16\pi\alpha^3}
    [&\frac{1}{4}(c^2-c_0^2)(c^2+c_0^2-4v^2-8\alpha\Delta)
    \nonumber  \\
    & + v^2(v^2+4\alpha\Delta)\ln(c/c_0)]
\label{E1part}
\end{align}

In the following, we study the possible TPT encoded in the ground state energy density.

{\sl (a). Band  insulator to Chern insulator transition}

In the range $ c < v $ and tuned by $\Delta$, one only have the $ E_0 $ part. It is easy to see a third-order non-analytical behaviour at $\Delta=0$,
\begin{align}
    E_\text{GS}=-\frac{|\Delta|^3}{6\pi v^2}+\cdots
\end{align}
  which has the dynamic exponent $ z=1 $.

{\sl (b). Band metal to Odd Chern metal transition}:

In the range $ c > v $ and tuned by $\Delta$, one need both $E_0$ and $E_1$ part.
It is easy to see the third-order non-analytical behaviour at $\Delta=0$ in the two parts gets canceled:
\begin{align}
    E_\text{GS}
    =-\frac{|\Delta|^3}{6\pi v^2}+\frac{|\Delta|^3}{6\pi v^2}+\cdots
\end{align}
Carefully treating the exact results $E_\text{GS}=E_0+E_1\Theta(c-c_0)$  tells $E_\text{GS}$ is smooth at $\Delta=0$,
which is consistent with numerical result's infinite order differentiable. This fact explains why
the 3rd order TPT across the green line changes to infinite order across the red dashed line in Fig.\ref{fig:Dirac0}.
Even so, as shown in Fig.13b, there is still a universal jump $ \Delta \sigma_H=v/c $, so it is still a TPT.

{\sl (c). Band  insulator to band metal transition}:

In the range $\alpha\Delta>0$ and tuned by either $\Delta$ or $c$.
It is easy to tell a second-order non-analytical behaviour at $\Delta=(c^2-v^2)/(4\alpha)$ when tuning $\Delta$,
\begin{align}
    E_\text{GS}
	\!=\!E_0-\frac{c^2\!-\!v^2}{4\pi c^2}(\Delta\!-\!\frac{c^2\!-\!v^2}{4\alpha})^2
	\Theta(c^2\!-\!v^2\!-\!4\alpha\Delta)
\end{align}
and also a second-order non-analytical behaviour at $c=\sqrt{v^2+4\alpha\Delta}$ when tuning $c$,
\begin{align}
    E_\text{GS}
	\!=\!E_0\!-\!\frac{\Delta}{4\pi \alpha^2}(c\!-\!\!\sqrt{v^2\!+\!4\alpha\Delta})^2
	\Theta(c^2\!-\!v^2\!-\!4\alpha\Delta)
\end{align}
   which has the dynamic exponent $ z=2 $.

{\sl (d). Chern insulator to Odd Chern metal transition}

In the range $\alpha\Delta<0$ and tuned by $c$.
It is easy to see a second-order non-analytical behaviour at $c=v$,
\begin{align}
    E_\text{GS}
    =E_0+\frac{\alpha\Delta}{4\pi|\alpha|^3}(c-v)^2\Theta(c^2-v^2)+\cdots
\end{align}
  which has the dynamic exponent $ z=2 $.

In summary, transitions (c) and (d) are second-order QPT with $ z=2 $;
(a) is third-order QPT with $ z=1 $. (b) is infinite-order without any non-analyticity.
These are consistent with the results achieved in a lattice in Sec.II-B.

\subsubsection{ Thermodynamic Quantities   }

{\sl (a). The density of states }

From the Hamiltonian \eqref{eq:Dirac}, the Matsubara Green's function
\begin{align}
    G(\mathbf{k},i\omega_n)
    =\frac{P_+(\mathbf{k})}{i\omega_n-\epsilon_+(\mathbf{k})}
    +\frac{P_-(\mathbf{k})}{i\omega_n-\epsilon_-(\mathbf{k})}
\end{align}
where $P_s$ are the projection operators onto $s=\pm$ bands
and $P_+ + P_-=1$.

\begin{figure}[tbhp]
    \centering
    \includegraphics[width=\linewidth]{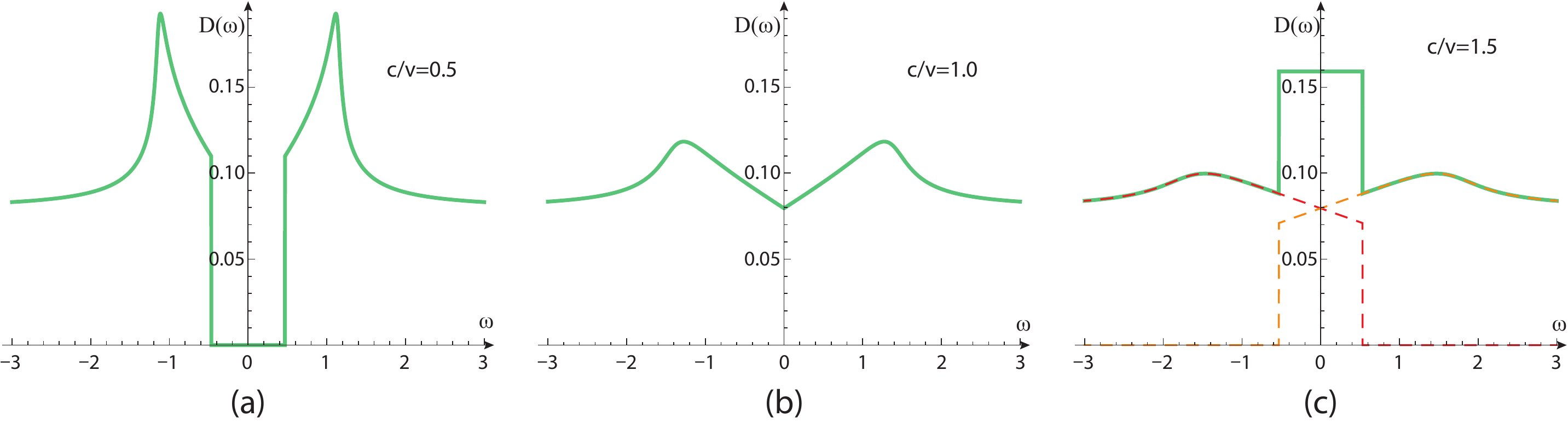}
    \caption{Density of states (DOS) of Eq.\eqref{eq:Dirac0} at different $c/v$ values:
    (a) $c/v=0.5$ in Chern Insulator, (b) $c/v=1.0$ on the QCP between Chern insulator and Odd Chern metal, (c) $c/v=1.5$ in Odd Chern metal.
    Here, we fixed $v=1$, $\alpha=-1$, $\Delta=1$ in Fig.\ref{fig:Dirac0}.
    Note that  $\epsilon_\pm(k)$ is quadratic in $k$ at the band minimum in Fig.\ref{fig:Dirac0_FS},
    which leads to a finite DOS at the band edge and $ z=2 $. So the DOS at $\omega=0$ in (b) is finite.
    The flat feature of total DOS near $\omega=0$
    is a result of truncation of $H(k)$ at the $k^2$ order. Indeed, the lattice calculations  on DOS in Fig.\ref{fig:DOS_L2}
    shows that it is not a constant anymore. It matches Fig.\ref{fig:DOS_L2}  very well near the low energy $ \omega \sim 0 $. }
\label{fig:DOS_C1}
\end{figure}

The total DOS is
\begin{align}
    D(\omega)
    =\int\frac{d^2q}{(2\pi)^2}[\delta(\omega-\epsilon_{+}(k))+\delta(\omega-\epsilon_{-}(k))]
\end{align}
It would be convenient to introduce the DOS for each band  $ D(\omega)=D_{+}(\omega)+ D_{-}(\omega) $:
\begin{align}
    D_{\pm}(\omega)=\int \frac{d^2\mathbf{k}}{4\pi^2}\delta(\omega-\epsilon_{\pm}(\mathbf{k})),
\end{align}
Due to the C-symmetry, $D_{+}(\omega)=D_{-}(-\omega)$.
\begin{align}
    D_{+}(\omega)=\frac{1}{4\pi|\alpha|}[1+\rho(\omega)]\Theta(\omega-\min\epsilon_+)
\label{DOSflat}
\end{align}
For some reason related to the C-symmetry,
$\rho(\epsilon)=-\rho(-\epsilon)$ is an odd function.
The odd property leads to flat feature of the total DOS near $\omega=0$
when $\min\epsilon_+<0<\max\epsilon_-$,
\begin{align}
    D(\omega)=\frac{1}{2\pi|\alpha|}, \textmd{ if } \min\epsilon_+<\omega<\max\epsilon_-
\label{DOS1}
\end{align}
The DOS is plotted in Fig.\ref{fig:DOS_C1}.

{\sl (b). The specific heat and Compressibility }

Here, we will make use of the DOS to evaluate the two conserved quantities, then compute the Wilson ratio.
The Helmholtz free energy density (taking $k_B=1$)
\begin{align}
    F(T)
    =-T\int\frac{d^2\mathbf{k}}{(2\pi)^2}
    \ln[2(1+\cosh(\epsilon_+(\mathbf{k})/T))]
\end{align}
where the zero temperature part of $F(T)$ is nothing but the ground-state energy density $E_\text{GS}$ calculated in Sec.III-A-2.
The specific heat (at constant volume) is given by
\begin{align}
    C_v(T)    =\int\frac{d^2\mathbf{k}}{(2\pi)^2}
            \frac{[\epsilon_+(\mathbf{k})]^2}
            {T^2[1+\cosh(\epsilon_+(\mathbf{k})/T)]}
\end{align}

The dynamic density-density response function is
\begin{align}
    \chi^{00}(q,i\omega_n)
    =-T\sum_{iv_m,k}\Tr[G(i\omega_n+iv_m,k+q) G(iv_m,k)]
\end{align}
Taking the $\omega_n\to0$ , then $q\to0$ limit in $\chi^{00}(q,i\omega_n)$ gives the (isothermal) uniform compressibility
\begin{align}
    \kappa_u(T)
    \!=\!\!\sum_{s=\pm}\!\int\!\! \frac{d^2k}{(2\pi)^2}\frac{\partial f(\epsilon_{s,k})}{\partial \epsilon_{s,k}}
    \!=\!\!\int\!\! \frac{d\omega D_+(\omega)}{T[1+\cosh(\omega/T)]}
\end{align}
and the zero temperature limit $\frac{\partial f(\epsilon)}{\partial \epsilon}=\delta(\epsilon)$
recovery the well-known relation between compressibility and DOS $\kappa_u(T)=D(0)$.

At low temperature $T\ll |\Delta|,v^2/|\alpha|, c^2/|\alpha|, |(c^2-v^2)/\alpha|$,
one can get a low temperature expansion of $C_v(T)$ and $\kappa_u(T)$ \cite{footnote3}.
We do not need to distinguish $\alpha\Delta>0$ or $\alpha\Delta<0$, but need to care if $\Delta=0$ or not.
By keeping the leading low $T$ dependence, assuming $|\alpha\Delta/v^2| < 1/2$ to simply the technical
calculations \cite{extendtolarge}, we have

{\em $ \Delta\neq 0 $ case in Fig.\ref{fig:Dirac0} }

At $c<c_0$,
we have $\tilde{\Delta}=\min_k \epsilon_+(k)>0$ and
$D(\omega)=0$ for $|\omega|<\tilde{\Delta}$ and $D(\tilde{\Delta})>0$ at the gap edge  ( Fig.\ref{fig:DOS_C1}a ),
thus $C_v(T)=2D(\tilde{\Delta})\tilde{\Delta}^2 T^{-1}e^{-\tilde{\Delta}/T}$.
Similarly, $\kappa_u(T)= 2D(\tilde{\Delta}) Te^{-|\tilde{\Delta}|/T}$.

At $c=c_0$,
we have $\tilde{\Delta}=\min_k \epsilon_+(k)=0$ and
$D(\omega)=D(0)+D'(0^+)|\omega|$ for small $|\omega|$ ( Fig.\ref{fig:DOS_C1}b ),
thus $C_v(T)=\frac{\pi^2}{3} D(0)T$. Here $D'(0^+)$ means the right derivatives of $D(\omega)$ at $0$.
Similarly, $\kappa_u(T)=D(0)+2\ln2 D'(0^+) T$;

At $c>c_0$,
we have $\tilde{\Delta}=\min_k \epsilon_+(k)<0$ and
$D(\omega)=\frac{1}{ 2 \pi |\alpha | } $ for $|\omega|<|\tilde{\Delta}|$ in Eq.\ref{DOS1} and $D(\tilde{\Delta})>0$ ( Fig.\ref{fig:DOS_C1}c ),
thus $C_v(T)=\frac{\pi^2}{3} D(0)T$. Similarly, $\kappa_u(T)=D(0)+2[ D(\tilde{\Delta})-D(0)] T e^{-| \tilde{\Delta} | /T } $.

{\em $ \Delta = 0 $ case in Fig.\ref{fig:Dirac0} }

At $c<c_0$,
we have $\tilde{\Delta}=\min_k \epsilon_+(k)=0$ and
$D(\omega)=D'(0^+)|\omega|$ for small $\omega$,
thus $C_v(T)=9\zeta(3)D'(0^+) T^2$ where $\zeta(3)\approx1.2021$.
Similarly, $\kappa_u(T)=2\ln2 D'(0^+) T$.

At $c=c_0$,
we have $\tilde{\Delta}=\min_k \epsilon_+(k)=0$ and $D(\omega)=D(0)+A|\omega|^{2/3}>0$ for small $\omega$,
thus $C_v(T)=\frac{\pi^2}{3} D(0) T$. Similarly, $\kappa_u(T)= D(0) + 1.1486 AT^{5/3}$.

At $c>c_0$, we have $\tilde{\Delta}=\min_k \epsilon_+(k)<0$ and
$D(\omega)=\text{const.}$ for $|\omega|<|\tilde{\Delta}|$, thus $C_v(T)=\frac{\pi^2}{3} D(0) T$.
Similarly,
$\kappa_u(T)=D(0)+2[D(\tilde{\Delta})-D(0)]Te^{-|\tilde{\Delta}|/T}$.

When taking only the leading term in the low-temperature specific heat and compressibility, we have:

When $\Delta\neq0$: at $c<c_0$, $C_v(T)=2D(\tilde{\Delta})\tilde{\Delta}^2 T^{-1}e^{-\tilde{\Delta}/T}$ and
$\kappa_u(T)= 2D(\tilde{\Delta}) Te^{-|\tilde{\Delta}|/T}$ ;
at $c\geq c_0$, $C_v(T)=\frac{\pi^2}{3} D(0)T$ and $\kappa_u(T)= D(0) $.

When $\Delta=0$: at $c<c_0$, $C_v(T)=9\zeta(3)D'(0^+) T^2$ and $\kappa_u(T)=2\ln2 D'(0^+) T$;
at $c\geq c_0$, $C_v(T)=\frac{\pi^2}{3} D(0) T$ and  $\kappa_u(T)= D(0) $.

{\sl (c). Wilson ratio}

The Wilson ratio is defined as $R_W=T\kappa_u/C_v$ which has
the following low temperature behaviours:

{\em  $\Delta\neq0$ case }

At $c<c_0$, $R_W=(T/\tilde{\Delta})^2$;
At $c \geq c_0$, $R_W=3/\pi^2$.

{\em  $\Delta = 0$ case }

At $c<c_0$, $R_W=2\ln2/(9\zeta(3))\approx0.1281$;
At $c \geq c_0$, $R_W=3/\pi^2$.

These results are consistent with those achieved directly on the lattice in Sec.II-C-3
and also listed in the last line of Table-I.

\subsection{ The bulk topological phase transitions near $|h/t|\sim 0$  (the $\alpha_x\alpha_y<0$ case). }

When $h\sim 0$ and near the two valleys $K_1=(\pi,0)$ and $K_2= (0, \pi)$, we have
\begin{align}
    H_1&\!=\![\Delta\!-\!\alpha (k_x^2-k_y^2)]\sigma_z\!-\!vk_x\sigma_x\!+\!vk_y\sigma_y\!-\!ck_y\sigma_0        \nonumber  \\
    H_2&\!=\![\Delta\!+\!\alpha (k_x^2-k_y^2)]\sigma_z\!+\!vk_x\sigma_x\!-\!vk_y\sigma_y\!+\!ck_y\sigma_0
\label{eq:Dirac12}
\end{align}
where $\Delta=-h$ and other parameters are the same as the $h\sim 4t$ case discussed in Sec.III-A.
Note the opposite sign of the velocities $v$ between $k_x$ and $k_y$
and opposite sign of $\alpha$ between $k_x^2$ and $k_y^2$ indicating $\alpha_x\alpha_y<0$.


  Due to the two valleys, we obtain four bands
\begin{align}
	\epsilon_{i,\pm}=\pm\sqrt{v^2k^2+[\Delta+\eta_i\alpha(k_x^2-k_y^2)]^2}+\eta_i c k_y\>,
\label{opposite}
\end{align}
where $\eta_i=(-1)^i$ and $i=1,2$. It can be compared to Eq.\ref{Dirac0eng} in the $\alpha_x\alpha_y > 0$ case.
So the two cases should have quite different physical properties.

When $\alpha\Delta<0$,
the two critical velocities $c_1=|v| < c_2=\sqrt{v^2-4\alpha\Delta}$. When $\sqrt{v^2-4\alpha\Delta}>c>|v|$,
Thus only $H_1$ becomes metal. When $c>\max(c_1,c_2)$,  both $H_1$ and $H_2$ become metal.

When $\alpha\Delta>0$,
the two critical velocities $c_1=\sqrt{v^2+4\alpha\Delta} > c_2=|v|$. When $\sqrt{v^2+4\alpha\Delta}>c>|v|$,
Thus only $H_2$ becomes metal. When $c>\max(c_1,c_2)$,  then both $H_1$ and $H_2$ become metal.

Another new feature of $\alpha_x\alpha_y<0$ is that
the FS extends to infinity when $c\geq\sqrt{2}|v|$.
The divergent $k_F$ suggests a FS collision between the two valleys,
which is consistent with the existence of phase C1 or C2 in the global lattice phase diagram Fig.\ref{fig:phaseLattice}.
Below, we will not discuss the C- phase, so restrict  $c<\sqrt{2}|v|$. If $c > \sqrt{2}|v|$, it leads to the class-3 TPT
discussed in \cite{weyl} and reviewed in Sec.II.

\begin{figure}[tbhp]
    \centering
    \includegraphics[width=0.7\linewidth]{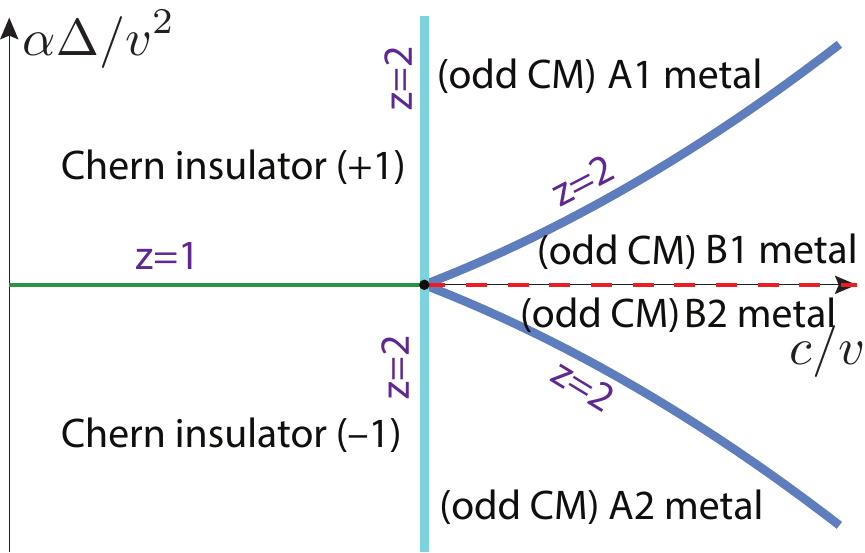}
    \caption{ The phase diagram of Eq.\eqref{eq:Dirac12}. It is the part near $ h/t \sim \pm 0 $ in the global phase diagram in Fig.\ref{fig:phaseLattice}. Similar to Fig.\ref{fig:Dirac0},
    Thick/thin/dashed line are 2nd/3rd/infinite order Topological phase transitions (TPTs) respectively.
    The consecutive Chern insulator to A1 Odd Chern metal, then to B1 Odd Chern metal transitions can be read from
    Fig.\ref{fig:Dirac12_FS}. The dashed line from the B1 Odd Chern metal to
    the B2 Odd Chern metal can be read from the B1 box and B1/B2 box on the right in Fig.\ref{fig:phaseLattice}.
    Similar to A1 Odd Chern metal to band metal transition in Fig.\ref{fig:Dirac0},
    it is also induced by the conic touching of the P- and H- FS and also infinite order. }
\label{fig:Dirac12}
\end{figure}

\begin{figure}[tbhp]
    \centering
   \includegraphics[width=\linewidth]{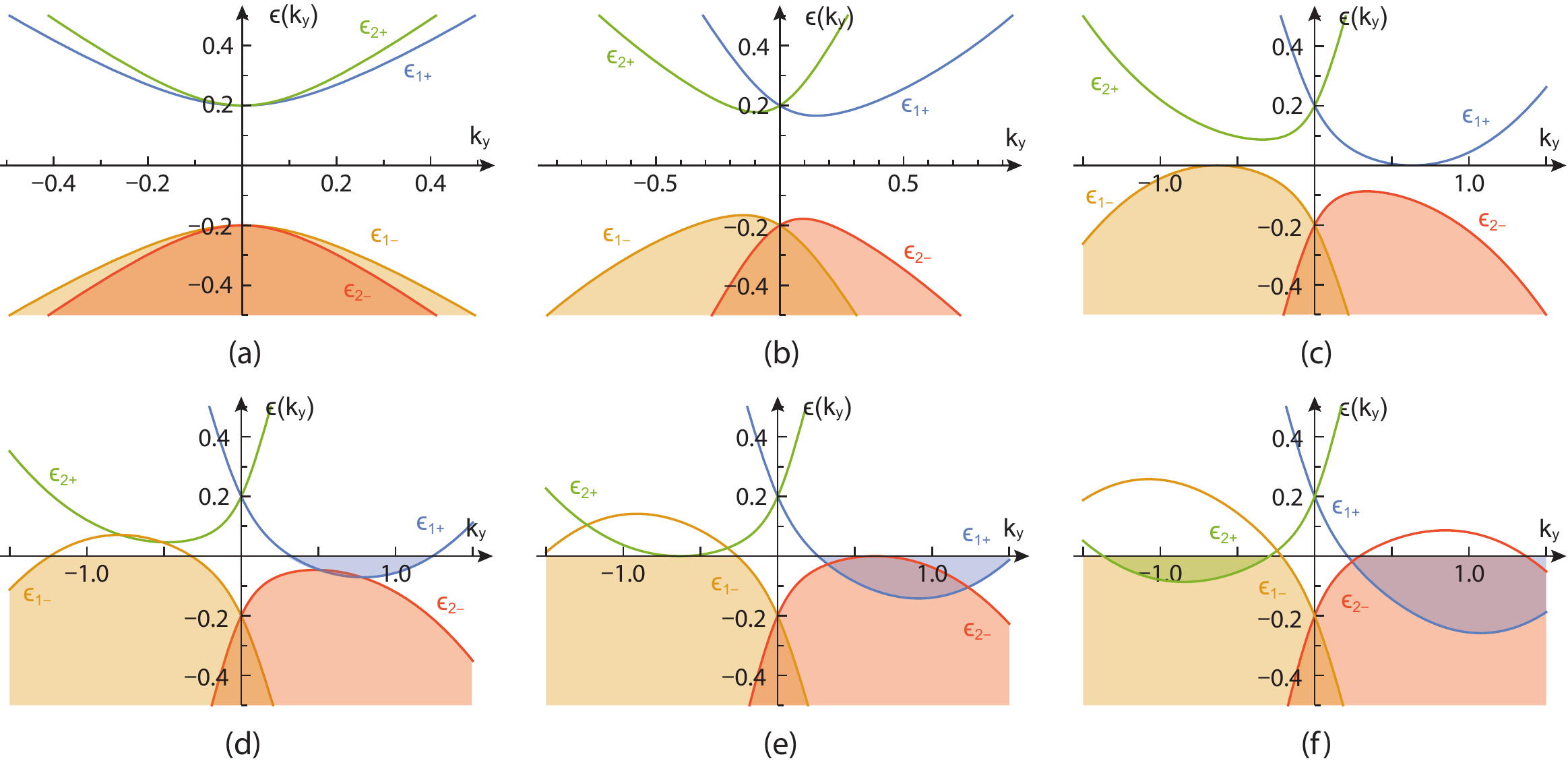}
    \caption{ The dispersion $\epsilon_{i,\pm}(k)$ as function of $k_y$
    with fixed $k_x=0$ and $v=1$, $\Delta=1/5$, $\alpha=-1/2 < 0 $, $c=0,0.5,1.0,\sqrt{1.4},1.3$.
    (a) both $H_1$ and $H_2$ has a direct gap at $k_y=0$; in Chern insulator phase.
    (b) Due to the opposite boost velocity in Eq.\ref{opposite}, the dispersion at $ K_1 $ and $ K_2 $ shift to the opposite directions.
    both $H_1$ and $H_2$ has an indirect gap at $k_y\neq 0$;
    (c) $H_1$ becomes gapless and shows two Fermi points at $k_y=\pm k_0$ with $ z=2 $,
    but $H_2$ still has an indirect gap. It corresponds to A1/B1 in Fig.\ref{fig:phaseLattice}.
    (d) $H_1$ becomes gapless and show two finite Fermi pockets,
    but $H_2$ becomes gapless and shows two Fermi points still at $k_y=\pm k_0$ also with $ z=2 $ after subtracting the non-critical $ H_1 $ part;
    (e) both $H_1$ and $H_2$ are gapless and show finite Fermi pockets \cite{Nooverlap}.
    It corresponds to B1 in Fig.\ref{fig:phaseLattice}.
    }
    \label{fig:Dirac12_FS}
\end{figure}

\subsubsection{ Hall conductance at zero and finite temperatures }

{\sl (a). Zero temperature }

When $c<\min(c_1,c_2)$,
the Hall conductance $\sigma_{H}$ of each valleys gives $\sigma_{H}^{(i)}=-\sgn(\Delta)/2$,
adding the two contributions together leads to
\begin{align}
    \sigma_{H}=\sigma_{H}^{(1)}+\sigma_{H}^{(2)}=-\sgn(\Delta)
\end{align}
which indicates $\sigma_{H}=-1$ for $\Delta>0$ and $\sigma_{H}=+1$ for $\Delta<0$.

If $c>\min(c_1,c_2)$, then we need to discus $\alpha\Delta>0$ or $\alpha\Delta<0$ separately.

{\em Case $\alpha\Delta<0$ }:

If $c_2>c>c_1=|v|$,  $H_1$ first develops an instability
while $H_2$ remains gapped.
\begin{align}
    \sigma_{H}^{(1)}&=-\sgn(\Delta)/2+\sgn(\Delta)\frac{c-v}{c}\Theta(v-c),\quad
      \nonumber   \\
    \sigma_{H}^{(2)}&=-\sgn(\Delta)/2
\end{align}

If $c>c_2$, both $H_1$ and $H_2$ are gapless,
however, $\sigma_{xy}^{(2)}$ remains the same as $c=0$ as dictated by Eq.\ref{nub}.
Thus, we arrive at
\begin{align}
    \sigma_{H}=\sgn(\Delta)\min(1,v/c)
\label{reduceHall}
\end{align}

{\em Case $\alpha\Delta>0$ }:

If $c_1>c>c_2=|v|$,  $H_2$ first develops an instability
while $H_1$ remains gapped.
\begin{align}
    \sigma_{H}^{(1)} & =-\sgn(\Delta)/2,\quad
     \nonumber   \\
    \sigma_{H}^{(2)} & =-\sgn(\Delta)/2+\sgn(\Delta)\frac{c-v}{c}\Theta(v-c)
\end{align}

If $c>c_1$, both $H_1$ and $H_2$ are gapless,
however, $\sigma_{xy}^{(1)}$ remains the same as $c=0$ as also dictated by Eq.\ref{nub}.
Thus, we also arrive at Eq.\ref{reduceHall}.

So we always have the Hall conductance Eq.\ref{reduceHall}, no matter in the insulating phase or
the metallic phase which is independent of many microscopic details.


\begin{figure}[tbhp]
    \centering
    \includegraphics[width=\linewidth]{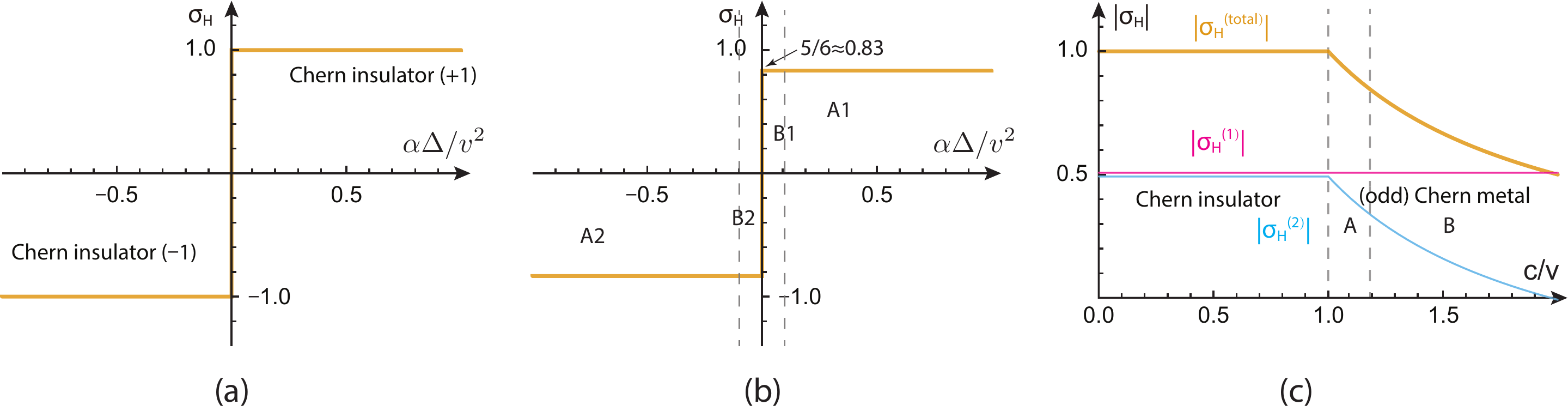}
    \caption{The Hall conductance of Eq.\ref{eq:Dirac12} at different $\alpha\Delta/v^2$ or $c/v$ values:
    varying $\alpha\Delta/v^2$ in (a) and (b):
    (a) fix $c/v=0.5$   (b) fix  $c/v=1.2$.
    There is a jump from B2 to B1 with the magnitude $ 2 \times v/c = 2 \times 5/6 $,
    but no change from B1 to A1 Odd Chern metal phase. However, as indicated in Fig.\ref{fig:Dirac12}, the former is infinite order,
    the latter is 2nd order with $ z=2 $.
    See also Fig.\ref{fig:Hall_T0}b on a lattice.
    Varying $c/v$ in (c) fix $\alpha\Delta/v^2=-0.1$,  $ \sigma^{1}_H  $ stays as a constant. But
    $ \sigma_H= \sigma^{1}_H+ \sigma^{2}_H = v/c $. }
    \label{fig:sigma_C2}
\end{figure}

{\sl (b). Finite temperature Hall conductance}

Similar to the $\alpha_x\alpha_y>0$ case in Eq.\ref{finiteTsigmaHh4},
the finite temperature Hall conductance at the valley $ i $ is
\begin{align}
    \sigma_{H}^{(i)}(T)
        &=\sigma_{H}^{(i)}(T=0)+\frac{1}{\pi}\int_{\mathbb{R}^2} d^2\mathbf{k}\>
            \Omega_{i,+}(\mathbf{k})
        \nonumber  \\
        &\times [f(\epsilon_{i,+}(\mathbf{k}))-\Theta(-\epsilon_{i,+}(\mathbf{k}))]
                    \nonumber  \\
		&=\mathrm{Ch}_{i,-}+\frac{1}{\pi}\int_{\mathbb{R}^2} d^2\mathbf{k}\>
			\Omega_{i,+}(\mathbf{k})f(\epsilon_{i,+}(\mathbf{k}))
\label{eq:sigma_xy(T)_12_2}
\end{align}
Thus, the total Hall conductance is:
\begin{align}
    \sigma_{H}(T)
        &=\sigma_{H}^{(1)}(T)+\sigma_{H}^{(2)}(T)
                \nonumber  \\
        &=\mathrm{Ch}_{1,-}+\mathrm{Ch}_{2,-} + \nu_{1}(T)+\nu_{2}(T)
\end{align}
where $\mathrm{Ch}_{i,-}$ and $\nu_{i}(T=0)$ has been studied in Sec.III-A-1.

At a low temperature $T$ which is the lowest energy scale,
one can get a low temperature expansion of $\nu_i(T)$.
Without loss of generality, we choose $\alpha\Delta<0$ case,
the two critical velocities $c_1=|v| < c_2=\sqrt{v^2-4\alpha\Delta}=\sqrt{v^2+4|\alpha\Delta|}$.
By keeping the leading low $T$ dependence and assuming $|\alpha\Delta/v^2|\ll 1/2$, we have

At $c<c_1$,
$H_1$ has a gap $2\tilde{\Delta}_1$ and $H_2$ has a bigger gap $2\tilde{\Delta}_2 > 2\tilde{\Delta}_1 $.
Thus,
$\nu_1(T)-\nu_1(T=0)\propto -\sgn(\Delta)Te^{-|\tilde{\Delta}_1|/T}$
and $\nu_2(T)-\nu_2(T=0)\propto -\sgn(\Delta)Te^{-|\tilde{\Delta}_2|/T}$;

At $c=c_1$,
$H_1$ is critical with $\tilde{\Delta}_1=0$ and $H_2$ still has a gap $2\tilde{\Delta}_2$.
Thus,
$\nu_1(T)-\nu_1(T=0)\propto -\sgn(\Delta)T$
and $\nu_2(T)-\nu_2(T=0)\propto -\sgn(\Delta)Te^{-|\tilde{\Delta}_2|/T}$;

At $c_2>c>c_1$,
$H_1$ is gapless and $H_2$ still has a gap $2\tilde{\Delta}_2$.
Thus,
$\nu_1(T)-\nu_1(T=0)\propto -\sgn(\Delta)T^2$
and $\nu_2(T)-\nu_2(T=0)\propto -\sgn(\Delta)Te^{-|\tilde{\Delta}_2|/T}$;

At $c=c_2$,
$H_1$ is gapless and $H_2$ is also critical with $\tilde{\Delta}_2=0$.
Thus,
$\nu_1(T)-\nu_1(T=0)\propto -\sgn(\Delta)T^2$
and $\nu_2(T)-\nu_2(T=0)\propto -\sgn(\Delta)T$;

At $c>c_2$,
both $H_1$ and $H_2$ are gapless.
Thus,
$\nu_1(T)-\nu_1(T=0)\propto -\sgn(\Delta)T^2$
and $\nu_2(T)-\nu_2(T=0)\propto -\sgn(\Delta)T^2$.

In summary, if we only need the leading behaviors of $\nu(T)=\nu_1(T)+\nu_2(T)$,
then we have:

At $c<c_1$, $\nu(T)-\nu(T=0)\propto -\sgn(\Delta)Te^{-|\tilde{\Delta}_1|/T}$;
at $c=c_1$, $\nu(T)-\nu(T=0)\propto -\sgn(\Delta)T$;
at $c_2>c>c_1$, $\nu(T)-\nu(T=0)\propto -\sgn(\Delta)T^2$;
at $c=c_2$, $\nu(T)-\nu(T=0)\propto -\sgn(\Delta)T$;
at $c>c_2$, $\nu(T)-\nu(T=0)\propto -\sgn(\Delta)T^2$.

\subsubsection{ Ground-state energy density and topological phase transitions }

When $c<\min(c_1,c_2)$, $\min\epsilon_{i,+}\geq0\geq\max\epsilon_{i,-}$,
both valley 1\&2's lower band is full occupied and 1\&2's upper band is complete empty,
thus the ground-state energy density is
$E_\text{GS}=(4\pi^2)^{-1}\int_{\mathbb{R}^2}d^2\mathbf{k} \> [\epsilon_{1-}(\mathbf{k})+\epsilon_{2-}(\mathbf{k})]$.
The integral is divergent, we need a cut-off $\Lambda$ for $k$.
Due to its odd property, the $ck_y$ part in Eq.\ref{opposite} drops off, one obtain:
\begin{align}
    E_0=-\frac{|\Delta|^3}{3\pi v^2}+E_\Lambda
\label{ELambda2}
\end{align}
where $E_\Lambda$ is $\Lambda$, $\alpha\Delta$ and $v$ dependent $C^3$ function.
Note that $E_\Lambda$ in the $\alpha_x\alpha_y<0$ case is dramatically different from that in the
$\alpha_x\alpha_y>0$ case listed in Eq.\ref{ELambda1}.

When $\max(c_1,c_2)>c>\min(c_1,c_2)$, one of $\epsilon_{1,-}$ and $\epsilon_{2,-}$ are partially filled,
then there is another parts $E_1$ and $E_\text{GS}=E_0+E_1$;
When $c>\max(c_1,c_2)$, both $\epsilon_{1,-}$ and $\epsilon_{2,-}$ are partially filled,
then there is also one more part $E_2$ and $E_\text{GS}=E_0+E_1+E_2$,
where $E_1$ and $E_2$ are slightly more complicated than that in the $\alpha_x\alpha_y>0$
case listed in Eq.\ref{E1part}:
\begin{align}
    E_1=\frac{-\pi}{32|\alpha|^3}
    [&8\alpha\Delta v^2+2v^4(\arcsin\frac{c}{\sqrt{2}v}-\frac{\pi}{4})
               \nonumber  \\
    & +c\sqrt{2v^2-c^2}(v^2-8\alpha\Delta-c^2)]                   \nonumber  \\
    E_2=\frac{-\pi}{32|\alpha|^3}
    [&4\alpha\Delta\sqrt{v^4-16\alpha^2\Delta^2}
                       \nonumber  \\
    &+2v^4(\arcsin\frac{c}{\sqrt{2}v}-\frac{1}{2}\arccos\frac{4\alpha\Delta}{v^2})
    \nonumber  \\
    &+c(v^2-8\alpha\Delta-c^2)\sqrt{2v^2-c^2}]
\end{align}

{\sl (a). Chern insulator (+1) to Chern insulator (-1) transition}:

In the range $ c< v $ and tuned by $\Delta$, one only have  the $E_0$ part which has
a third-order non-analytical behaviour at $\Delta=0$,
\begin{align}
    E_\text{GS}=-\frac{|\Delta|^3}{3\pi v^2}+\cdots
\end{align}
which has the dynamic exponent $ z=1 $.

{\sl (b). B1 Odd Chern metal to B2 Odd Chern metal transition}:

In the range $ c > v $ and the driven parameter is $\Delta$.
Now we have both $E_0$ and $E_1 + E_2$ part, the third-order non-analytical term at $\Delta=0$ from the two parts gets canceled,
\begin{align}
    E_\text{GS}= -\frac{|\Delta|^3}{3\pi v^2}+ \frac{|\Delta|^3}{3\pi v^2} + \cdots
\end{align}
which explainers why
the 3rd order TPT across the green line changes to infinite order across the red dashed line in Fig.\ref{fig:Dirac12}.
Even so, as shown in Fig.17b, there is still a universal jump $ \Delta \sigma_H=2v/c $, so it is still a TPT.

{\sl (c).  Odd Chern A metal to Odd Chern B metal transition }

In the range $\alpha\Delta>0$ and tuned by either $\Delta$ or $c$,
one finds a second-order non-analytical behaviour at $\Delta=(c^2-v^2)/(4\alpha)$ tuned by $\Delta$,
\begin{align}
    E_\text{GS}&=\frac{v^2-c^2}{4\pi\alpha\sqrt{c^2(2v^2-c^2)}}[\Delta-(c^2-v^2)/(4\alpha)]^2
	\nonumber  \\
&\times \Theta(c^2-v^2-4\alpha\Delta)+\cdots
\end{align}
and also a second-order non-analytical behaviour at $c=\sqrt{v^2+4\alpha\Delta}$ tuned by $ c $,
\begin{align}
    E_\text{GS}&=
-\frac{\Delta\sqrt{v^2+4\alpha\Delta}}{4\pi \alpha\sqrt{v^2-4\alpha\Delta}} (c-\sqrt{v^2+4\alpha\Delta})^2
    \nonumber  \\
& \times \Theta(c^2-v^2-4\alpha\Delta)+\cdots
\end{align}
  which has the dynamic exponent $ z=2 $.

{\sl (d). Chern insulator to A Odd Chern metal transition}:

In the range $\alpha\Delta>0$ and tuned by $c$, one finds a second-order non-analytical behaviour at $c=v$,
\begin{align}
    E_\text{GS}=E_0-\frac{(c-v)^2\Delta}{4\pi\alpha^2}\Theta(c-v)+\cdots
\end{align}
which has the dynamic exponent $ z=2 $.

In summary, (c) and (d) are 2nd-order with $ z=2 $;
(a) is 3rd-order with $ z=1 $; (b) is infinite-order.

\subsubsection{ Thermodynamic quantities  }

{\sl (a). The density of states }

From the Hamiltonian \eqref{eq:Dirac}, the Matsubara Green's function
\begin{align}
    G_i(\mathbf{k},i\omega_n)
    =\frac{P_{i,+}(\mathbf{k})}{i\omega_n-\epsilon_{i,+}(\mathbf{k})}
    +\frac{P_{i,-}(\mathbf{k})}{i\omega_n-\epsilon_{i,-}(\mathbf{k})}
\end{align}
where $P_{i,s}$ are the projection operators onto $s=\pm$ bands.

\begin{figure}[tbhp]
    \centering
    \includegraphics[width=0.9\linewidth]{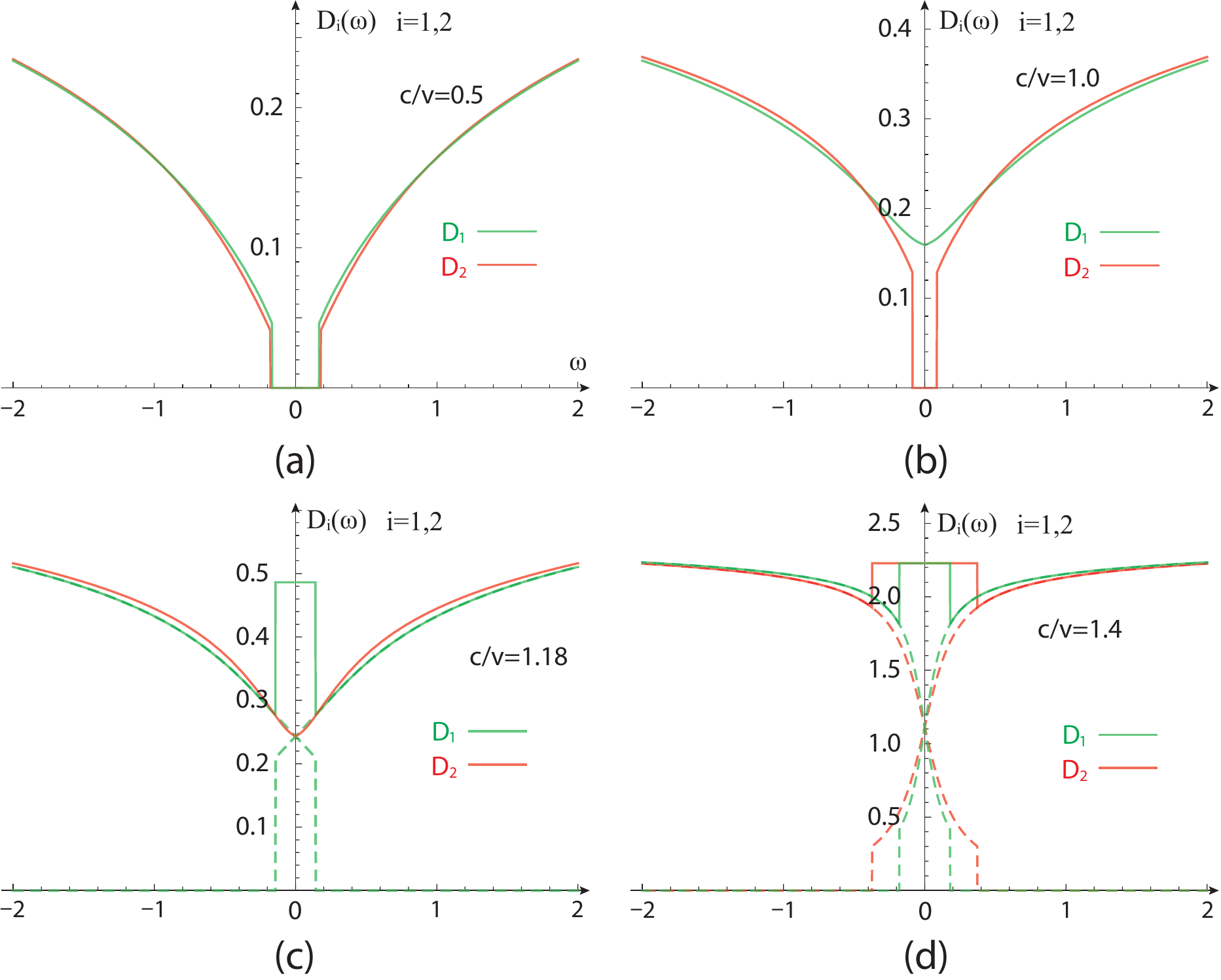}
    \caption{The Density of states $ D_1 (\omega ) $ and $ D_2(\omega ) $ of Eq.\eqref{eq:Dirac12} at different $c/v$ values:
    (a) $c/v=0.5$, (b) $c/v=1.0$, (c) $c/v=\sqrt{1.4}\approx1.18$, (d) $c/v=1.4$.
    Here, we fixed $v=1$, $\alpha=1/2$, $\Delta=1/5$. In (c) and (d),
    the central plateau DOS value is given in Eq.\ref{centralDOS12}.
    Again, similar to Fig.\ref{fig:DOS_C1}, the flat feature of total DOS near $\omega=0$
    is a result of truncation of $H(k)$ at $k^2$ order,
    and the lattice DOS shown in Fig.\ref{fig:DOS_L2} is not a constant anymore.
    The total DOS $ D(\omega)= D_1 (\omega ) +  D_2(\omega ) $ is just the sum of the two.  }
    \label{fig:DOS_C2}
\end{figure}

The total DOS for each flavor is
\begin{align}
    D_i(\omega)
    =\int\frac{d^2q}{(2\pi)^2}[\delta(\omega-\epsilon_{i,+}(k))+\delta(\omega-\epsilon_{i,-}(k))]
\end{align}
It would be convenient to introduce the DOS for each band $ D_{i}(\omega)=D_{i,+}(\omega)+D_{i,-}(\omega) $:
\begin{align}
    D_{i,\pm}(\omega)=\int \frac{d^2\mathbf{k}}{4\pi^2}\delta(\omega-\epsilon_{i,\pm}(\mathbf{k})),
\end{align}
Due to the C-symmetry, $D_{i,+}(\omega)=D_{i,-}(-\omega)$.
Similar to Eq.\ref{DOSflat}, we also have
\begin{align}
        D_{i,+}\!(\omega)\!=\!\frac{1}{4\pi|\alpha|}\sqrt{\!\frac{c^2}{2v^2\!-\!c^2}}
        [1\!+\!\rho_i(\omega)]\Theta(\omega\!-\!\min\epsilon_{i,+}\!)
\end{align}
where $\rho_i(\omega)=-\rho_i(-\omega)$ is an odd function.
When $\min\epsilon_{i,+}<0<\max\epsilon_{i,-}$, the odd property leads to the
flat feature of total DOS $ D(\omega)=D_{1}(\omega)+D_{2}(\omega)  $  for $H_i$ near $\omega=0$
( Fig.\ref{fig:DOS_C2} ):
\begin{align}
    D_{i}(\omega)\!=\!\frac{1}{2\pi|\alpha|}\sqrt{\frac{c^2}{2v^2\!-c^2}},
    \min\epsilon_{i,+}<\omega<\max\epsilon_{i,-}
\label{centralDOS12}
\end{align}
 where as alerted below Eq.\ref{opposite}, we restrict $ c < \sqrt{2} v $, so that the DOS remain finite.
 Otherwise, one must resort to the lattice calculations in Sec.II.

{\sl (b). The specific heat and  compressibility }


The Helmholtz free energy density $ F(T) =F_1(T)+F_2(T) $:
\begin{align}
    & F_i(T)=E_{i,\text{GS}} + \int\frac{d^2\mathbf{k}}{(2\pi)^2}
    \Big(|\epsilon_{i,+}(\mathbf{k})|
    \nonumber   \\
    &-T\ln[2(1+\cosh(\epsilon_{i,+}(\mathbf{k})/T))]
    \Big)
\end{align}
where the zero temperature part of $F(T)$ is nothing but
the ground-state energy density $E_\text{GS}$ calculated in Sec.III-B-2.
The specific heat (at constant volume) is given by $ C_v(T)=C_{v,1}(T)+C_{v,2}(T) $:
\begin{align}
    C_v(T) =\int\frac{d^2\mathbf{k}}{(2\pi)^2}
            \sum_{i=1,2}\frac{[\epsilon_{i,+}(\mathbf{k})]^2}
            {T^2[1+\cosh(\epsilon_{i,+}(\mathbf{k})/T)]}
\end{align}

Similarly,
the (isothermal) uniform compressibility $ \kappa_u(T)= \kappa_{u,1}(T)+\kappa_{u,2}(T) $ is
\begin{align}
    \kappa_u(T)
    & =\sum_{i,s}\int \frac{d^2k}{(2\pi)^2}\frac{\partial f(\epsilon_{i,s}(k))}{\partial \epsilon_{i,s}(k)}
    =\int \frac{d\omega D(\omega)}{2T(1+\cosh(\omega/T))}
\end{align}
where $ D_i(\omega), D(\omega)  $ are listed in Eq.\ref{centralDOS12}.

Its zero temperature limit $\frac{\partial f(\epsilon)}{\partial \epsilon}=\delta(\epsilon)$
recovers the well-known relation between the compressibility and the DOS $\kappa_u(T)=D(0)$.

At low temperature $T$ which is the lowest energy scale,
one can get a low temperature expansion of $C_v(T)$ and $ \kappa_u(T) $.
Without loss of generality, we choose $\alpha\Delta<0$ case in Fig.\ref{fig:Dirac12},
the two critical velocities $c_1=|v| < c_2=\sqrt{v^2-4\alpha\Delta}=\sqrt{v^2+4|\alpha\Delta|} $.
By keeping the leading low $T$ dependence and assuming $|\alpha\Delta/v^2| <  1/2$, we have

{\em $\Delta\neq0$ cases:}

At $c<c_1$, $H_1$ has a gap $2\tilde{\Delta}_1>0$ and $H_2$ has a bigger gap $2\tilde{\Delta}_2> 2\tilde{\Delta}_1  $,
$D_i(\omega)=0$ for $|\omega|<\tilde{\Delta}_i$ ( Fig.\ref{fig:DOS_C2}a ).
Thus,
$C_{v,i}(T)=2D_{i}(\tilde{\Delta}_i)\tilde{\Delta}_i^2 T^{-1}e^{-|\tilde{\Delta}_i|/T}$,
so $C_{v}(T)= 2D_{1}(\tilde{\Delta}_1)\tilde{\Delta}_1^2 T^{-1}e^{-|\tilde{\Delta}_1|/T}$.
Similarly,
$\kappa_{u,i}(T)= 2D_i(\tilde{\Delta}_i) T e^{-\tilde{\Delta}_i/T}$,
so $\kappa_{u}(T)= 2D_1(\tilde{\Delta}_1) T e^{-\tilde{\Delta}_1/T}$. Both $C_{v}(T) $ and $\kappa_{u}(T) $
are dominated by the node 1.

At $c=c_1$, $H_1$ is critical with $\tilde{\Delta}_1=0$ and $H_2$ still has a gap $2\tilde{\Delta}_2>0$,
$D_1(\omega)=D_{1}(0)+D_{1}'(0^+)|\omega|$ for small $|\omega|$,
$D_2(\omega)=0$ for $|\omega|<\tilde{\Delta}_2$ ( Fig.\ref{fig:DOS_C2}b ).
Thus,
$C_{v,1}(T)=\frac{\pi^2}{3}D_1(0)T$
and $C_{v,2}(T)=2D_2(\tilde{\Delta}_2)\tilde{\Delta}_2^2T^{-1}e^{-|\tilde{\Delta}_2|/T}$,
so $C_{v}(T)=\frac{\pi^2}{3}D_1(0)T$.
Similarly,
$\kappa_{u,1}(T)=D_1(0)+2\ln2 D'_1(0^+) T$ and
$\kappa_{u,2}(T)=2D_2(\tilde{\Delta}_2) T e^{-\tilde{\Delta}_2/T}$,
so $\kappa_{u}(T)=D_1(0)+2\ln2 D'_1(0^+) T$. Both $C_{v}(T) $ and $\kappa_{u}(T) $
are still dominated by the node 1.

At $c_2>c>c_1$,
$H_1$ is gapless with $\tilde{\Delta}_1<0$ and $H_2$ still has a gap $2\tilde{\Delta}_2 > 0 $,
$D_1(\omega)= \frac{1}{2\pi|\alpha|}\sqrt{\frac{c^2}{2v^2-c^2}}  $ in Eq.\ref{centralDOS12}   for $|\omega|<|\tilde{\Delta}_1|$,
$D_2(\omega)=0$ for $|\omega|<\tilde{\Delta}_2$.
Thus,
$C_{v,1}(T)=\frac{\pi^2}{3}D_{1}(0)T$
and $C_{v,2}(T)=2D_{2}(\tilde{\Delta}_2)\tilde{\Delta}_2^2T^{-1}e^{-|\tilde{\Delta}_2|/T}$,
so $C_{v}(T)=\frac{\pi^2}{3}D_1(0)T$. Similarly,
$\kappa_{u,1}(T)=D_1(0)+2[D_1(\tilde{\Delta}_1)-D_1(0)]Te^{-|\tilde{\Delta}_1|/T}$
and $\kappa_{u,2}(T)=2D_2(\tilde{\Delta}_2)Te^{-|\tilde{\Delta}_2|/T}$,
so $\kappa_{u}(T)=D_1(0)
+2[D_1(\tilde{\Delta}_1)-D_1(0)]Te^{-|\tilde{\Delta}_1|/T}
+2D_2(\tilde{\Delta}_2)Te^{-|\tilde{\Delta}_2|/T}$.
However, the relative magnitude of $|\tilde{\Delta}_1|$ and $|\tilde{\Delta}_2|$ depends on $c$, so one can not tell which is
smaller in general.

At $c=c_2$,
$H_1$ is gapless with $\tilde{\Delta}_1<0$ and $H_2$ becomes critical with $\tilde{\Delta}_2=0$,
$D_1(\omega)= \frac{1}{2\pi|\alpha|}\sqrt{\frac{c^2}{2v^2-c^2}}  $ in Eq.\ref{centralDOS12} for $|\omega|<|\tilde{\Delta}_1|$,
$D_2(\omega)=D_2(0)+D_2'(0^+)|\omega|$ for small $|\omega|$ ( Fig.\ref{fig:DOS_C2}c ).
Thus, $C_{v,1}(T)=\frac{\pi^2}{3}D_{1}(0)T$ and $C_{v,2}(T)=\frac{\pi^2}{3}D_{2}(0)T$,
so $C_{v}(T)=\frac{\pi^2}{3}D(0)T$. Similarly,
$\kappa_{u,1}(T)=D_1(0)+2[D_1(\tilde{\Delta}_1)-D_1(0)]Te^{-|\tilde{\Delta}_1|/T}$
and $\kappa_{u,2}(T)=D_2(0)+2\ln2 D_2'(0)T$,
so $\kappa_{u}(T)=D(0)+2\ln2 D_2'(0)T$.

At $c>c_2$,
$H_1$ is gapless with $\tilde{\Delta}_1<0$ and $H_2$ is also gapless with $\tilde{\Delta}_2<0$,
$D_1(\omega)= \frac{1}{2\pi|\alpha|}\sqrt{\frac{c^2}{2v^2-c^2}}  $ in Eq.\ref{centralDOS12}
for small $|\omega|<|\tilde{\Delta}_1|$,
$D_2(\omega)= \frac{1}{2\pi|\alpha|}\sqrt{\frac{c^2}{2v^2-c^2}}  $ in Eq.\ref{centralDOS12}
for small $|\omega|<|\tilde{\Delta}_2|$ ( Fig.\ref{fig:DOS_C2}d ).
Thus,
$C_{v,1}(T)=\frac{\pi^2}{3}D_{1}(0)T$ and $C_{v,2}(T)=\frac{\pi^2}{3}D_{2}(0)T$,
so $C_{v}(T)=\frac{\pi^2}{3}D(0)T$.
Similarly,
$\kappa_{u,1}(T)=D_1(0)+2[D_1(\tilde{\Delta}_1)-D_1(0)]Te^{-|\tilde{\Delta}_1|/T}$
and $\kappa_{u,2}(T)=D_2(0)+2[D_2(\tilde{\Delta}_1)-D_2(0)]Te^{-|\tilde{\Delta}_2|/T}$.
so $\kappa_{u}(T)= D(0)+2[D_2(\tilde{\Delta}_1)-D_2(0)]Te^{-|\tilde{\Delta}_2|/T}$.

{\em $\Delta=0$ cases:}

In this case, $c_1=c_2= c $ and $H_1$ and $H_2$ have exactly the same spectrum.

At $c<v$,
$D_i(\omega)=D'_{i}(0)|\omega|$ for small $|\omega|$,
thus $C_{v,1}(T)=C_{v,2}(T)=9\zeta(3)D'_{i}(0^+)T^2$
and $C_{v}(T)=9\zeta(3)D'(0^+)T^2$.
Similarly, $\kappa_{u,1}(T)=\kappa_{u,2}(T)=2\ln2 D_{i}'(0)T$
and $\kappa_{u}(T)=2\ln2 D'(0)T$.

At $c=v$ and $\Delta=0$, $D_i(\omega)=D_i(0)+A_{i}(0)|\omega|^{2/3}$ for small $|\omega|$,
thus $C_{v,1}(T)=C_{v,2}(T)=\frac{\pi^2}{3}D_{i}(0)T$
and $C_{v}(T)=\frac{\pi^2}{3}D(0)T$.
Similarly, $\kappa_{u,1}(T)=\kappa_{u,2}(T)=D_{i}(0)+1.1486 A_{i} T^{5/3}$
and  $\kappa_{u}(T)=D(0)+1.1486 A T^{5/3}$.

At $c>v$ and $\Delta=0$, $D_i(\omega)=D_i(0)$ for small $|\omega|$,
$C_{v,1}(T)=C_{v,2}(T)=\frac{\pi^2}{3}D_{i}(0)T$ and $C_{v}(T)=\frac{\pi^2}{3}D(0)T$.
Similarly, $\kappa_{u,1}(T)=\kappa_{u,2}(T)=D_{i}(0)+2[D_i(\tilde{\Delta}_i)-D_i(0)]Te^{-|\tilde{\Delta}_i|/T}$
and  $\kappa_{u}(T)=D(0)+2[D(\tilde{\Delta})-D(0)]Te^{-|\tilde{\Delta}|/T}$.

If we only consider the leading low temperature behaviors of
$C_{v}(T)=C_{v,1}(T)+C_{v,2}(T)$ and $\kappa_{u}(T)=\kappa_{u,1}(T)+\kappa_{u,2}(T) $,
one can summarize these results  as:

{\em When $\Delta\neq0$ }:
at $c<c_1$, $C_{v}(T)= 2D_{1}(\tilde{\Delta}_1)\tilde{\Delta}_1^2 T^{-1}e^{-|\tilde{\Delta}_1|/T}$ and
$\kappa_{u}(T)= 2D_1(\tilde{\Delta}_1) T e^{-\tilde{\Delta}_1/T}$;
at $c\geq c_1$, $C_{v}(T)=\frac{\pi^2}{3}D(0)T$ and $\kappa_{u}(T)=D(0) $.

{\em When $\Delta=0$ }: at $c<v$, $C_{v}(T)=9\zeta(3)D'(0^+) T$ and $\kappa_{u}(T)=2\ln2 D'(0)T$;
at $c\geq v$, $C_{v}(T)=\frac{\pi^2}{3}D(0)T^2$ and $\kappa_{u}(T)=D(0) $.

{\sl (c). The Wilson ratio}

The Wilson ratio is defined as $R_W=T\kappa_u/C_v$ which has the
following low temperature behaviours.

{\em $\Delta\neq0$ case: }

When $c<c_0$,  $R_W=(T/\tilde{\Delta})^2$; when $c\geq c_0$, $R_W=3/\pi^2$.

{\em $\Delta=0$ case: }

When $c<c_0$, $R_W=2\ln2/(9\zeta(3))\approx0.1281$; when $c\geq c_0$, $R_W=3/\pi^2$.

These results are consistent with those achieved directly on the lattice in Sec.II-C-3
and also listed in the last line of Table-I.

\section{ The chiral edge properties in a strip geometry }

  Following the approach used in the bulk properties, we will first study the edge properties from the lattice system, then
  investigate them from the continuum effective theory, then contrast the two complementary approaches.

\subsection{ Edge states from the microscopic lattice theory }

For the periodic boundary condition in the $y$-direction
and open boundary condition in the $x$-direction,
$k_y$ is a good quantum number,
the Hamiltonian in the mixed $ ( i, k_y ) $ representation becomes
\begin{align}
    &H\! =\!\!\sum_{k_y,i,j}\!
            \!c_{i,k_y}^\dagger\! \{
            [-(h\!+\!2t\cos k_y)\sigma_z
	    \!\!+\!2t_s\sin k_y\sigma_y
	    \!\!+\!2t_b\sin k_y \sigma_0]\delta_{i,j}
            \nonumber   \\
    &  +(t\sigma_z\!-\!it_s\sigma_y)\delta_{i,j+1}
            +(t\sigma_z\!+\!it_s\sigma_y)\delta_{i,j-1}
            \} c_{j,k_y}
\end{align}

For the periodic boundary condition in the $x$-direction
and open boundary condition in the $y$-direction,
$k_x$ is a good quantum number,
the Hamiltonian in the mixed $ ( k_x, i ) $ representation becomes:
\begin{align}
	&H\! =\!\!\sum_{k_x,i,j}\!
        c_{k_x,i}^\dagger \{
        [-(h+2t\cos k_x)\sigma_z+t_s\sin k_x\sigma_x]\delta_{i,j}
                              \nonumber   \\
       & \!+\!(t\sigma_z\!-\!it_s\sigma_y\!-\!it_b\sigma_0)\delta_{i,j+1}
        \!+\!(t\sigma_z\!+\!it_s\sigma_y\!+\!it_b\sigma_0)\delta_{i,j-1}
        \} c_{k_x,j}
\end{align}

\begin{figure}[tbhp]
    \centering
    \includegraphics[width=\linewidth]{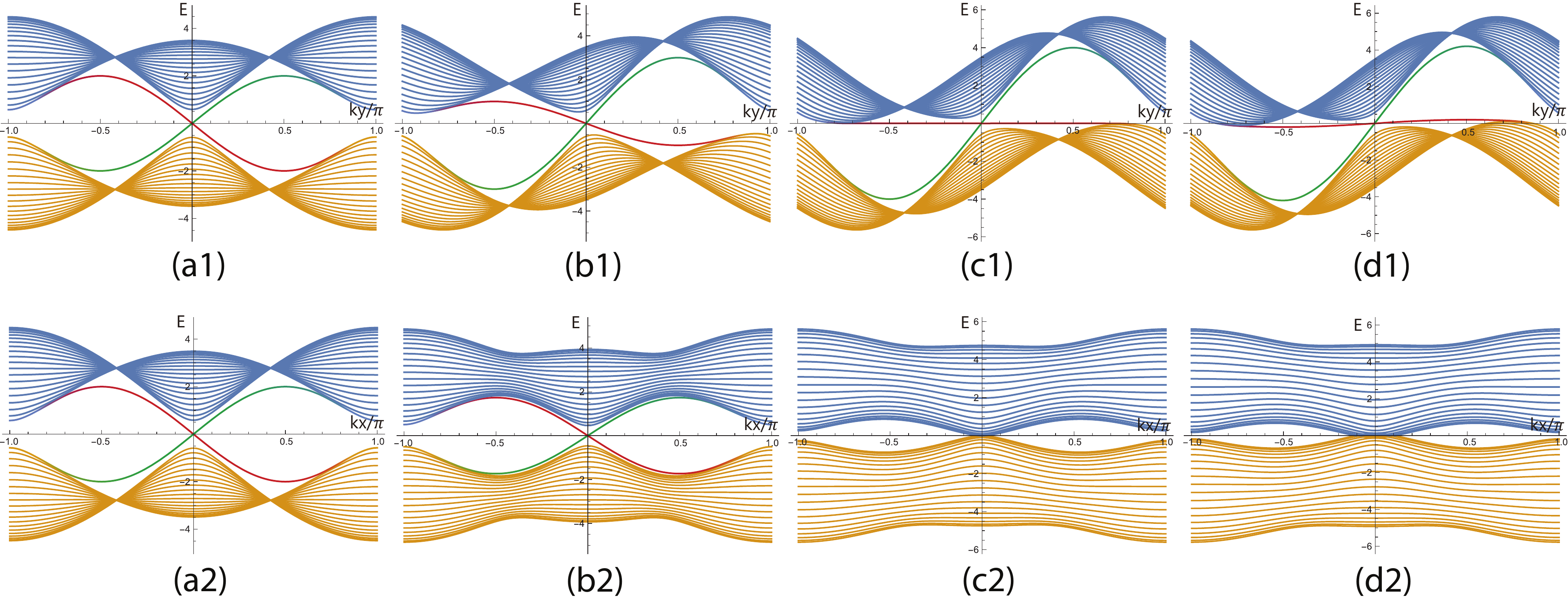}
    \caption{The edge states of the lattice Hamiltonian in a strip geometry. We fixed $h=-0.5$.
    From left to right, the parameter $t_b/t_s$ is $0,0.5,1.0,1.1$, respectively.	
    (Top) Longitudinal injection: With the periodic boundary condition in the $y$-direction and the open
    boundary condition in the $x$-direction. The edge modes always exist, but undergoes the edge reconstruction at $ t_s/t =1 $.
    The two edges move along the opposite directions when $ t_s/t < 1 $ in (a1) and (b1) in the CI,
    then  one edge becomes flat at $ t_s/t= 1 $ in (c1),
    then two edges move along the {\em same } direction when  $ t_s/t > 1 $ in (d1) in the odd CM,
	(Bottom) Transverse injection:
    Exchanging the role of $ x $ and $ y $ direction. The edge modes exist only when $ t_s/t < 1 $ in (a2) and (b2),
    but squeezed out at $ t_s/t= 1 $ in (c2) where the direct bulk gap closes, completely disappear when  $ t_s/t > 1 $ in (d2) in the odd CM.
    Although the edge modes show quite different behaviours in the line 1 and the line 2,
    there seems a one to one Longitudinal/Transvese edge-edge correspondence between them.
    In both figures, one can shift $k\to k+\pi$ to reach $h=+0.5$ results.
    See also Fig.\ref{fig:xEdge} and Fig.\ref{fig:yEdge} for the continuum calculations. See also Fig.S1 for the expanded figure. }
\label{fig:Edge_Lattice}
\end{figure}

\begin{figure}[tbhp]
    \centering
    \includegraphics[width=\linewidth]{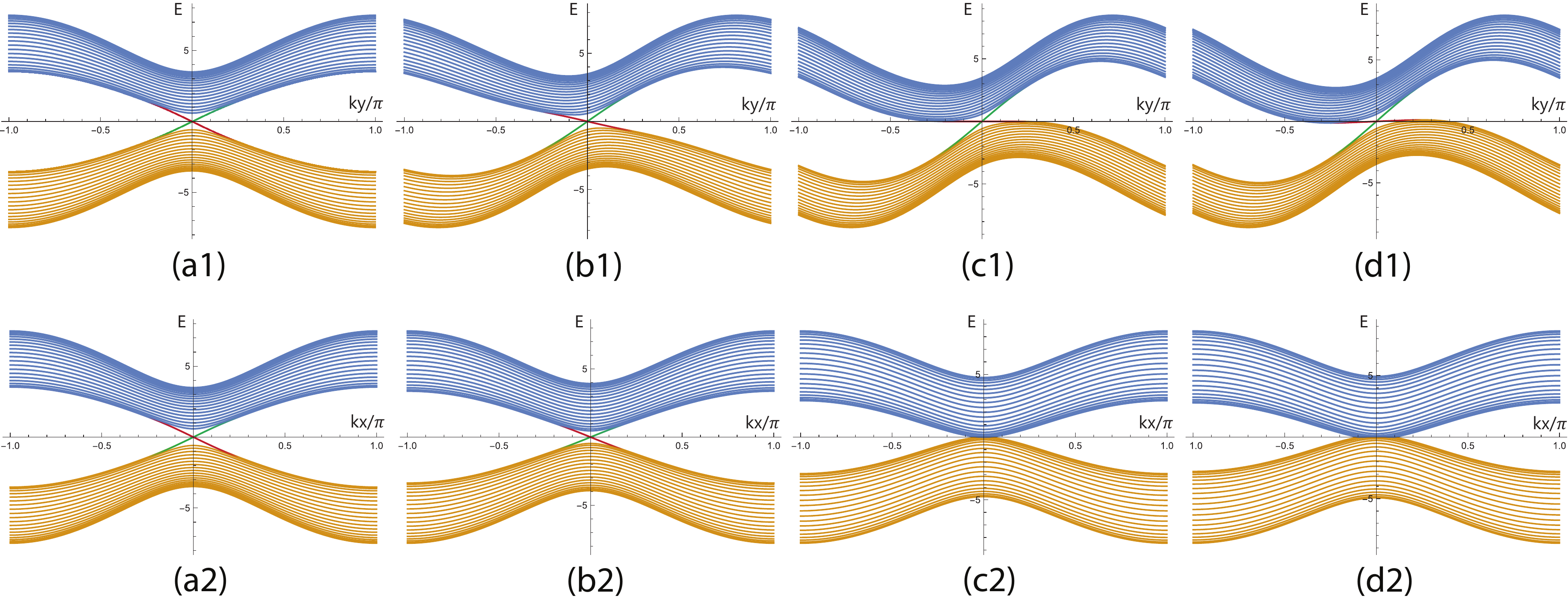}
    \caption{ The same situation as in Fig.\ref{fig:Edge_Lattice} except
	 at a fixed $h=-3.5$. In both figures, one can shift $k\to k+\pi$ to reach $h=+3.5$ results.
     It shows qualitatively the same edge TPTs as those in Fig.\ref{fig:Edge_Lattice}. See also Fig.S2 for the expanded figure.  }
\label{fig:Edge_Lattice2}
\end{figure}

In the Fig.\ref{fig:Edge_Lattice} and \ref{fig:Edge_Lattice2}, we show the numerical results on the lattice edge states.

\subsubsection{ Interpretation of the Hall conductance in terms of the edge states, enriched bulk-edge and new L/edge-T/edge correspondences }

The bulk-edge correspondence in  a static frame is also enriched under the injection or in a moving sample:
Relative to the injection or boost, there is a longitudinal or transverse edge, so the original  bulk-edge correspondence
is enriched to the bulk to longitudinal/or transverse edge correspondence, then the longitudinal edge to transverse edge
correspondence.

 In the longitudinal injecting case, the edge state always exist, so its contribution to $ \sigma_H $ remains
 quantized as $ \sigma_H= Ch_{-} =1 $ throughout the TPT at (c1). However, there is no bulk contributions in the Chern insulator,
 but the bulk starts to contribute $ \sigma_H=\nu_b= v/c -1  $ in the Odd Chern metal phase.
 The two parts lead back to Eq.\ref{eq:Ch+nu} evaluated in the bulk:
\begin{align}
    \sigma_{H}& =\frac{1}{2\pi}\int_{\mathbb{R}^2} \Omega_-(\mathbf{k})d^2\mathbf{k}
                +\frac{1}{\pi}\int_{\mathbb{R}^2} \Omega_+(\mathbf{k})\Theta(-\epsilon_{+}(\mathbf{k}))d^2\mathbf{k}
                 \nonumber   \\
              &  ={\rm Ch}_-+\nu_b
\end{align}

 In the transverse injecting case, the edge states exist only before the TPT at (c2),
 so its contribution to $ \sigma_H $ remains
 quantized as $ \sigma_H= Ch_{-} $ only before the TPT at (c2). There is no bulk contributions in the Chern insulator before (c2).
 However, after (c2), the edge states emerge into the bulk and disappear.
 All the contributions come from the bulk  as $ \sigma_H= v/c < 1 $.


\subsection{ Solving the edge states from the effective theory in the continuum  }

We will solve the edge states with
the periodic boundary condition in the $y$-direction and
the open boundary condition in the $x$-direction,
then vice versa.

\subsubsection{ Solving the edge states in the longitudinal injection }

We will solve the model in a strip geometry \cite{topoSF}
with the periodic boundary condition in the $y$-direction
and open boundary condition in the $x$-direction.
The continuum Hamiltonian in the mixed $ ( \partial_x, k_y) $ representation:
\begin{align}
    H(\partial_x,k_y)&=
	(\Delta-\alpha_x\partial_x^2)\sigma_z-iv_x\partial_x\sigma_x
	\nonumber   \\
	&\phantom{=}
	+\alpha_y k_y^2\sigma_z+v_yk_y\sigma_y -ck_y\sigma_0
\label{H1dxky}
\end{align}
The problem will first be studied in the $k_y=0$ limit
where $H_{1D}
=(\Delta-\alpha_x\partial_x^2)\sigma_z-iv_x\partial_x\sigma_x$,
and then extended to finite $k_y\neq0$.
Due to the C-symmetry,
the edge mode $\psi(x,k_y=0)$ is expected to exist at zero energy,
therefore $H_{1D}\psi=0$.
Multiplying both side by $\sigma_z$ gives
$[(\Delta-\alpha_x\partial_x^2)+v_x\partial_x\sigma_y]\psi=0$.
Choosing $\sigma_y\psi_\pm=\pm\psi_\pm$ with
$\psi_\pm=\tfrac{1}{\sqrt{2}}(\phi,\pm i\phi)^\intercal$,
or $\psi_\pm=\phi\chi_\pm$ with
\begin{align}
\chi_\pm=\tfrac{1}{\sqrt{2}}(\begin{smallmatrix}1\\\pm i\end{smallmatrix})
\label{chipml}
\end{align}
then the coupled differential equations can be reduced to
a second order homogeneous ordinary differential equation
\begin{equation}
[(\Delta-\alpha_x\partial_x^2)\pm v_x\partial_x]\phi=0\>.
\label{PDE-x}
\end{equation}
Substituting the ansatz $\phi\propto e^{-\lambda x}$
leads to a character equation
$\alpha_x\lambda^2\pm v_x\lambda-\Delta=0$ with the solutions:
\begin{align}
\lambda_{1,2}^{(s)}&=\frac{-sv_x\pm\sqrt{v_x^2+4\alpha_x\Delta}}{2\alpha_x},
\label{lambda12x}
\end{align}
where $s=\pm$ corresponding to the choice of $\pm$ in Eq.\ref{PDE-x}.
Thus, the general solution can be written as
\begin{equation}
\phi=c_1 e^{-\lambda_1x}+c_2 e^{-\lambda_2x}\>.
\label{edgegeneralx}
\end{equation}
Due to the C-symmetry, one only need to consider one edge.
Choosing the left edge and imposing the wave function to vanish
at $x=0$ and $x=+\infty$ requires $c_1=-c_2$ and $\Re\lambda_{1,2}>0$.
Thus, we have $\phi_L(x)\propto (e^{-\lambda_1x}-e^{-\lambda_2x})$,
where $\lambda_{1,2}$ denote either $\lambda_{1,2}^{(+)}$ or $\lambda_{1,2}^{(-)}$ in Eq.\ref{lambda12x}
whichever have positive real parts.

The condition for $\Re\lambda_{1,2}>0$ is analyzed as:

If $v_x>0$, $\alpha_x>0$, $\Delta<0$,
the localized edge state is
$\psi_L(x)\propto(e^{-\lambda_1^{(-)}x}-e^{-\lambda_2^{(-)}x})\chi_-$.

If $v_x>0$, $\alpha_x<0$, $\Delta>0$,
the localized edge state is
$\psi_L(x)\propto(e^{-\lambda_1^{(+)}x}-e^{-\lambda_2^{(+)}x})\chi_+$.

If $v_x<0$, $\alpha_x>0$, $\Delta<0$,
the localized edge state is
$\psi_L(x)\propto(e^{-\lambda_1^{(+)}x}-e^{-\lambda_2^{(+)}x})\chi_+$.

If $v_x<0$, $\alpha_x<0$, $\Delta>0$,
the localized edge state is
$\psi_L(x)\propto(e^{-\lambda_1^{(-)}x}-e^{-\lambda_2^{(-)}x})\chi_-$.

Otherwise, there is not any localized edge state.

In short, the left localized edge state exists when $\alpha_x\Delta<0$
and $\psi_L=+\sgn(v_x\Delta)\psi_L$.
The C-symmetry indicates that
the right localized edge state also exists when $\alpha_x\Delta<0$
and $\sigma_y\psi_R=-\sgn(v_x\Delta)\psi_R$.

When extending the results at $ k_y=0 $ to finite $k_y$ case,
one can just replace $\Delta\to\Delta+\alpha k_y^2$ in Eq.\ref{lambda12x},
and then $\psi_L(x,y)\propto\psi_L(x)e^{ik_yy}$
is an eigenstate of $H(\partial_x,k_y)$.
[It is obvious to see that the exact spectrum is the same
as that achieved by treating $k_y$-dependent part perturbatively.
Note that $\chi_\pm$ is not eigenstate of $\alpha_xk_x^2\sigma_z$.]
Therefore, we get the edge effective Hamiltonian
including both left $\psi_L$ and right $\psi_R$ edge:
\begin{align}
    H_\text{edge}(k_y)&
	\!=\!\sgn(v_x\Delta)[(v_y\!-\!c)k_y\psi_L^\dagger \psi_L
    \!-\!(v_y\!+\!c)k_y\psi_R^\dagger \psi_R]
\label{edgey}
\end{align}
which shows the dispersion relations for the edge states at the open $x$-boundary are
\begin{align}
    \epsilon_L(k_y)&=+\sgn(v_x\Delta)(v_y-c)k_y,   \quad
\nonumber   \\
    \epsilon_R(k_y)&=-\sgn(v_x\Delta)(v_y+c)k_y
\label{pmedge}
\end{align}
Meanwhile, the bulk spectrum is continuous,
which is given by
\begin{align}
    \epsilon_\text{bulk}^{\pm}(k_y\!)
	\!=\!\pm\!\sqrt{\!
		(\!\Delta\!+\!\alpha_xk_x^2\!\!+\!\alpha_yk_y^2)^2
		\!\!+\!v_x^2k_x^2\!+\!v_y^2k_y^2}
	\!-\!ck_y
\end{align}
where $k_x$ is a continuous real parameter.

The edge state extends in a finite regime around $k_y=0$
which can be estimated by
when the energy of edge state first enter into the bulk spectrum.
Solving $\min_{k_x}\epsilon_\text{bulk}^{+}(k_y)=\epsilon_L(k_y)$
gives the max $k_x$.
Thus when $\alpha\Delta<0$,
the edge state survives in the regime
$|k_y|<\sqrt{-\Delta/\alpha}$ which is also independent of boost $c$.

\begin{figure}[tbhp]
    \centering
    \includegraphics[width=\linewidth]{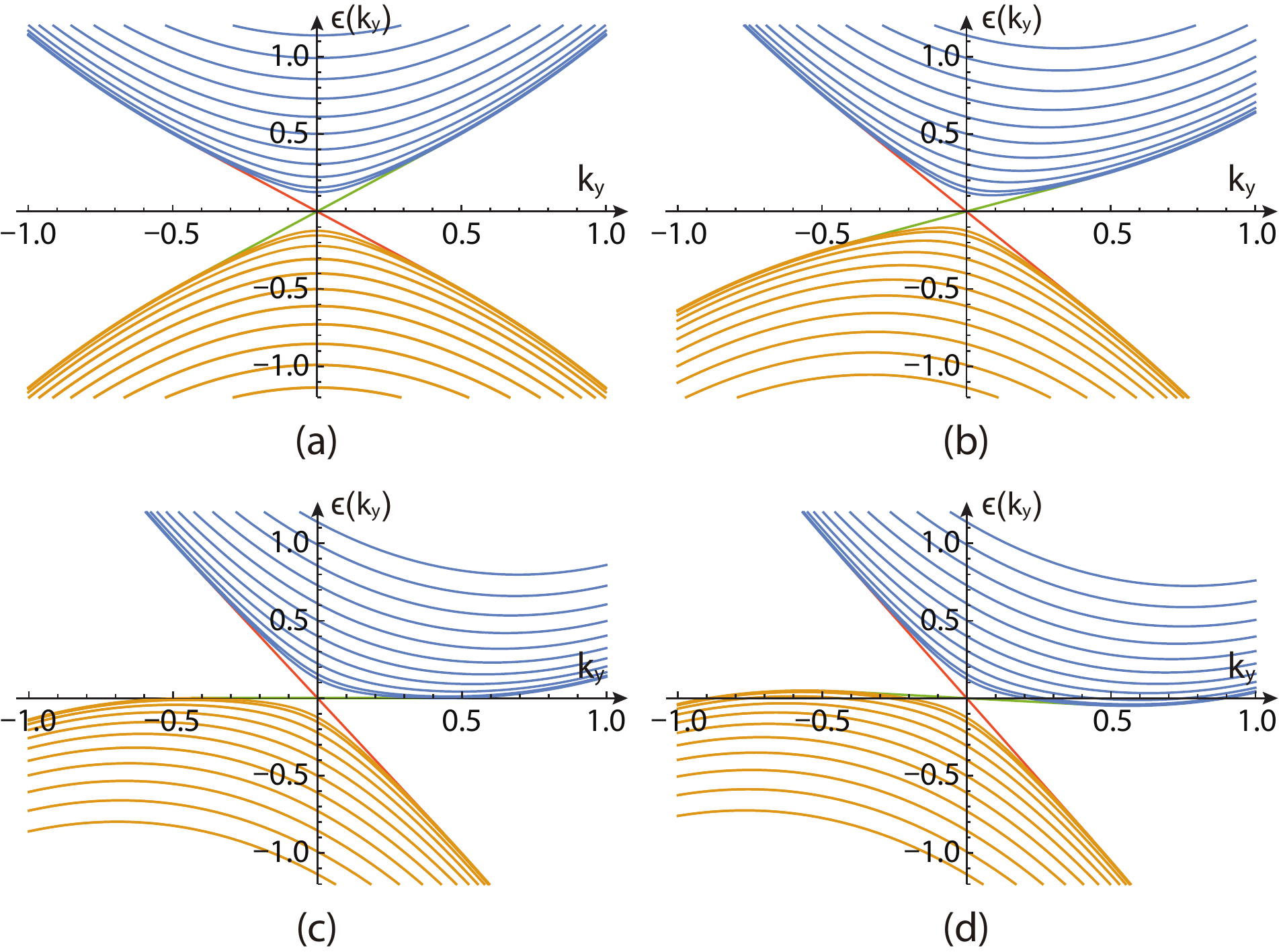}
    \caption{Longitudinal (L-) edge: The spectrum of the Hamiltonian $H_{1D}(\partial_x,k_y)$ in Eq.\ref{H1dxky}
    with periodic boundary condition in the $y$-direction and open
    boundary condition in the $x$-direction.
	From left to right, the parameter $c$ is $0,0.5,1.0,1.1$, respectively.
    Other parameters are fixed as $v_x=v_y=1$, $\alpha_{x,y}=1$, $\Delta=-0.1$.
 The left/right edge mode is highlighted in the red/green color.
 The results are consistent with those on the lattice in Fig.\ref{fig:Edge_Lattice},
 \ref{fig:Edge_Lattice2} with the same boundary conditions. }
\label{fig:xEdge}
\end{figure}

Which shows one edge state becomes zero slope at the QPT,
then reverses its slope after.
We plot bulk states and edge states in Fig.\ref{fig:xEdge}.

Putting $ c=0 $ in Eq.\ref{edgey} recovers the edge theory without the injection.
Then directly performing a Galileo boost along the edge leads to Eq.\ref{edgey}.

\subsubsection{ Solving the edge states in the transverse injection  }

Similarly, we can also consider the model in a strip geometry \cite{topoSF}
with the periodic boundary condition in the $x$-direction
and open boundary condition in the $y$-direction.
The continuum Hamiltonian in the mixed $ ( k_x, \partial_y ) $ representation:
\begin{align}
    H(k_x,\partial_y)&=
    (\Delta-\alpha_y \partial_y^2)\sigma_z
	-iv_y\partial_y\sigma_y+ic\partial_y\sigma_0
    \nonumber   \\
   &\phantom{=}
    +\alpha_x k_x^2\sigma_z+v_xk_x\sigma_x
\label{H1dkxy}
\end{align}
The problem will first be studied in the $k_x=0$ limit
where $H_{1D}=(\Delta-\alpha_y\partial_y^2)\sigma_z
-iv_y\partial_y\sigma_y+ic\partial_y\sigma_0$,
and then extended to finite $k_x\neq0$.
Due to the C-symmetry,
the edge mode $\psi(k_x=0,y)$ is expected to exist at zero energy,
thus $H_{1D}\psi=0$.
Multiplying both side by $\sigma_z$ gives
$[\Delta+(ic\sigma_z-v_y\sigma_x)\partial_y-\alpha\partial_y^2]\psi=0$.
Choosing $(ic\sigma_z-v_y\sigma_x)\psi_\pm=\pm\sqrt{\smash[b]{v_y^2-c^2}}\psi_\pm$
with $\psi_\pm=\phi\chi_{\pm}$ where the two-component spinor is
\begin{equation}
\chi_{\pm}=\tfrac{1}{\sqrt{2}}
	(\begin{smallmatrix}
	    1\\\xi_\pm\\
	\end{smallmatrix}), ~~~\xi_\pm=(ic\mp\sqrt{\smash[b]{v_y^2-c^2}})/v_y
\label{spinortran}
\end{equation}
which depends on the transverse boost $ c $ explicitly.
Then the coupled differential equation also can be reduced to
a second order ordinary differential equation
\begin{equation}
    (\Delta \pm\sqrt{\smash[b]{v_y^2-c^2}}\partial_y
	-\alpha_y\partial_y^2)\phi=0,
\label{PDE-y}
\end{equation}
where $\pm$ corresponding to the choice of $\psi_\pm$.
Substituting the ansatz $\phi=e^{-\lambda y}$ leads to
$\alpha_y\lambda^2\pm
\sqrt{\vphantom{v_x}\smash[b]{v_y^2-c^2}}\lambda-\Delta=0$
with the roots:
\begin{align}
    \lambda_{1,2}^{(s)}
	& =\frac{
	-s\sqrt{\vphantom{V_x^o}\smash[b]{v_y^2-c^2}}
	\pm\sqrt{\vphantom{v_x}\smash[b]{v_y^2-c^2+4\alpha_y\Delta}}}
	{2\alpha_y},
\label{lambda12y}
\end{align}
where $s=\pm$ corresponds to the choice of $\pm$ in Eq.\ref{PDE-y}.
Thus, the general solution can be written as
\begin{equation}
\phi=c_1 e^{-\lambda_1y}+c_2 e^{-\lambda_2y}
\label{edgegeneraly}
\end{equation}

Due to the C-symmetry, one only need consider one edge.
Choosing the bottom edge, imposing the wave function to vanish at $y=0$ and $y=+\infty$
requires $c_1=-c_2$ and $\Re\lambda_{1,2}>0$.
Thus, we have $\phi_B(y)\propto (e^{-\lambda_1y}-e^{-\lambda_2y})$
where $\lambda_{1,2}$ denote $\lambda_{1,2}^{(+)}$ or $\lambda_{1,2}^{(-)}$ in Eq.\ref{lambda12y}
whichever have positive real parts.

The condition for $\Re\lambda_{1,2}>0$ is analyzed as:

If $v_y^2>c^2$, $\alpha_y>0$, $\Delta<0$,
the localized edge state is
$\psi_B(y)\propto(e^{-\lambda_1^{(-)}y}-e^{-\lambda_2^{(-)}y})\chi_-$;

If $v_y^2>c^2$, $\alpha_y<0$, $\Delta>0$,
the localized edge state is
$\psi_B(y)\propto(e^{-\lambda_1^{(+)}y}-e^{-\lambda_2^{(+)}y})\chi_+$;

Otherwise, there is not any localized edge state.

In short,
the Bottom localized edge state exists when $\alpha_y\Delta<0$ and $c^2<v_y^2$,
and $(ic\sigma_z-v_y\sigma_x)\psi_B=+\sgn(\Delta)\sqrt{\smash[b]{v_y^2-c^2}}\psi_B$.
The C-symmetry indicates that the Top localized edge state exists when $\alpha_y\Delta<0$ and $c^2<v_y^2$,
and $(ic\sigma_z-v_y\sigma_x)\psi_T=-\sgn(\Delta)\sqrt{\smash[b]{v_y^2-c^2}}\psi_T $.

When extending to finite $k_x$ case,
one need replace $\lambda_{1,2}\to\lambda_{1,2}(k_x)$
and $\chi_\pm\to\chi_\pm(k_x)$,
then redo the eigenvalue problem.
[After tedious algebra, we find it gives the same spectrum
as treated the $k_x$-dependent part as perturbation.]
Alternatively, one may just take $\psi_B(x,y)=\psi_B(y)e^{ik_xx}$
and use the first order perturbation theory.
Due to $\langle \chi_{\pm}|\sigma_x|\chi_{\pm}\rangle=\mp\sqrt{\smash[b]{v_y^2-c^2}}/v_y$
and $\langle \chi_{\pm}|\sigma_z|\chi_{\pm}\rangle=0$, we find
the edge effective Hamiltonian including both the bottom $\psi_B $ and the top $\psi_T $ edge state
\begin{align}
    H_\text{edge}(k_x)
	=-\sgn(v_y\Delta)v_x\sqrt{1-c^2/v_y^2}k_x
	[\psi_B^\dagger \psi_B-\psi_T^\dagger \psi_T]
\label{edgex}
\end{align}
which indicates the dispersion relations for the edge states at the $y$-boundary:
\begin{align}
    \epsilon_B(k_x)& =-\sgn(v_y\Delta)v_x\sqrt{1-c^2/v_y^2}k_x,   \quad
\nonumber   \\
    \epsilon_T(k_x) & =+\sgn(v_y\Delta)v_x\sqrt{1-c^2/v_y^2}k_x
\label{squarerootedge}
\end{align}
Meanwhile, the bulk spectrum is continuous,
which is given by
\begin{align}
    \epsilon_\text{bulk}^{\pm}(k_x\!)
	\!=\!\pm\!\sqrt{\!
		(\!\Delta\!+\!\alpha_xk_x^2\!\!+\!\alpha_yk_y^2)^2
		\!\!+\!v_x^2k_x^2\!+\!v_y^2k_y^2}
	\!-\!ck_y
\end{align}
where $k_y$ is a continuous real parameter.
The edge state usual extent in a finit regime around $k_x=0$,
which can be estimated by
when the energy of edge state first enter into the bulk spectrum.
Solving $\min_{k_y}\epsilon_\text{bulk}^{+}(k_x)=\epsilon_B(k_x)$
gives the max $k_x$.
Thus when $\alpha\Delta<0$ and $|v_y|>|c|$,
the edge state survives in the regime
$|k_x|<\sqrt{(c^2/v_y^2-1)\Delta/\alpha}$,
and the regime is decrease as increasing the boost $c$.

\begin{figure}[!tbhp]
    \centering
    \includegraphics[width=\linewidth]{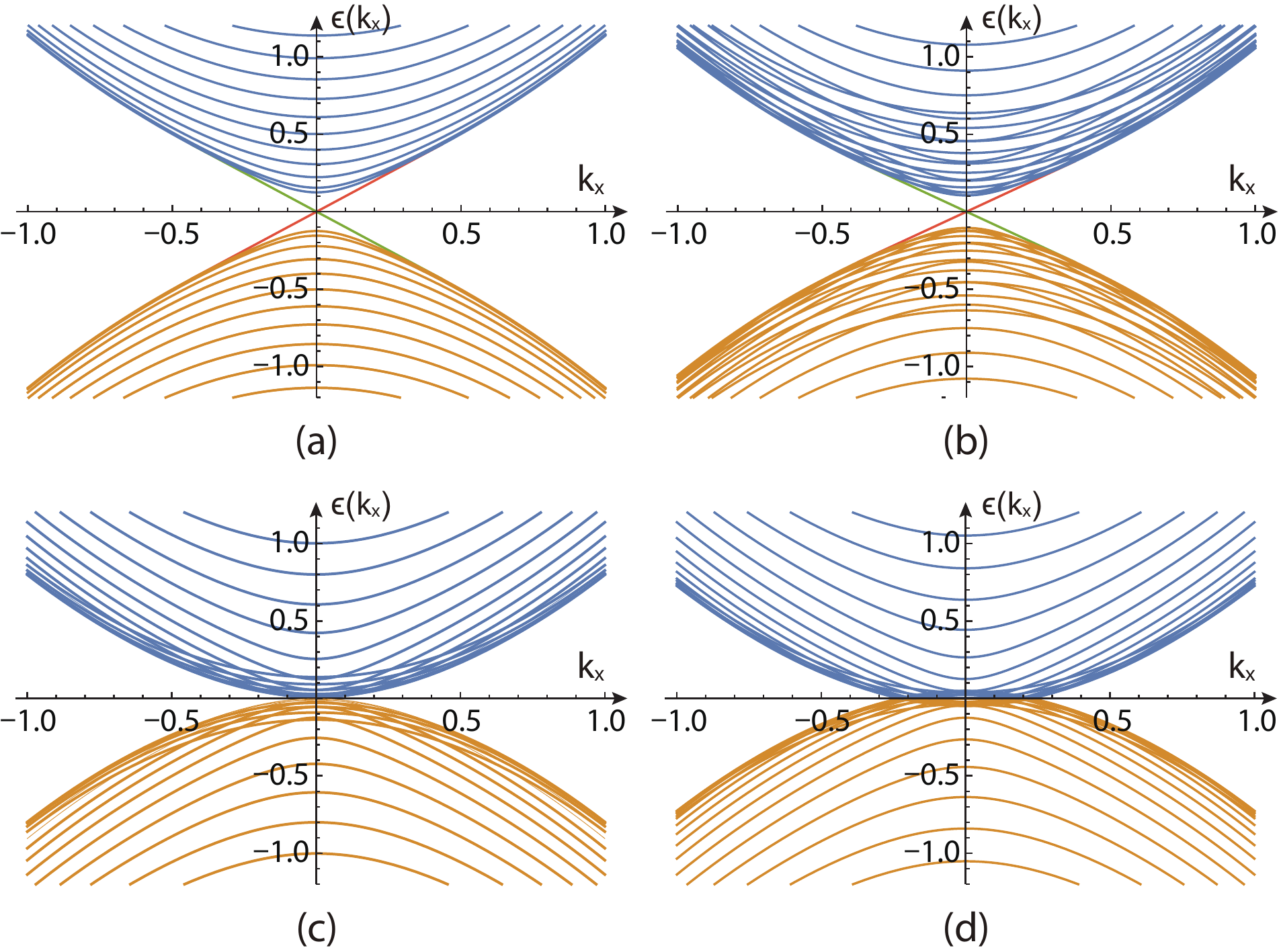}
    \caption{The transverse (T-) edge: spectrum of the Hamiltonian $H_{1D}(k_x,\partial_y)$ in Eq.\ref{H1dkxy}.
    with periodic boundary condition in the $x$-direction and open
    boundary condition in the $y$-direction.
	From left to right, the parameter $c$ is $0,0.5,1.0,1.1$, respectively.
    The T-edge disappears at the same time as the bulk TPT with its velocity vanishing as $ \sqrt{v^2-c^2} $.
    Other parameters are fixed as $v_x=v_y=1$, $\alpha_{x,y}=1$, $\Delta=-0.1$.
    The Top/Bottom edge mode is highlighted in the red/green color.
    The results are consistent with those on the lattice in Fig.\ref{fig:Edge_Lattice},
    \ref{fig:Edge_Lattice2} with the same boundary conditions.  }
    \label{fig:yEdge}
\end{figure}

Which shows edge states do not exist anymore after the bulk QPT.
We plot bulk states and edge states in Fig.\ref{fig:yEdge}.

Putting $ c=0 $ in Eq.\ref{edgex} recovers the edge theory without the injection.
However, in contrast to Eq.\ref{edgey}, one may not achieve Eq.\ref{edgex}
by directly performing a Galileo boost. Naively, a direct boost may lead to $ \sqrt{ v^2 + c^2 } $
instead of $ \sqrt{ v^2 - c^2 } $. To achieve Eq.\ref{edgex}, one may still need to solve the bulk + the transverse boundary condition
as done here
and find that one must substitute $ c \to ic $ in the naive Galileo boost results to reach the correct transverse boot result.
While the $ i $ in the substitution in $ c \to ic $ stands for the decay of edge mode along the boost direction.
 Eq.\ref{edgex} indicates that the two
 perpendicular motions: the spinor edge wave propagation along the x-edge and the boost along the y- axis are coupled to each
 other through the SOC which breaks the GI explicitly.
 Indeed, its 2-component spinor $ \chi_{\pm} $ 
 does depend on the transverse boost  sensitively.

\subsection{ The bulk-edge correspondence from the continuum edge theory }

  We will discuss the bulk-edge correspondence near $h\sim -4t$ and $ h \sim 0 $ respectively.

\subsubsection{ The bulk-edge correspondence near  $h\sim -4t$ case }


In the $h\sim -4t$ case, there is only one valley at $K_0=(0,0)$, the bulk effective Hamiltonian is:
\begin{align}
    H_0&\!=\![\Delta+\alpha (k_x^2\!+\!k_y^2)]\sigma_z
	\!+\!vk_x\sigma_x\!+\!vk_y\sigma_y\!-\!ck_y\sigma_0,
\end{align}
where $\Delta=-4t-h$, $\alpha=t$, $v=2t_s$, $c=2t_b$,

Since we choose $t,t_s,t_b>0$, thus
$h<-4t$ leads to $\Delta > 0$ and $\alpha\Delta>0$, no edge states.
$h>-4t$ leads to $\Delta < 0$ and $\alpha\Delta<0$. Edge state is possible.

{\em (a) Open boundary condition in the $x$-direction: longitudinal injection }

When $h<-4t$, there is no edge state;

When $h>-4t$, there is always one localized edge state which
is given by the continuum theory near $K_0=(0,0)$ and  exists near $K_{0y}=0$.

{\em (b) Open boundary condition in the $y$-direction: transverse injection }

If $\alpha\Delta<0$ and $v^2>c^2$, then $H_0$ has one localized edge state.

Thus, $h<-4t$, there is no edge state;

When $h>-4t$, if $ c^2 < v^2 $, there is one localized edge state which is given
by the continuum theory near $K_0=(0,0)$ and exists near $K_{0x}=0$.
If $ c^2 > v^2 $, there is still no edge state.

These results in (a) and (b) are consistent with those on a lattice reached in Sec.IV-A.

\subsubsection{The bulk-edge correspondence near  $h\sim 0$ case }


In the $h\sim 0 $ case, there are two valleys at $K_1$, $K_2$, the bulk effective Hamiltonian are:
\begin{align}
    H_1&\!=\![\Delta-\alpha (k_x^2-k_y^2)]\sigma_z\!-\!vk_x\sigma_x\!+\!vk_y\sigma_y\!-\!ck_y\sigma_0  \nonumber  \\
    H_2&\!=\![\Delta+\alpha (k_x^2-k_y^2)]\sigma_z\!+\!vk_x\sigma_x\!-\!vk_y\sigma_y\!+\!ck_y\sigma_0
\end{align}
where $\Delta=-h$, $\alpha=t$, $v=2t_s$, $c=2t_b$, written in the generic form of Eq.\ref{eq:Dirac}:
\begin{align}
    H_1&\!=\!(\Delta\!+\!\alpha_{1x}k_x^2+\alpha_{1y}k_y^2)\sigma_z
    \!+\!v_{1x}k_x\sigma_x\!+\!v_{1y}k_y\sigma_y\!+\!c_1k_y\sigma_0  \nonumber  \\
    H_2&\!=\!(\Delta+\alpha_{2x}k_x^2\!+\!\alpha_{2y}k_y^2)\sigma_z
    \!+\!v_{2x}k_x\sigma_x\!+\!v_{2y}k_y\sigma_y\!+\!c_2k_y\sigma_0
    \label{eq:H1H2n1}
\end{align}
  which leads to the relations:
\begin{align}
    \alpha_{1x}\!=\!-\alpha,\> \alpha_{1y}\!=\!+\alpha,\>
    v_{1x}\!=\!-v,\> v_{1y}\!=\!+v,\> c_1\!=\!-c   \nonumber  \\
    \alpha_{2x}\!=\!+\alpha,\> \alpha_{2y}\!=\!-\alpha,\>
    v_{2x}\!=\!+v,\> v_{2y}\!=\!-v,\> c_2\!=\!+c
\end{align}
If $\alpha_{1x}\Delta>0$, then $\alpha_{2x}\Delta<0$, $\alpha_{1y}\Delta<0, \alpha_{2y}\Delta>0$,
which means only one valley may have one edge state for a given type of boundary.
Since we choose $t,t_s,t_b>0$, thus $h>0$ leads to $\Delta<0$ and $\alpha\Delta<0$;
$h<0$ leads to $\Delta>0$ and $\alpha\Delta > 0$.

{\em (a) Open boundary condition in the $x$-direction: longitudinal injection }

If $\alpha\Delta>0$, then $\alpha_{1x}\Delta<0$ which means $H_1$ has one localized edge state;

If $\alpha\Delta<0$, then $\alpha_{2x}\Delta<0$ which means $H_2$ has one localized edge state.

Thus, there is always one  edge state.
When $h>0$, there is one edge state given by the continuum theory near $K_2=(0,\pi)$ and exists near $K_{2y}=\pi$.
This is expected, because it is smoothly connected to the edge state near $ h< 4t $ where there is only one valley at $
K_3= ( \pi, \pi ) $ and one edge states exists near $K_{3y}=\pi$.

When $h<0$, there is one  edge state given by the continuum theory near $K_1=(\pi,0)$ and  exists near $K_{1y}=0$.
This is expected, because it is smoothly connected to the edge state near $ h > - 4t $ where there is only one valley at $
K_0= ( 0, 0 ) $ and one edge states exists near $K_{0y}= 0 $.

{\em (b) Open boundary condition in the $y$-direction: transverse injection }

If $ c^2 > v^2$, then there is not any localized edge state.
If $ c^2 < v^2 $ and $\alpha\Delta>0$, then $\alpha_{2y}\Delta<0$ which means $H_2$ has one localized edge state;
if $ c^2 < v^2$ and $\alpha\Delta<0$, then $\alpha_{1y}\Delta<0$ which means $H_1$ has one  localized edge state.

Thus, there is one edge state only if $t_b< t_s $.

When $h>0$, there is one edge state  given by the continuum theory near $K_1=(\pi,0)$
and exists near $K_{1x}=\pi$.
This is expected, because it is smoothly connected to the edge state near $ h< 4t $ where there is only one valley at $
K_3= ( \pi, \pi ) $ and one edge states exists near $K_{3x}=\pi$.

When $h<0$, there is one edge state given by the continuum theory near $K_2=(0,\pi)$ and
exists near $K_{2x}=0$.
This is expected, because it is smoothly connected to the edge state near $ h > - 4t $ where there is only one valley at $
K_0= ( 0, 0 ) $ and one edge states exists near $K_{0x}= 0 $.

These results in (a) and (b) are consistent with those on a lattice reached in Sec.IV-A.
\section{ Gauge invariant current: bulk properties }

 Injecting currents into the system could result in the following form:
\begin{equation}
  H_{inj,x}= - [ \kappa_{b1} \sum_{i} J_{ix} + i \sum^{\infty}_{n=2} (t_{bn}/n)  c^{\dagger}_i c_{i+nx} ] + h.c.
\label{boostLattice}
\end{equation}
  where $ J_{ix} $ is the NN gauge-invariant current, the other terms $ n=2,3,\cdots $ are NNN, NNNN,....
  $ \kappa_{b1} $ is dimensionaless and $ t_{bn}, n \geq 2 $ carry the same dimension as the hopping.

  To capture the physics, one only need to include the NN and the $ n= 2 $ NNN term ( which can also be called higher order current ).
  We still take the ``divide and conquer''  strategy to treat the two terms separately and differently.
  As shown in Fig.19-22, for the $ n=1 $ case, the bulk TPT and the edge reconstruction happens at the same time.
  But this coincidence could be due to the $ n=1 $ case which maybe a fine tuning phase.
  It is not protected by any symmetry.
  So they can split in a more general case. If so, the edge reconstruction must happen earlier than the bulk TPT, not the other way around.
  Then, there must be an intermediate phase between the bulk TPT and the edge reconstruction, we call such a phase an odd Chern insulator phase
  which has the same bulk properties as the ordinary Chern insulator, but different edge properties:
  Its longitudinal edge modes satisfy the exotic relation $ v_L v_R > 0 $, its transverse edge modes satisfy
  the conventional relation $ v_L v_R < 0 $.
  In the longitudinal  edge, the edge mode undergoes its own edge TPT with an longitudinal edge dynamic exponent $ z_L=3 $
  instead of being  exactly flat for the $ n=1 $ even inside the Chern insulator before the bulk TPT.
  So there is an even enriched surface TPT before the bulk TPT.
  As to be shown in Sec.VII and Sec.VIII, this case also provide another example of odd Chern metal.
  The BM also leads to a non-vanishing AHE. Remarkably, the $ \sigma_H $ jump from the odd Chern metal to the BM remains
  the same  universal non-integer number as $ n=1 $.

\subsection{ The NN gauge-invariant current and the NNN current term  }
Due to the commutation relation
\begin{align}
    [c_{i,\sigma}^\dagger c_{i,\sigma},\!\sum_{j}\!
    c_{j,\alpha}^\dagger T_{\alpha\beta} c_{j+\delta,\beta}]
    \!=\!c_{i,\alpha}^\dagger T_{\alpha\beta} c_{i+\delta,\beta}
    \!-\!c_{i-\delta,\alpha}^\dagger T_{\alpha\beta} c_{i,\beta}
\end{align}
where the summation over spin indices $\sigma,\alpha,\beta$
is implied.

Given a general Hubbard Hamiltonian with SOC on a square lattice
\begin{align}
    H_0& =\sum_{j\delta} [c_{j,\alpha}^\dagger T_{\alpha\beta}^\delta c_{j+\delta,\beta}+h.c.]
    +\sum_{j}U n_j^2-\mu \sum_{j}n_j
  \nonumber  \\
    &-h\sum_j(n_{j\uparrow}-n_{j\downarrow})
\label{general_H}
\end{align}
where $\delta=\hat{x},\hat{y}$ and $U$ is Hubbard interaction.
The quantum anomalous Hall model Eq.\ref{QAH0} is just a special case of the general Hamiltonian in Eq.\eqref{general_H},
with $T^x=-t_0\sigma_z+it_{s0} \sigma^x$ and $T^y=-t_0 \sigma_z + it_{s0} \sigma^y$.

Combining the Heisenberg equation of motion and the continuity equation
\begin{align}
    &\frac{d}{dt}n_i=i[H_0,n_i]
        =-\nabla\cdot \mathbf{J}_i   \nonumber  \\
    &i[H_0,n_i]=\sum_{\alpha\beta\delta}
    [i(c_{i,\alpha}^\dagger T_{\alpha\beta}^\delta c_{i+\delta,\beta}
    -c_{i-\delta,\alpha}^\dagger T_{\alpha\beta}^\delta c_{i,\beta})+h.c.]  \nonumber \\
    &\nabla\cdot \mathbf{J}_i=J_{i}^x-J_{i-x}^x+J_{i}^y-J_{i-y}^y
\end{align}
Thus one can identify the gauge-invariant current as
\begin{align}
    J_{i}^x&=-ic_{i,\alpha}^\dagger T_{\alpha\beta}^x c_{i+x,\beta}+h.c.  \nonumber \\
    J_{i}^y&=-ic_{i,\alpha}^\dagger T_{\alpha\beta}^y c_{i+y,\beta}+h.c.
\label{gaugeinvcurrent}
\end{align}
 which is different from the injecting current Eq.\ref{eq:Boosted_H}.

 The total number conservation follows:
\begin{align}
    \frac{d}{dt}\sum_i n_i
    =\sum_i(J_{i}^x-J_{i-x}^x+J_{i}^y-J_{i-y}^y)=0
\end{align}
However, $\sum_i J_{i}^x\neq0$ and $\sum_i J_{i}^y\neq0$.

Consider a new Hamiltonian $H_{inj}=H_0- \sum_i (\kappa_{b1,x}J_{i}^x+\kappa_{b1,y}J_{i}^y)$,
\begin{align}
    H_{inj}
    &=\sum_{j\delta} [\kappa_\delta e^{i\phi_\delta}c_{j,\alpha}^\dagger T_{\alpha\beta}^\delta c_{j+\delta,\beta}+h.c.]
    +\sum_{j}U n_j^2    \nonumber \\
    &-\mu \sum_{j}n_j-h\sum_j(n_{j\uparrow}-n_{j\downarrow})
\end{align}
where $\kappa_\delta=\sqrt{1+\kappa_{b1,\delta}^2}$ and $\phi_{\delta}=\arg(1+i\kappa_{b1,\delta}), tan \phi_{\delta}= \kappa_{b1,\delta} $.
The phase factor $e^{i\phi_\delta}$ can be transformed away via:
\begin{equation}
  \tilde{c}_{j}= e^{-i(j_x\phi_{x}+j_y\phi_{y})}c_{j}
\label{tildebasis}
\end{equation}
  This is expected based on the fact that the gauge-invariant current has the same structure as the hopping and the SOC term, so can be absorbed
  by a transformation like Eq.\ref{tildebasis}, but leave the interaction $ U $ and the chemical potential $ \mu $
  un-touched. This absorbtion  does not happen for the injecting current Eq.\ref{eq:Boosted_H}.

  Now we study the Hamiltonian in the  $ \tilde{c}_{j} $ basis and incorporate the NNN $ n=2 $ boost term
\begin{align}
	H_\text{inj}
	=\tilde{H}_0+\sum_{i}[i(t_{b}/n) \tilde{c}_i^\dagger \tilde{c}_{i+ny}+h.c.]
\label{eq:Boosted_Hn}
\end{align}
 where $ t= t_0 \sqrt{1 + \kappa^2_{b1} },  t_s= t_{s0} \sqrt{1 + \kappa^2_{b1} } $. For simplicity, we assume $ t_0=t_{s0} $, so
\begin{equation}
  t=t_s= t_{0}\sqrt{ 1 + \kappa^2_{b1} }
\label{tts}
\end{equation}

  Then in the $ \tilde{c}_{j} $ basis, the $ n=2 $ term $ t_b $ becomes effectively as:
\begin{equation}
  t_b= \frac{ 1-\kappa^2_{b1} }{ 1 +\kappa^2_{b1}  } t_{b2}
\label{tbform}
\end{equation}
 where $ t_b $ has the same sign as $ t_{b2} $ only when $ 0 \leq \kappa_{b1} < 1 $, vanishes at $ \kappa_{b1}=1 $,
 but becomes opposite to $ t_{b2}  $ after $ \kappa_{b1} > 1 $. This sign change reflects the underlying lattice effects.
 Setting $ \kappa_{b1}=0 $ reduces to the pure $ t_{b2} $ case without any NN current term.

 In the following, we set $ U=0 $ and $ \mu=0 $ and study the effects of this NNN boost term.
 We treat $ t_b/t $ as an independent parameters to tune various TPTs in Fig.\ref{fig:phaseLattice2}.

\subsection{The lattice theory}

In the momentum space, the Hamiltonian Eq.\ref{eq:Boosted_Hn} becomes ( dropping $ \tilde{} $  for notational simplicity )
\begin{align}
	H_\text{inj}
	& =\sum_{k}c_k^\dagger\{-[h+2t(\cos k_x+\cos k_y)]\sigma_z
	+2t_s\sin k_x\sigma_x
   \nonumber \\
   & +2t_s\sin k_y\sigma_y-2(t_b/n)\sin (nk_y)\sigma_0\}c_k
\label{eq:Boosted_Hk_n}
\end{align}
   where the two competing scales are listed in Eq.\ref{tbform} and Eq.\ref{tts} respectively.

Diagonalization of Eq.\eqref{eq:Boosted_Hk_n} leads to two bands
\begin{align}
	&E_\pm(\mathbf{k})
		=-2(t_b/n) \sin (nk_y)
        \nonumber \\
        &\pm\!\!\sqrt{[h\!+\!2t(\cos k_x\!+\!\cos k_y)]^2\!+\!4t_s^2(\sin^2\!k_x\!+\!\sin^2\!k_y)}
\end{align}
Since $E_+(\mathbf{k})\geq E_-(\mathbf{k})$ always holds for a fixed $\mathbf{k}$,
we will call the $E_+$ the upper band and the $E_-$ the lower band.
When $t_b$ is sufficiently small,
it is in an insulating phase;
When $t_b$ is sufficiently large,
it is in a metallic phase, with hole FS given by $E_-(\mathbf{k})=0$
and electronic FS given by $E_+(\mathbf{k})=0$.

\begin{figure}[!tbhp]
    \centering
    \includegraphics[width=0.8\linewidth]{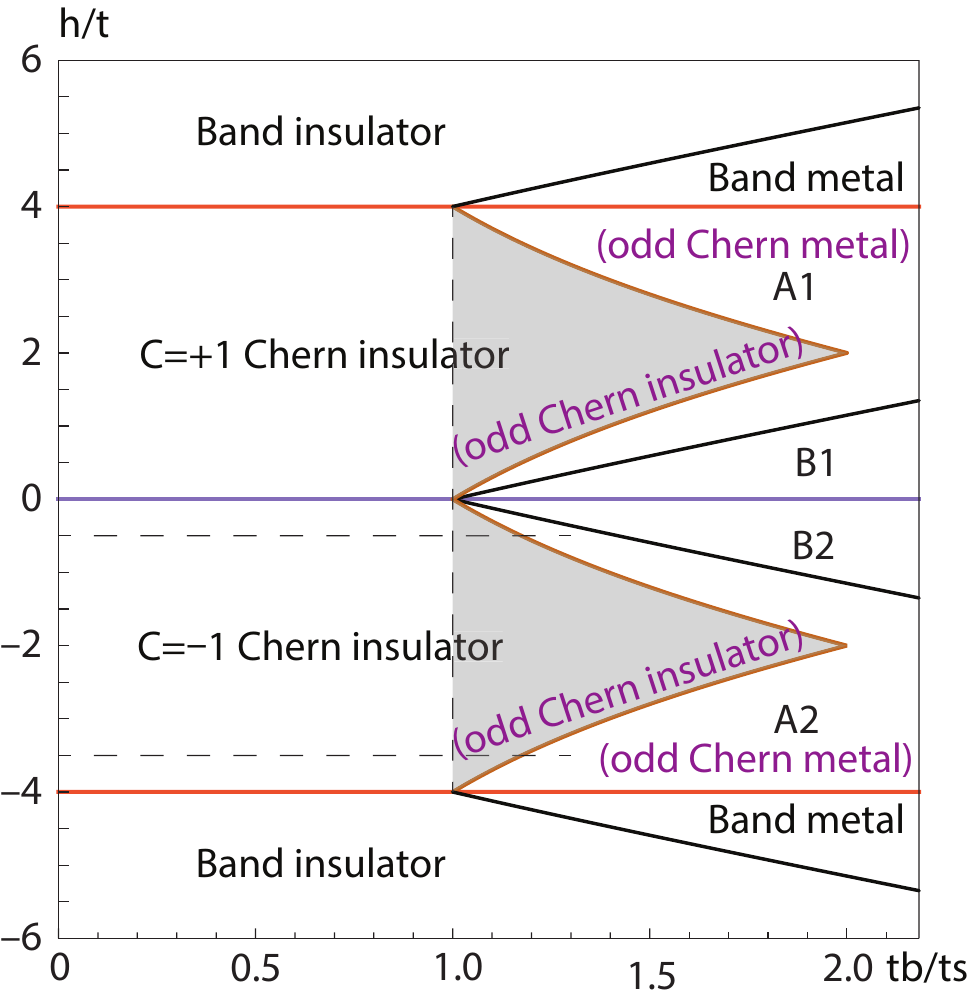}
    \caption{The global phase diagram of the Lattice Hamiltonian \eqref{eq:Boosted_Hk_n}
    with $n=2$ and $t=t_s$ under the $ n=1 $ gauge-invariant injecting current.
    Setting $ t_b=0 $ recovers dropping the $n=2$ term.
    The horizontal dashed line $ h=-0.5 $ corresponds to the edge mode in Fig.\ref{fig:Edge_Lattice_n2}.
    The new phase is the odd Chern insulator phase in the shaded lobe.
    The vertical dashed line indicates the edge TPT with the parallel edge dynamic exponent $ z_L=3 $.
    It   happens earlier than the orange solid line which is the bulk TPT
    with the bulk dynamic exponent $ z=2 $ from the
    bulk Chern insulator to Odd Chern metal also shown in Fig.\ref{fig:Edge_Lattice_n2}b1.
	Now the band metal has a nonzero Hall conductance,
	A1(A2) phase has one electronic/hole Fermi surface (FS) with positive(negative) Hall conductance.
	B1(B2) phase has two electronic/hole FSs with positive(negative) Hall conductance in Fig.\ref{fig:Dirac12_n2}.
	While the C1 and C2 phases in Fig.\ref{fig:phaseLattice} do not exist anymore when $n>1$.
    The phase boundary is related by the Mirror reflection with respect to $ h \leftrightarrow \pm 4-h $
    with the mirror symmetric (MS) point at $ h=\pm 2 $.
    However, the MS only holds at the phase boundary when $ t=t_s $. Compared to Fig.\ref{fig:phaseLattice}.   }
\label{fig:phaseLattice2}
\end{figure}

 In fact, Sec.II-IV correspond to the $n=1$ case. Here we will briefly explore the $n=2$ cases.
%
The first critical $t_b$ (let us call it $t_{b,c1}$) are determined by the global minimization problem
$\min_k E_+(\mathbf{k};t_b)=0$.
When $t_b>t_{b,c1}$, the energy bands overlap and Fermi surface ( FS) start to appear.
It was shown in Sec.II that when $n=1$, the critical $t_{b,c1}=t_s$ for $4t>h>0$ ( Fig.\ref{fig:phaseLattice} ).
However, when $n>1$, $t_{b,c1}$ is more complicated.
For example, when $n=2$, the critical $t_{b,c1}> t_s$ for $4t>h>0$.
When $h\sim0$, the second critical $t_b$ (let us call it $t_{b,c2}$ ) are determined by the local minimization problem
$\min_k E_+(\mathbf{k};t_b)=0$ with $k$ near $(0,\pi)$ and $(\pi,0)$.
When $t_b>t_{b,c2}$, the energy bands overlap and two FSs start to appear.
Unlike the $n=1$ case, if $t>t_{b,c2}$ these two Fermi surfaces do not collide with each other,
so no C phases in Fig.\ref{fig:phaseLattice} exist here.
But when $t_b>t_{b,c2}$, the FS can collide with itself, namely become extensive in $k_x$ to cover the entire $k_x\in[-\pi,\pi]$.
The extensive FS is signaled by the divergence of $k_F$ in continuum theory.
The global phase diagram of the Lattice Hamiltonian \eqref{eq:Boosted_Hk_n} with $n=2$ is shown in Fig.\ref{fig:phaseLattice2}


\subsubsection{ The universal conductance jump of zero temperature Hall conductance   }
\begin{figure}[tbhp]
    \centering
    \includegraphics[width=0.8\linewidth]{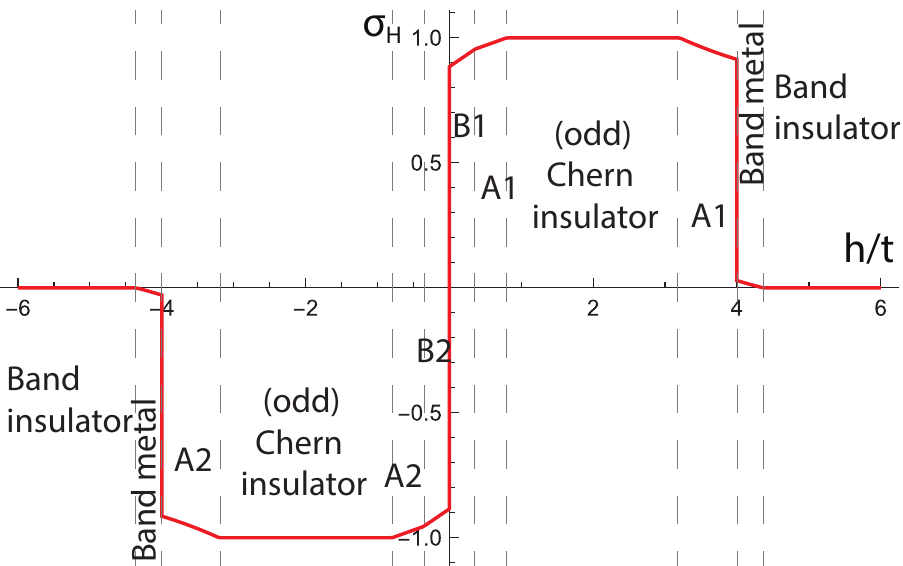}
    \caption{The Hall conductance $\sigma_H$ with a higher order boost $n=2$ as a function of $h$
    with fixed $t=1$, $t_s=1$, and $t_b/t_s=1.3$. There is an intermediate odd Chern insulator.
    The $\sigma_H$ of band metal is not zero,
    and $\sigma_H$ of the odd Chern metal A1 or B1 is not $\pm t_s/t_b$. However,
    the jump in $\sigma_H$ from A1 odd Chern metal to the band metal remains $ t_s/t_b $,
    that from B2 to B1 odd Chern metal  remains $ 2 t_s/t_b $.
    Compared to Fig.\ref{fig:Hall_T0}b, $\sigma_H$ is no longer a simple step function when $n>1$.
    However, all the jumps stay the same as the $ n=1 $ case, indicating its universal feature.  }
\label{fig:Hall-1_n2}
\end{figure}

On the lattice scale, the value of Hall conductance $ \sigma_H $ of the insulator phases of $n>1$ case is the same as those in the $n=1$ case;
but its value in the metal phase of $n>1$ case is different from that in the $n=1$ case at least in the following ways:
1) The $ \sigma_H $  in the band metal phase is not zero anymore in the former,
but it is identically zero in the latter.
2) The $ \sigma_H $ in  the Odd Chern metal phase is not $\pm t_s/t_b$ anymore in the former,
but it is $\pm t_s/t_b$ in the latter.
We show an example of Hall conductance as a function of $h$ in Fig.\ref{fig:Hall-1_n2}.

\subsection{The continuum limit}
In the momentum space, Eq.\eqref{eq:Boosted_Hk_n} becomes
\begin{align}
	H(\mathbf{k})&=-[h+2t(\cos k_x+\cos k_y)]\sigma_z
	+2t_s\sin k_x\sigma_x
 \nonumber \\
   & +2t_s\sin k_y\sigma_y-2(t_b/n)\sin (nk_y)\sigma_0
\end{align}
when $h\sim 4t$, low-energy excitations exist near $ \mathbf{K}_3= (\pi,\pi)$
\begin{align}
	H_3(\mathbf{K}_3+k)&=-[h-4t+t(k_x^2+k_y^2)]\sigma_z-2t_sk_x\sigma_x
\nonumber \\
  & -2t_sk_y\sigma_y-(-1)^n2t_bk_y\sigma_0
\end{align}
when $h\sim 0$, low-energy excitations exist near both $\mathbf{K}_1=(\pi,0)$ and $\mathbf{K}_2=(0,\pi)$
\begin{align}
    H_1(\mathbf{K}_1 +\mathbf{k})&=-[h+t(k_x^2-k_y^2)]\sigma_z
    -2t_sk_x\sigma_x+2t_sk_y\sigma_y
    \nonumber \\
    &-2t_bk_y\sigma_0
    \nonumber  \\
	H_2(\mathbf{K}_2+\mathbf{k})&=-[h-t(k_x^2-k_y^2)]\sigma_z
	  +2t_sk_x\sigma_x-2t_sk_y\sigma_y
                        \nonumber \\
    &-(-1)^n 2t_bk_y\sigma_0
\end{align}
when $h\sim -4t$, low-energy excitations exist near $ \mathbf{K}_0 = (0,0)$
\begin{align}
	H_0(\mathbf{k})&=-[h+4t-t(k_x^2+k_y^2)]\sigma_z
  \nonumber \\
	 & +2t_sk_x\sigma_x+2t_sk_x\sigma_x-2t_bk_y\sigma_0
\end{align}
Thus only even or odd nature of $n$ is important.

The continuum theory with just one valley is identical to that in the injecting case with $ n=1 $.
So we only need to look at the continuum theory near the two valleys $K_1$ and $K_2$:
\begin{align}
    H_1&=[\Delta-\alpha (k_x^2-k_y^2)]\sigma_z-vk_x\sigma_x+vk_y\sigma_y-ck_y\sigma_0   \nonumber  \\
    H_2&=[\Delta+\alpha (k_x^2-k_y^2)]\sigma_z+vk_x\sigma_x-vk_y\sigma_y-ck_y\sigma_0
\label{eq:Dirac12_n2}
\end{align}
where $\Delta=-h, \alpha= -t, v=2 t_s $  and $ c= 2 t_b $.
Note the different sign of velocities $v$ between $k_x$ and $k_y$
and different sign of $\alpha$ between $k_x^2$ and $k_y^2$.
Most importantly, the Doppler shifts are identical in the two nodes,
in contrast to Eq.\ref{opposite}  where they are opposite in the two nodes.

\begin{figure}[tbhp]
    \centering
   \includegraphics[width=\linewidth]{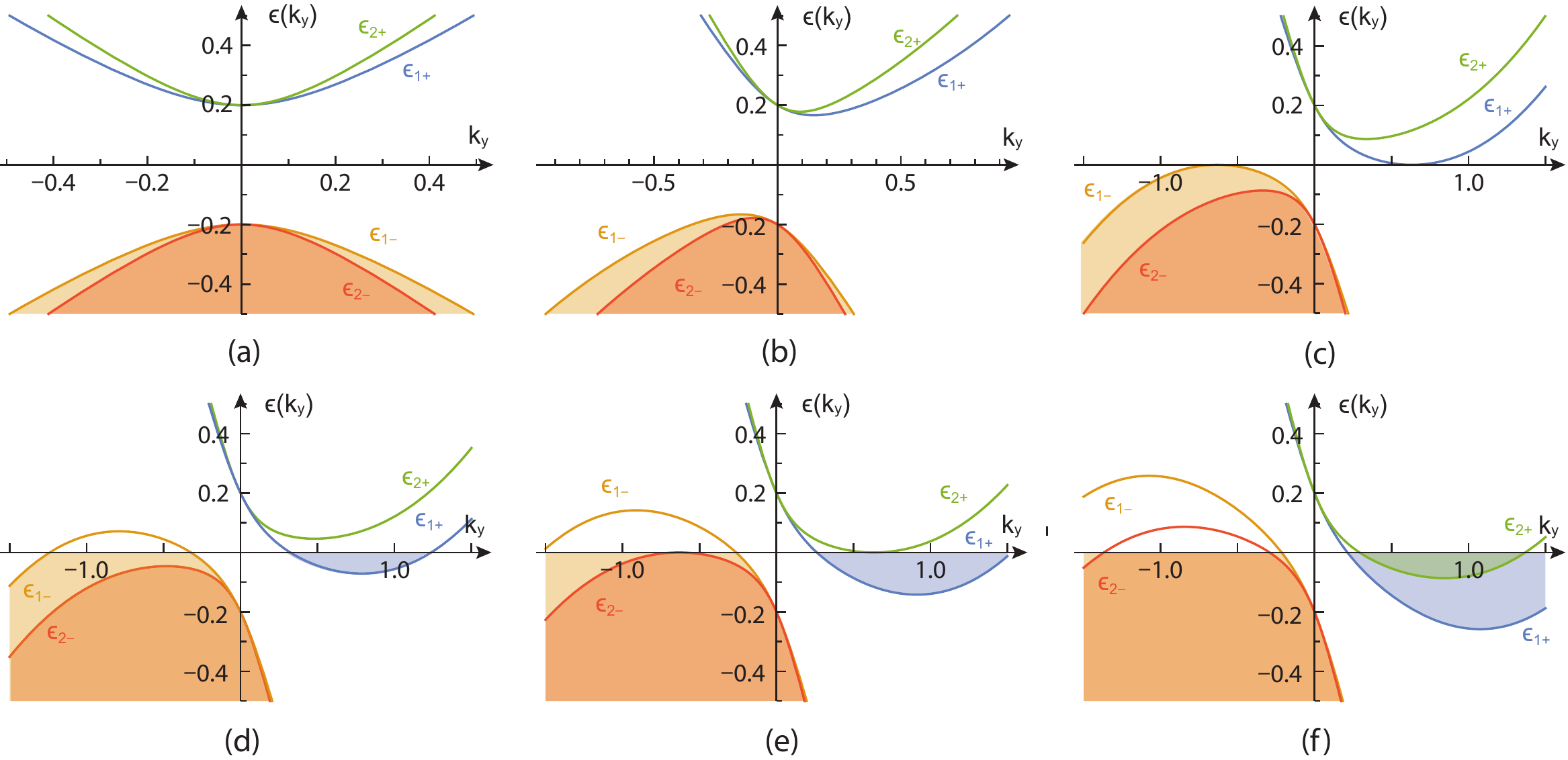}
    \caption{$\epsilon_{i,\pm}(k)$ with even $n$ as function of $k_y$
    with fixed $k_x=0$ and $v=1$, $\Delta=1/5$, $\alpha=-1/2$, $c=0,0.5,1.0,\sqrt{1.4},1.3$.
    (a) both $H_1$ and $H_2$ has a direct gap at $k_y=0$; in the Chern Insulator phase
    (b) Due to the same sign of the boost velocity in Eq.\ref{samesign}, $H_1$ and $H_2$ shift to the same direction, so has an indirect gap at $k_y\neq 0$;
    (c) $H_1$ becomes gapless and shows two Fermi points at $k_y=\pm k_0$ with $ z=2 $,
    but $H_2$ still has an indirect gap. This is in the A1 phase in Fig.\ref{fig:phaseLattice2}.
    (d) $H_1$ becomes gapless and show finite Fermi pockets,
     $H_2$ also becomes gapless and shows two Fermi points still at $k_y=\pm k_0$  with $ z=2 $ after subtracting the non-critical contributions
    from $ H_1 $;
    (e) both $H_1$ and $H_2$ are gapless and show finite Fermi pockets. This is in the B1 phase in Fig.\ref{fig:phaseLattice2}.
    Compared to Fig.\ref{fig:Dirac12_FS}.
    $ H_1 $ and $ H_{2} $ may collide with each other ( namely, hit the BZ boundary, not shown in Fig.23  )
    instead of  colliding with each other as in  Fig.\ref{fig:Dirac12_FS}. }
\label{fig:Dirac12_n2}
\end{figure}

Due to the extra two degree of freedom,
we obtain the four bands
\begin{align}
	\epsilon_{i,\pm}=\pm\sqrt{v^2k^2+[\Delta+(-1)^i\alpha(k_x^2-k_y^2)]^2}+ c k_y\>,
\label{samesign}
\end{align}
where $i=1,2$.
Due to  $c_1=c_2$, the electron Fermi momentum of $\epsilon_{1,2}$ will have the same sign.
We show  the evolution of the FS of $H_1$ and $H_2$ in Fig.\ref{fig:Dirac12_n2}.

Another important feature of $ \alpha_x\alpha_y<0 $ is that
the FS extends to infinity when $c\geq\sqrt{2}|v|$.
The divergent $k_F$ hints the FS is extensive in $k_x$, which covers the entire $k_x\in[-\pi,\pi]$.

\section{The topological phases in the gauge invariant current: edge properties  }

 Following the approach used in the Sec.IV for the $ n=1 $ case, we will first study the edge properties from the lattice system, then
 investigate them from the continuum effective theory, then contrast the two complementary approaches.

\subsection{New bulk-edge correspondence  from the lattice theory}

For the periodic boundary condition in the $y$-direction
and open boundary condition in the $x$-direction,
$k_y$ is a good quantum number,
the Hamiltonian in the mixed $ ( i, k_y ) $ representation becomes
\begin{align}
    H\!&=\!\!\sum_{k_y,i,j}\!
            c_{i,k_y}^\dagger \{
            [-(h+2t\cos k_y)\sigma_z\!
            \nonumber   \\
            & +\!2t_s\sin k_y\sigma_y\!+\!t_b\sin 2k_y \sigma_0]\delta_{i,j}
                   \nonumber   \\
            &+(t\sigma_z\!-\!it_s\sigma_y)\delta_{i,j+1}
            +(t\sigma_z\!+\!it_s\sigma_y)\delta_{i,j-1}
            \} c_{j,k_y}
\end{align}

For the periodic boundary condition in the $x$-direction
and open boundary condition in the $y$-direction,
$k_x$ is a good quantum number,
the Hamiltonian in the mixed $ ( k_x, i ) $  representation becomes:
\begin{align}
    H\!&=\!\!\sum_{k_x,i,j}\!
        c_{k_x,i}^\dagger \{
        [-(h+2t\cos k_x)\sigma_z+t_s\sin k_x\sigma_x]\delta_{i,j}
        \nonumber   \\
       & +(t\sigma_z\!-\!it_s\sigma_y)\delta_{i,j+1}
        +(t\sigma_z\!+\!it_s\sigma_y)\delta_{i,j-1}
        \nonumber   \\
	   &-it_b\sigma_0(\delta_{i,j+2}-\delta_{i,j-2})/2
        \} c_{k_x,j}
\end{align}

\begin{figure}[tbhp]
    \centering
    \includegraphics[width=\linewidth]{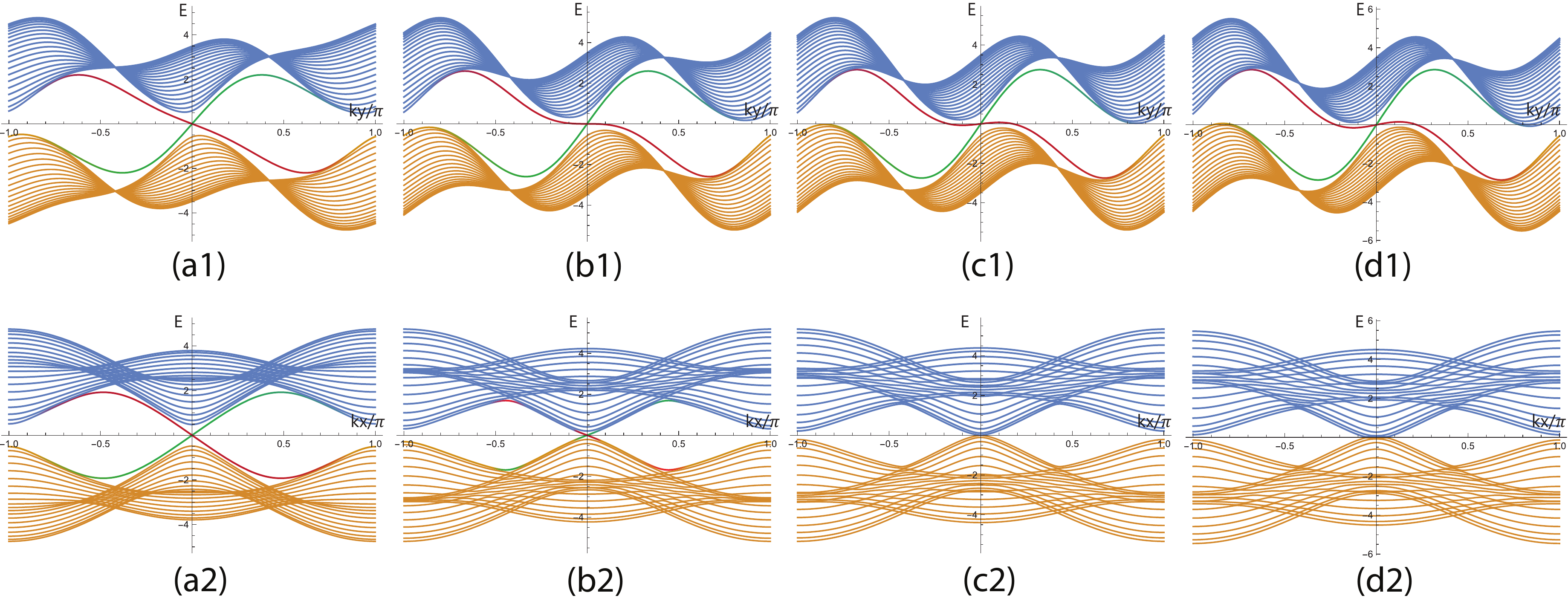}
    \caption{The edge state of the lattice Hamiltonian Eq. \eqref{eq:Boosted_Hk_n}.
    From (a) to (d), the parameter $t_b/t_s$ is $0.5,1.0,1.17,1.3$, respectively. We fixed $h=-0.5$.
    (Top) Longitudinal boost: With periodic boundary condition in the $y$-direction and open
    boundary condition in the $x$-direction. The edge modes always exist in this case.
    The two edge mods move in the opposite direction near $ k_y=0 $ in (a1) Chern Insulator where $ t_b/t_s < 1 $,
    then one edge mode's slope vanishes in (b1) where $ t_b/t_s = 1 $
    with the edge dispersion $ \omega \sim k^3_y $, namely the longitudinal edge dynamic exponent $ z_L=3 $.
    then the two edge modes move along the same direction near $ k_y=0 $ in (c1) odd Chern insulator where $ t_b/t_s = 1.17 > 1 $.
    At the same time, the system's (in-direct) gap vanishes which corresponds to the $ z=2 $ bulk TPT from the
    $ C=-1 $ odd Chern insulator to A2 Odd Chern metal in Fig.\ref{fig:phaseLattice2}.
    It gets to the Odd Chern metal phase in (d1) where $ t_b/t_s = 1.3 > 1.17 $, the two edge modes still move along the same direction.
	(Bottom) Transverse boost:
    Exchanging the role of $ x-$ and $y-$ direction.
    The edge mode exists upto (c2) where $ t_b/t_s = 1.17 > 1 $. So the odd CI between (b2)  and (c2) still has the transverse edge mode.
    At (c2), the system's direct gap vanishes which corresponds to the TPT from the
    $ C=-1 $ odd Chern insulator to A2 Odd Chern metal in the bulk in Fig.\ref{fig:phaseLattice2}.
    It is in the Odd Chern metal phase in (d2) where $ t_b/t_s = 1.3 > 1.17 $, no edge mode.
    The T-edge disappears at the same time as the bulk TPT with its velocity still vanishing  $ \sqrt{v^2-c^2} $ as in Fig.\ref{fig:yEdge}.
    One can shift $k\to k+\pi$ to reach $h=+0.5$ results. See also Fig.S3 for the expanded figure.	}
\label{fig:Edge_Lattice_n2}
\end{figure}

\begin{figure}[tbhp]
    \centering
    \includegraphics[width=\linewidth]{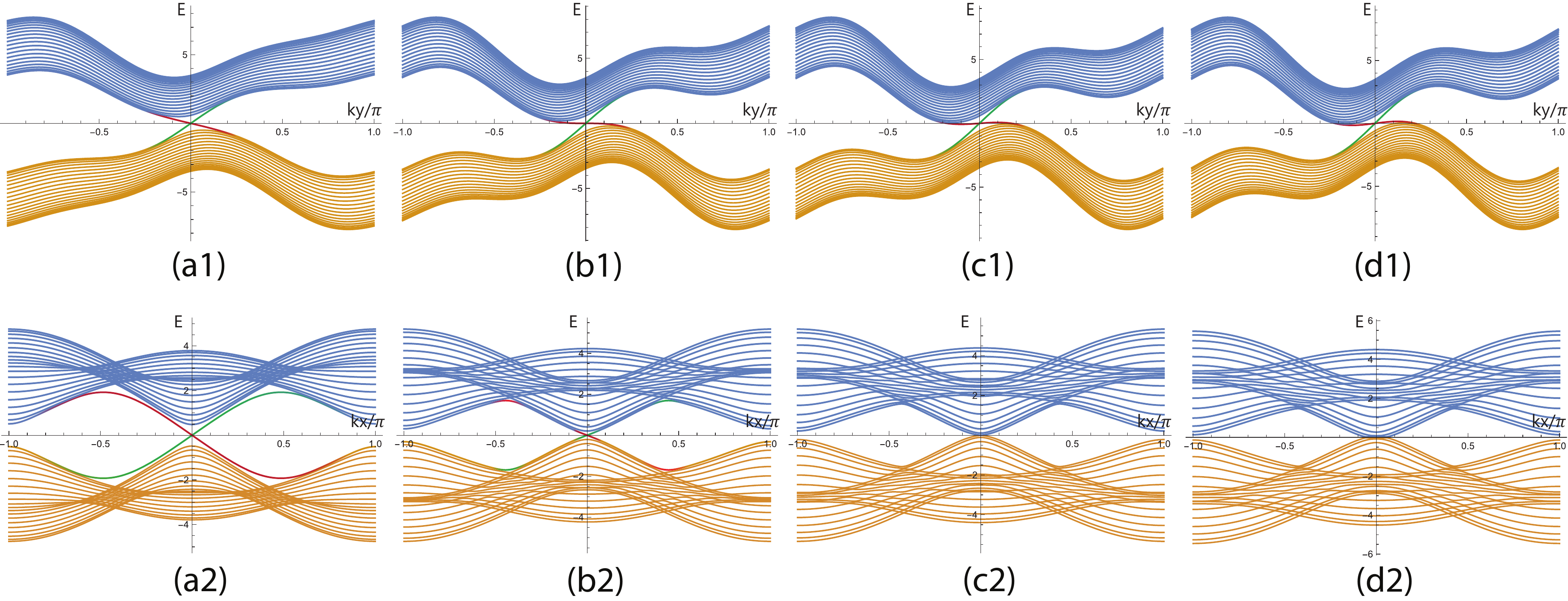}
    \caption{ The same situation as Fig.\ref{fig:Edge_Lattice_n2} except $ h=-3.5 $ which is Mirror reflected image of $ h=-0.5 $.
     As alerted in  Fig.\ref{fig:phaseLattice2},
     despite the bulk phase boundary in Fig.\ref{fig:phaseLattice2} has such a Mirror symmetry at $ t=t_s $,
     it is not persevered in the presence of the strip boundaries.
     One can shift $k\to k+\pi$ to reach $h=+3.5$ results  which is Mirror reflected image of $ h=0.5 $.
     It shows qualitatively the same edge TPTs, odd CI and odd CM as those in Fig.\ref{fig:Edge_Lattice_n2}.
     See also Fig.S4 for the expanded figure. }
\label{fig:Edge_Lattice2_n2}
\end{figure}

We show the numerical results on the lattice edge states in the
Fig.\ref{fig:Edge_Lattice_n2} and \ref{fig:Edge_Lattice2_n2}.

When comparing the $n=2$ case with previous $n=1$ case,
We discover several new surface TPT and novel bulk-edge correspondence:

{\em (a) Longitudinal injection }

We first choose periodic boundary condition in the $y$-direction
and open boundary condition in the $x$-direction.
For $n=1$ case and $t_b/t_s=1$, the bulk is critical.
The $n=1$ edge mode is almost flat due to the cancelation of $ \omega \sim t_s\sin k_y-t_b\sin k_y=0$.
So the TPT happens in the edge and the bulk at the same time.
However, for $ n=2 $, the TPT splits into two with the odd Chern insulator intervening between:
the reconstruction  in the edge always happens earlier than in the bulk.
When $t_b/t_s<1$, the two edge modes in the Left and Right move along the opposite direction.
When $t_b/t_s>1$, the two edge modes in the Left and Right move along the same direction.
At the QCP $t_b/t_s=1 $, the $n=2$ edge mode is not flat, due to the edge dispersion relation
\begin{align}
 \omega & \sim t_s\sin k_y-(t_b/2)\sin 2k_y
 \nonumber   \\
 &=(t_s-t_b\cos k_y)\sin k_y \sim k_y^3
\label{z3edge}
\end{align}
which is shown in Fig.\ref{fig:Edge_Lattice_n2}b1.
So the CI to odd CI transition indeed happens at $t_b/t_s=1$ with the dynamic exponent $ z_L=3 $ due to the longitudinal  edge reconstruction.
The bulk remains gapped despite the edge mode reconstruction at the QCP .
Then as $t_b/t_s $ increase further, the bulk gap closes and also undergoes a TPT at $ t_b/t_s =1.17 $ in Fig.\ref{fig:Edge_Lattice_n2}c.
It corresponds to nothing but the bulk TPT from the $ C=1 $ odd Chern insulator to the A1 Odd Chern metal in Fig.\ref{fig:phaseLattice2} and
Fig.\ref{fig:Dirac12_n2}c.

{\em (b) Transverse injection: }

Then we choose the periodic boundary condition in the $x$-direction
and open boundary condition in the $y$-direction.
For $n=1$ case and $t_b/t_s=1$, the bulk is critical.
The $n=1$ edge mode is also squeezed away.
So the TPT happens in the edge and the bulk at the same time.
However, for $ n=2 $, one can find the edge mode of the odd CI still exists at $t_b/t_s=1$ in Fig.\ref{fig:Edge_Lattice_n2}-b2.
So the transverse edge mode of the odd CI survives always before the bulk gap closing.
So we conclude that the odd CI has the longitudinal edge modes satisfying the exotic  $ v_L v_R > 0 $, but
the transverse edge modes satisfying the conventional $ v_L v_R < 0 $.
Eq.\ref{edgex}. break down and need to be replaced by more refined  edge theory by incorporating high order
derivatives in the continuum theory.


\subsection{ New bulk-edge correspondence  from the continuum theory }

Because $ h/t \sim -4 $ case with only one valley is similar to the injecting case
with $ n=1 $ on the long-wave length limit,
so we only need to focus on the $h\sim 0$ case  with two valleys  when $n=2$.

 When $h\sim 0$,  the effective Hamiltonians near the two valleys $K_1$, $K_2$ are:
\begin{align}
    H_1&\!=\![\Delta-\alpha (k_x^2-k_y^2)]\sigma_z\!-\!vk_x\sigma_x\!+\!vk_y\sigma_y\!-\!ck_y\sigma_0
    \nonumber  \\
    H_2&\!=\![\Delta+\alpha (k_x^2-k_y^2)]\sigma_z\!+\!vk_x\sigma_x\!-\!vk_y\sigma_y\!-\!ck_y\sigma_0
\end{align}
where $\Delta=-h$, $\alpha=t$, $v=2t_s$, $c=2t_b$.
In fact, all the 4 nodes suffer the same sign of Doppler shifts when $ n=2 $ case.

 It can be rewritten in the generic form of Eq.\ref{eq:H1H2n1}
\begin{align}
    H_1&=(\Delta+\alpha_{1x}k_x^2+\alpha_{1y}k_y^2)\sigma_z+v_{1x}k_x\sigma_x+v_{1y}k_y\sigma_y
    \nonumber   \\
    &-c_1k_y\sigma_0  \nonumber  \\
    H_2&=(\Delta+\alpha_{2x}k_x^2+\alpha_{2y}k_y^2)\sigma_z+v_{2x}k_x\sigma_x+v_{2y}k_y\sigma_y
    \nonumber   \\
    &-c_2k_y\sigma_0
    \label{eq:H1H2n2}
\end{align}
 which leads to the relations
\begin{align}
    \alpha_{1x}=-\alpha,~ \alpha_{1y}=+\alpha,~ v_{1x}=-v,~ v_{1y}=+v,~ c_1=+c   \nonumber \\
    \alpha_{2x}=+\alpha,~ \alpha_{2y}=-\alpha,~ v_{2x}=+v,~ v_{2y}=-v,~ c_2=+c
\end{align}
If $\alpha_{1x}\Delta>0$, then $\alpha_{2x}\Delta<0$, $\alpha_{1y}\Delta<0$, and $\alpha_{2y}\Delta>0$,
which means only one  of the valleys may have one edge state for a given type of boundary.
Further discussions on the long-wavelength limit to the quadratic order  are the same as Sec.IV-C-2.
For example, if one keeps only upto the quadratic order,
then both $n=1$ and $n=2$ gives zero slope when $t_b/t_s=1$.

However, to check against the new bulk-edge correspondence discovered on the lattice theory in Sec.VI-A,
one may need to go to higher order derivatives in the continuum edge theory.
For example, to find the dynamic exponent $ z_L=3 $ in the longitudinal boost,
one needs to go to at least the cubic order in Eq.\ref{z3edge}.
 Similarly, one need to push Eq.\ref{edgex} to higher order derivatives to describe the edge states evolution
 in Fig.\ref{fig:Edge_Lattice_n2} and \ref{fig:Edge_Lattice2_n2}
 under the transverse boost.
More works need here to discern the fine structures of the edge states by going to high order in the momentum.

\section{ The classification of even/odd Chern metal and band metal }

So far, we only consider the case where the Hamiltonian respects the $\mathcal{C}$-symmetry, but breaks the $\mathcal{P}$-symmetry,
namely, odd  under the  $\mathcal{P}$. We call it the odd Chern metal.
In the Appendix A and B, we discuss the complimentary cases: respects the $\mathcal{P}$-symmetry, but
breaks the $\mathcal{C}$-symmetry. We call it the even Chern metal which will be shown to show dramatically different behaviours than the
odd Chern metals. In this section, we classify Chern metals as even and odd. In the next section, we discuss
the general case which breaks both the $\mathcal{C}$-symmetry and $\mathcal{P}$-symmetry.

 The generic Hamiltonian in a material consists of two parts
\begin{align}
    H(k)=\epsilon_0(k)\sigma_0 + H_\text{QAH}(k),
\label{QAHevenodd}
\end{align}
where the $H_\text{QAH}(k)$ part respects both $\mathcal{C}$-symmetry in Eq.\ref{Csymmetry} and $\mathcal{P}$-symmetry.
An even function  $\epsilon_0(k)$ breaks the $\mathcal{C}$-symmetry, but keeps  the $\mathcal{P}$-symmetry,
In the appendix A and B, we consider the two typical even cases respectively:
(I) $\epsilon_0(k)=-\mu=-t_b$ and (II) $\epsilon_0(k)=-2t_b(1-\cos k_y)$.
More general case $\epsilon_0(k)=-2t_b(1-\cos k_x+ 1-\cos k_y)$ can be straightforwardly extended to.
To break both the $\mathcal{C}$-symmetry and $\mathcal{P}$-symmetry,
we  choose $\epsilon_0(k)=-2t_b(\sin k_y+1-\cos k_y)$ as a concrete example to discuss in Sec.VIII.

Varying the strength of $\epsilon_0(k)$ drives an Chern insulator to a Chern metal with Fermi surfaces.
Both carry non-vanishing Chern number.
If keeping $\mathcal{C}$-symmetry, tentatively, we name the Chern metal as ``odd'' Chern metal;
If keeping $\mathcal{P}$-symmetry, tentatively, we name the Chern metal as ``even'' Chern metal.
If no symmetries is kept, then it could be either odd-like or even-like Chern metal.

In the following, we will only focus on exploring the differences between odd and even Chern metal
in the phase diagram and the Hall conductance from both the bulk and edge picture.

\subsection{ Phase diagram: TPT from (odd) Chern Insulator to odd/even Chern metal due to the
competition between the P-even and P-odd component }

In the global phase diagram, due to its topological protection, the Chern insulator remains the same
in both P-breaking and P-preserving deformation. Even the band insulator remains the same.
However, the metallic phases may differ.
The big differences between odd and even Chern metals can be best seen
around the critical points at $h=\pm 4t,0$.
In the odd Chern metal, as shown in Fig.2 ( also Fig.11 and Fig.15 ) and Fig.23,
the QPTs at $h=\pm 4t$ and $0$ ( Dirac theory ) with $ z=1 $ is stable,
which means the TPT from the CI to the odd CM needs a sufficiently large enough $t_b$ even near $h=\pm 4t$ and $0$.
The energy scale of the critical $t_b$ is comparable to the Fermi velocity of the Dirac cone,
which is of the energy scale $t_s$. As demonstrated in the previous sections,
there is a universal non-integer jump from the odd Chern metal to the band metal.
The longitudinal edge modes in the odd CI and odd CM on the two opposite side of a sample  move along the same direction.
These salient bulk and edge properties make it very easy to distinguish an odd CM from the BM.

In the even Chern metal, as shown in  Fig.\ref{even2phase}a,
the critical points at $h= 4t$ with the Dirac point at $ ( \pi, \pi) $ and $h=0$ ( with the two Dirac points at $ (\pi,0) $ and $ (0, \pi) $ )
with $ z=1 $ is not stable,
even a small $t_b$ drives a TPT from the  CI  to even CM near $h= 4t$ and $0$.
Then the energy scale of transition $t_b$ is comparable with the gap of the Dirac cone,
which is of the energy scale $t$ or $h\pm 4t$. This has been demonstrated in the appendix B.
However, the critical points at $h= -4t$ with the Dirac point at $ ( 0, 0) $ remains stable.
So due to C- symmetry breaking by the P- preserving energy dispersion,
the CM at $ 0 < h/t < 4 $ behave very differently than that at $ -4 < h/t < 0 $:
the former is essentially the same as BM, the latter is the true CM with edge modes well separated from the bulk.
So the naively thought  even Chern metal at $ 0 < h/t < 4 $ turns out to be the same phase as BM ( Table II ).

\subsection{ Universal Hall conductance jump: bulk picture }

It was well-known that there is an integer (quantized) Hall conductance jump between
Chern insulator and band insulator, or different Chern insulators ( see also Fig.2 ).
As shown in the previous sections and elucidated further in the appendix C,
a non-integer (unquantized but universal $ v/c $ ) Hall conductance jump from odd or odd-like Chern metal to itself or to a band metal.
As shown in the appendix A, B and Sec.VIII, there may be an integer (quantized) Hall conductance jump from even or
even-like Chern metal to itself or to a band metal.

In the bulk picture, the Universal Hall conductance jump across the two metallic phases
( either even/odd Chern metal or a band metal ) at $ h=h_c $  can be understood as follows:
it is the sum of the upper and lower bands $\Delta\sigma=\Delta\sigma_- + \Delta\sigma_+$, where
\begin{align}
    \Delta\sigma_s(h_c)
    &=\lim_{h\to h_c^+}\frac{1}{2\pi}\int_{k~\text{filled}} d^2k\>\Omega_s(k;h)
    \nonumber  \\
    &-\lim_{h\to h_c^-}\frac{1}{2\pi}\int_{k~\text{filled}} d^2k\>\Omega_s(k;h)
\label{Delta_sigma}
\end{align}
 where $ s= \pm $ stands for the upper/lower band respectively.

\begin{figure}[tbhp]
    \centering
    \includegraphics[width=\linewidth]{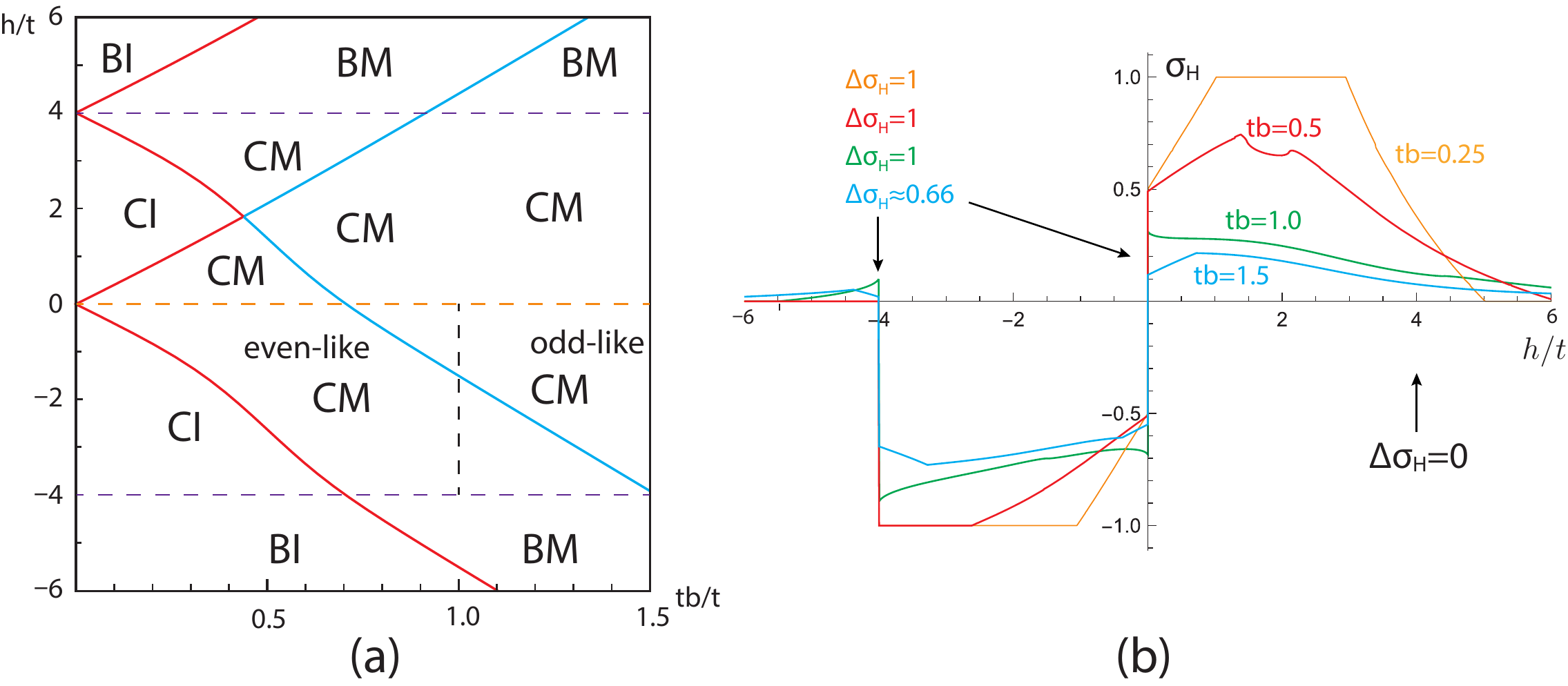}
    \caption{(a) The global phase diagram of the Lattice Hamiltonian \eqref{eq:H-Even-3} of the even-odd mixing Chern metal.
    The even-like CM or odd-like CM exist when $ -4 < h/t < 0 $ with an edge reconstruction between the two with the
    dynamic longitudinal exponent $ z_L=2 $ in Eq.\ref{z2edgemetal}. In the even CM $ v_L v_R < 0 $,
    in the odd CM $ v_L v_R > 0 $. At the $ z_L=2 $ edge reconstruction point, $ v_L v_R = 0 $.
    While all the CM at $ 0 < h/t < 4 $ are essentially the same
    as the BM despite superficially their bands have a non-vanishing Chern number, practically also larger AHE.
    Due to this superficial and practical difference, we still keep the symbol CM in the figure, but they really belong to the same phase.
    (b) The Hall conductance as a function of $h/t$ for various fixed values of $t_b/t=0.25,0.50,1.00,1.50$.
    The other parameters are $ t=t_s=1 $.
    The Hall conductance only shows a jump at $h/t=-4 $ and $ h/t=0$.
    For example,
    the $t_b/t=0.25$ curve shows a unit jump near $h/t=0$ from the even-like CM to a BM
    and also a unit jump near $h/t=-4$ from the CI to a BI which always has $ \sigma_H=0 $;
    On the other hand, $t_b/t=1.5$  curve shows a universal non-integer jump near $h/t=0$ and also near $h/t=-4$ from the odd-like CM to  a BM.
    Note that due to the absence of even CM to even CM transition here, the factor of 2  in the odd CM
    to odd CM transition in Fig.\ref{fig:Hall-1_n2}  does not appear here.
    Near $ h/t \sim 0^{-} $, it is easy to reach the even-like CM  from the CI due to $ t_b/t \to 0^{-} $, but it still need $ t_b/t=1 $
    to reach the odd-like CM.   }
\label{evenoddphase}
\end{figure}

As shown in the appendix D, When $h\to h_c$, $\Omega_s(k;h)$ will show a singularity or a Berry phase at the corresponding Dirac points.
\begin{align}
 h\to +4|t|, &\frac{1}{2\pi}\Omega_s(\mathbf{k};h)
\to +(s/2) \mathrm{sgn}(4|t|-h)\delta (\mathbf{k}-\mathbf{K}_3)   \nonumber  \\
&+F(\mathbf{k}),  \nonumber  \\
 h\to 0,~~~~~ & \frac{1}{2\pi}\Omega_s(\mathbf{k};h)
\to (s/2) \mathrm{sgn}(h)[\delta (\mathbf{k}-\mathbf{K}_1)   \nonumber  \\
& +\delta (\mathbf{k}-\mathbf{K}_2)]+F(\mathbf{k}),   \nonumber  \\
 h\to -4|t|, & \frac{1}{2\pi}\Omega_s(\mathbf{k};h)
\to -(s/2) \mathrm{sgn}(4|t|+h)\delta (\mathbf{k}-\mathbf{K}_0)   \nonumber  \\
& +F(\mathbf{k}),
\label{Delta_sigmaexpress}
\end{align}
where $\mathbf{K}_3=(\pi,\pi),\mathbf{K}_1=(\pi,0), \mathbf{K}_2=(0,\pi)$ and $\mathbf{K}_0=(0,0)$ as listed in Sec.III.

Note that the limit of $\Omega_s(\mathbf{k};h)$ has both a discontinuous  $\delta$-function
and a continuous part $F(\mathbf{k})$. The latter is a non-universal function which depends on the microscopic details,
but it has no contribution to the Universal Hall conductance jump.
When tuning $h$ across $h_c$, there are three situations:

1) The FS includes the singularity of $\Omega_s(k;h)$;

2) The FS excludes the singularity of $\Omega_s(k;h)$;

3) The FS does not include (or exclude) the singularity of $\Omega_s(k;h)$ when $h\neq h_c$, but approaches it as $h \to h_c$.

In the case 1) and 2) which belongs to the even Chern metal case, one can interchange the limit with the integral in
Eq.\ref{Delta_sigma} which is an integration over the $\delta$-functions.
Then a simple counting tells the Universal Hall conductance jump is zero or an integer which is in the same class after modula an integer $ Z $.

For the case 3) which belongs to the odd Chern metal case, one can not interchange limit with the integral in Eq.\ref{Delta_sigma},
so one must do the integral first, then take the limit.
This order gives a non-integer number which is found to be a universal number $v/(2c)$ with $c>v$ for each singularity point,
where $c$ and $v$ are the two parameters of the boosted continuum Dirac theory in Eq.24.
For the specific lattice Hamiltonian Eq.\ref{QAHevenodd} $v/(2c)=t_s/(2t_b)$.

\subsection{ Universal Hall conductance jump: the edge picture and the bulk-edge correspondence in the even/odd Chern metal }

Because as the transverse  edge mode behave similarly ( see Fig.26 and Fig.27 )
in all the cases: it disappears at the same time as the bulk TPT with the edge velocity vanishing as $ \sqrt{v^2-c^2} $.
So we only consider the longitudinal case in the edge mode.
Near the QPTs $h=\pm 4t$ or $0$ and  $ t_b=0 $, there might be edge states near the momentum
where the direct band gap closes  at $ \mathbf{K}_3=(\pi,\pi), \mathbf{K}_1=(\pi,0), \mathbf{K}_2=(0,\pi), \mathbf{K}_0=(0,0) $.
When $h/t<-4$, there is no edge states;
when $-4<h/t<0$, there is an edge state near $k_y=0$;
when $0<h/t<+4$, there is an edge states near $k_y=\pi$;
when $h/t>+4$, there is no edge states.
To judge the Universal Hall conductance jump from the edge picture,
one need to identify the relation between the Fermi energy and the edge states near the two minima $k_y=0$ and $k_y=\pi$.

We reach the following conclusion:
When tuning $h$ across the critical points $h=\pm 4t$ or $0$,
If the Fermi energy cuts edge states near the two minima in one phase, but not the other,
then there is a Universal Hall conductance jump. Otherwise the Hall conductance jump is zero.
As $-4<h/t<0$ goes to $0<h/t<+4$, because the change near the minimum $k_y=0$ to near the minimum $k_y=\pi$ in the momentum space,
one may need to keep track of the changing of edge states near both minima.

One may also tell the difference between an integer jump in the even Chern metal and a non-integer jump in the odd Chern metal
by the following criterion:
In the even $v_L v_R<0$, in the odd $v_L v_R>0$ when $h\to\pm 4t$ or $0$ ( Table II ),
where $v_L$ and $v_R$ denote the velocities of the left and right edge modes respectively.
Recall that $v=\partial \epsilon_{k_y}/\partial k_y|_{\epsilon_{k_y}=\epsilon_F=0}$.

\begin{widetext}
\begin{center}
\begin{table}[!tbhp]
    \setlength{\tabcolsep}{10pt}
        \caption{ The classification of unquantized Anomalous Hall effect. The Bulk $ \sigma_H $  jump is defined across the TPT from
        the corresponding phase to the band metal phase. As shown in Table III, the integer jump from the BM to the even Chern metal is 
        equal to that of Chern number near $ h=-4t $, but only one half of   that of Chern number near $ h=0 $.  }
        \begin{tabular}{c|c|c|c|c|c}
        \hline\hline
        AHE metal       & Band metal   & Even Chern metal   &  Odd Chern metal \\  \hline
        Symmetry          &   No         &  $ C$- No, $ P $-Yes      &     $C $-Yes, $ P $-No                \\
        Chern number     &     0 or $ \pm 1 $      &  $ \pm 1 $         &  $ \pm 1  $    \\
        Bulk $ \sigma_H $  jump     &     0        &  Integer          &  non-integer      \\
        Longitudinal Edge      &  no edge or floating edge  & edge velocities $ v_L v_R < 0 $     &  edge velocities  $ v_L v_R > 0 $     \\
        \hline\hline
        \end{tabular}
\end{table}
\end{center}
\end{widetext}
    Where 
    we did not list the quantized AHE phases: CI and odd CI which has
    the L/T edge $ v_L v_R < 0 $ and L/T edge $ v_L v_R >/< 0 $ respectively.
    The trivial BI is not listed either.
    For the bulk TPT or edge TPT between any of the two phases, see the Table I.

    The Time reversal symmetry is always spontaneously broken. The bulk jump $ \sigma_H $ is from the BM to the even or odd CM.
    Note that (1) All the three metals have no transverse edges. (2) Even or odd CM must have a non-vanishing
    Chern number for the relevant energy bands,  but the band metal may or may not have a non-vanishing  Chern number. So Chern number becomes an irrelevant topological number in this classification of gapless metallic phases.

    We did not list the even-like CM or Odd-like CM to be examined in the next section which breaks both C- and P-, because they
    qualitatively belong to either even or odd CM.
    At the edge QCP $t_b/t_s=1 $ from the even-like to odd-like Chern metal, the edge mode is not flat,
    due to the edge dispersion relation
\begin{align}
 \omega & \sim  t_s\sin k_y-t_b (\sin k_y + 1 -\cos k_y )
                      \nonumber   \\
  &  = -t_b ( 1- \cos k_y) \sim - k_y^2 < 0
\label{z2edgemetal}
\end{align}
  which leads to the dynamic longitudinal edge exponent $ z_L=2 $ from the even-like CM to the odd-like CM ( Fig.\ref{evenlikeedge} and
  Fig.\ref{oddlikeedge} ). It can be contrasted to the longitudinal edge exponent $ z_L=3 $ from the Chern insulator to
  the odd Chern insulator in Fig.26 and Fig.27.
  The former is inside the gapless metallic phase, the latter is inside the gapped insulating phase.
  However, both have the same bulk, but different edges. At both QCPs, $ v_L v_R =0 $ which dictates the boundary.

 The BM in this table is nothing but the previously well studied one
 contributing to the un-quantized AHE \cite{AHE2,Haldane2004,AHE3}. The even CM with the conventional edge modes
 $ v_L v_R < 0 $  may probably have appeared in the context of bilayer Kagome metals \cite{bilayerkagome} or twisted bilayer graphene \cite{TBG}. Unfortunately, neither bulk nor edge properties of this possible even Chern metal, its distinctions with the band metal
 have been discussed in any depth in these {\sl ab initio } density functional theory calculations.
 The odd CM with the exotic edge modes $ v_L v_R > 0 $
 for $ n=1 $ in Sec.II-IV and for $ n=2 $ in Sec. V-VI are discovered  and investigated systematically here.
 While the itinerant metal contributing the AHE due to the Berry phase acquired by electrons moving
 in the non-coplanar spin texture from the monopoles  in the real space in a Ferromagnet \cite{AHE1}
 does not fall into this Table which is only for non-interacting fermions.

In the appendix A and B, we demonstrate the above general statements and Table II for two specific examples of the even Chern metal.
In the next section, we do it for a concrete example of the generic case in any material which is a mixing of even-odd Chern metal,
but still show either even-like or odd-like Chern metal features.

\begin{figure}[tbhp]
    \centering
    \includegraphics[width=0.9\linewidth]{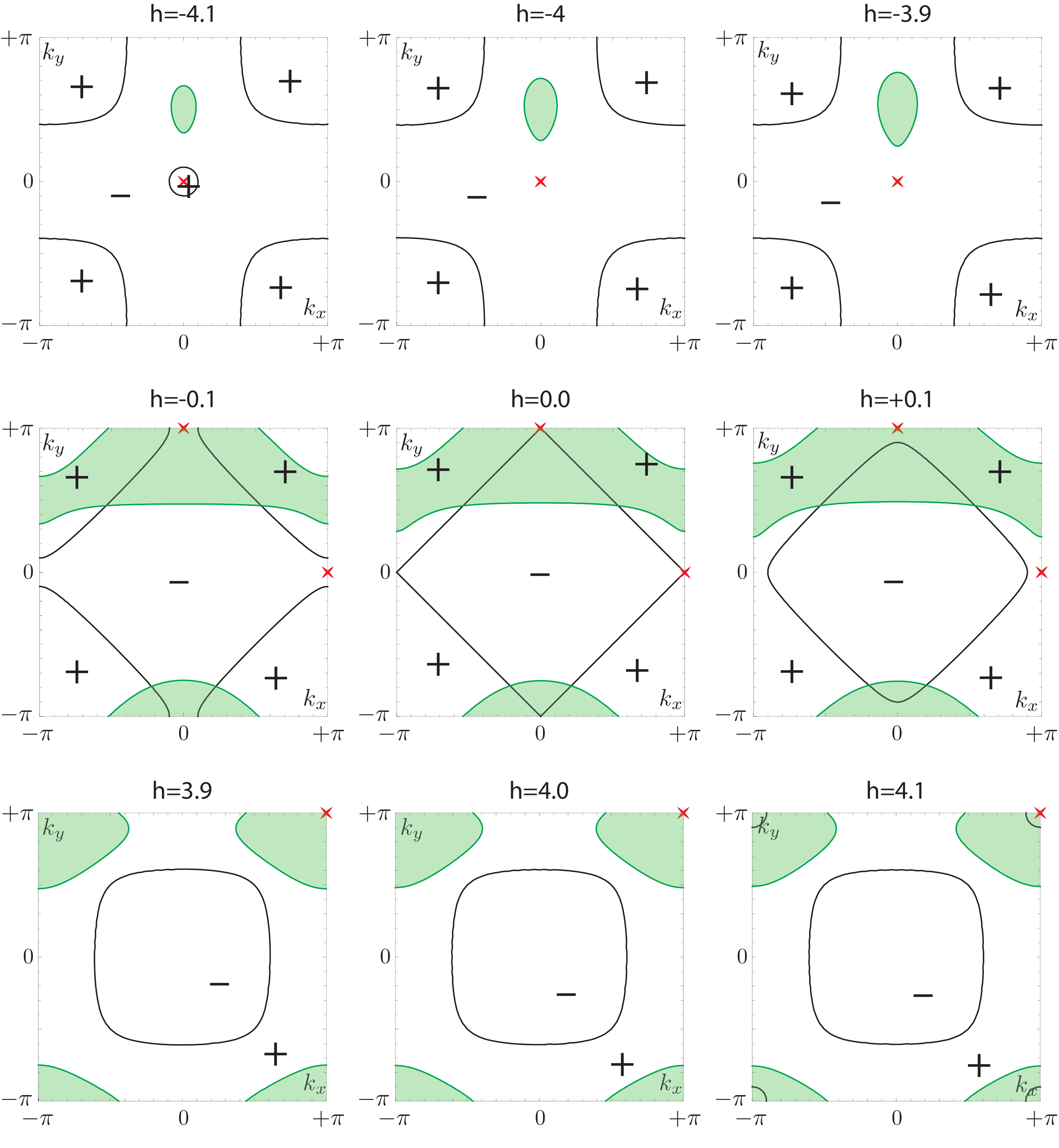}
    \caption{ The Berry curvature $\Omega_{+}(\mathbf{k})$ and the Fermi surfaces (FS) of the even-like Chern metal,
    it leads to $ \Delta \sigma_{+} $ in the Table III.
    Other parameters are $t=1$, $t_s=1$, $t_b=0.75 < 1 $.
    The black curve is the contour of $\Omega_+(\mathbf{k})=0$,
    which separates the positive part denoted by ``+'' from the negative part denoted by ``-''.
    The red $\times$ denotes the singular part of $\Omega_{+}(\mathbf{k})$ listed below Eq.\ref{Delta_sigma}.
    The green line denotes the electron FS.
    Top: near $h/t=-4$,
    the upper band FS always excludes the singularity of $\Omega_{+}(\mathbf{k})$ leading to  $ \Delta \sigma_{+}=0 $,
    thus the Universal Hall conductance jump is 1;
    Middle: near $h/t=0$: the upper band FS only encloses one of the two singularities  of $\Omega_{+}(\mathbf{k})$ leading to  $ \Delta \sigma_{+}=-1 $,
    thus the Universal Hall conductance jump is also 1;
    Bottom: near $h/t=+4$, the upper band FS always enclose the singularity of $\Omega_{+}(\mathbf{k})$  leading to  $ \Delta \sigma_{+}= 1 $,
    thus the Universal Hall conductance jump is 0 ( Table III ).
    The lower band is always occupied, so always encloses the singularity of $\Omega_{-}(\mathbf{k})$,  it leads to $ \Delta \sigma_{-} $ in the Table III.
    These features are essentially the same as those in the even Chern metal case shown in Fig.\ref{even2berry},
    so the metallic phase in  $ -4t < h < 0 $ can be called even-like Chern metal.}
\label{evenlikeberry}
\end{figure}

\begin{figure}[tbhp]
    \centering
    \includegraphics[width=\linewidth]{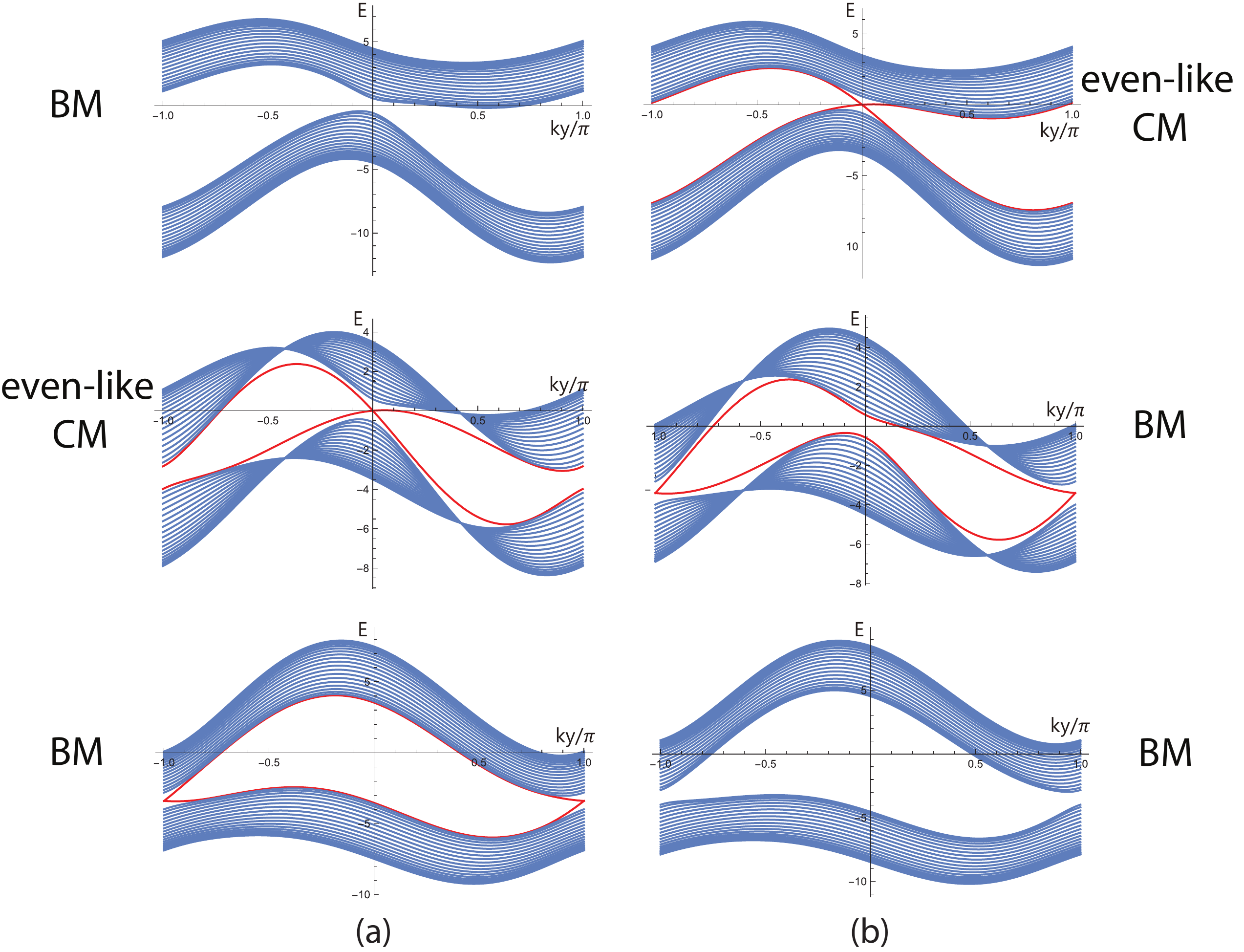}
    \caption{The longitudinal edge structure of the even-like Chern metal  at different $h/t$ values.
    Other parameters are $t=1$, $t_s=1$, $t_b=0.85< 1 $.
    Top (a) $h/t=-4.5$, no edge state, bulk FS, just a  band metal
    (b) $h/t=-3.5$, an edge state at Fermi energy near $ k_y=0 $, bulk FS, even-like Chern metal.
    There is TPT from (a) to (b) and an associated unit Hall conductance jump due to the useful edge mode in (b).
    Middle (a) $h/t=-0.5$, an edge state at Fermi energy near $ k_y=0 $, even-like Chern metal (b) $h/t=+0.5$,
    there is an edge state near $ k_y= \pi $, but it is well below the Fermi energy. So despite there is an edge mode floating around the
    vast majority of bulk states. It is a BM.
    There is a TPT from (a) to (b) and an associated unit Hall conductance jump due to the useful edge mode in (a).
    Bottom (a) $h/t=+3.5$, there is an edge state near $ k_y= \pi $, but it is well below the Fermi energy, a BM (b) $h/t=+4.5$, no edge state, a BM.
    There is no Hall conductance jump.
    Inside the even-like Chern metal, the edge modes at the Fermi energy near $ k_y=0 $ satisfy $v_Lv_R<0$,
    thus the Hall conductance jump is an integer.
    These features are essentially the same as those in the even Chern metal  shown in Fig.\ref{even2edge},
    so can be called even-like Chern metal.   }
\label{evenlikeedge}
\end{figure}


\section{ The even-like and odd-like Chern metals }

In any real material, there exist both the even and odd component. So
in this section, we study the even-odd mixing case $\epsilon_0(k)=-2t_b(\sin k_y+1-\cos k_y)$
which break both the $\mathcal{C}$-symmetry and $\mathcal{P}$-symmetry.

The total Hamiltonian takes the form:
\begin{align}
    H(k)=-2t_b(\sin k_y+1-\cos k_y)+H_\text{QAH}(k)
\label{eq:H-Even-3}
\end{align}
The two energy bands are:
\begin{align}
E_\pm(k)=\pm E_\text{QAH}(k)-2t_b(\sin k_y+1-\cos k_y)
\end{align}
At the QPT from the Chern metal phase to the band metal,
there is no Hall conductance jump at $h/t=4$,
but a Hall conductance jump at $h/t=-4$ and $0$.
To exam the Hall conductance jump, by tracing the location of the FS,
we find it can still be classified into  the even-like or odd-like Chern metals.

\begin{figure}[tbhp]
    \centering
    \includegraphics[width=0.9\linewidth]{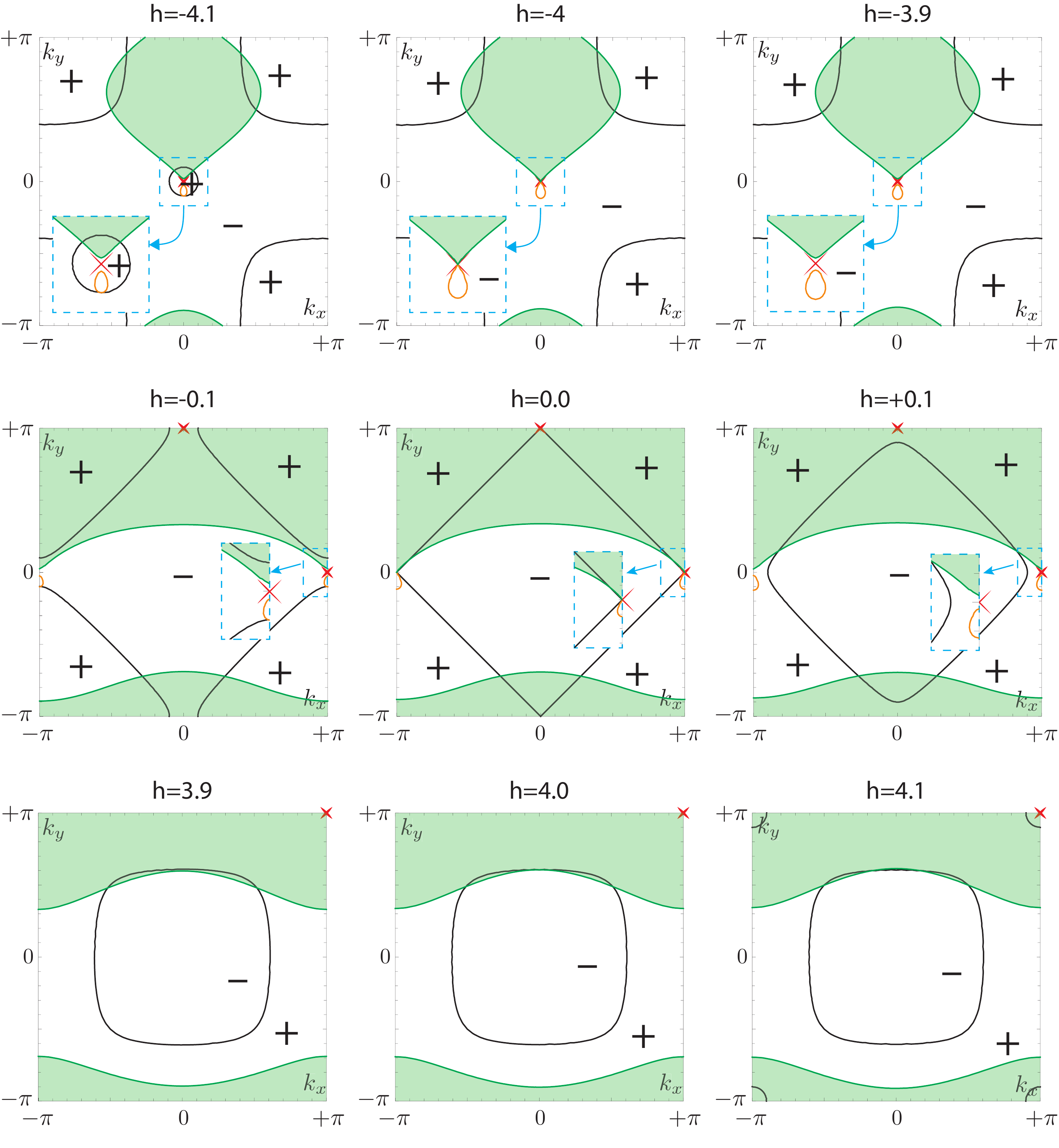}
    \caption{Berry curvature $\Omega_{+}(\mathbf{k})$ and the FS of the odd-like Chern metal:
    Other parameters are $t=1$, $t_s=1$, $t_b=1.25 > 1 $.
    The black curve,  the red $\times$ and the green line denote the same information as Fig.\ref{even1berry}.
    However, the main difference than Fig.\ref{evenlikeberry} is that the lower band is not always occupied, then
    the orange line denotes the hole FS of the lower band which is small and magnified in the figure.
    Top: near $h/t=-4$:
    the upper band FS inclined to enclose the singularity of $\Omega_{+}(\mathbf{k})$,
    it leads to $ \Delta \sigma_{+} $ in the first line in  Table III.
    the lower band FS inclined to exclude the singularity of $\Omega_{-}(\mathbf{k})$,
    it leads to $ \Delta \sigma_{-} $ in the first line in the Table III.
    thus the Universal Hall conductance jump is a non-integer;
    Middle: near $h/t=0$:
    the upper band FS inclined to enclose one of the two singularities of $\Omega_{+}(\mathbf{k})$,
    the lower band FS inclined to exclude one of the two singularities of $\Omega_{-}(\mathbf{k})$,
    thus the Universal Hall conductance jump is a non-integer;
    Bottom: near $h/t=+4$: the upper band FS always enclose the singularity of $\Omega_{+}(\mathbf{k})$,
    the lower band is full occupied,  thus the Hall conductance jump is 0. As shown in Table IV,
    Fig.\ref{oddlikeedge} and appendix C, the metallic phase in  $ -4t < h < 0 $ can be called odd-like Chern metal.}
\label{oddlikeberry}
\end{figure}

\begin{figure}[tbhp]
    \centering
    \includegraphics[width=\linewidth]{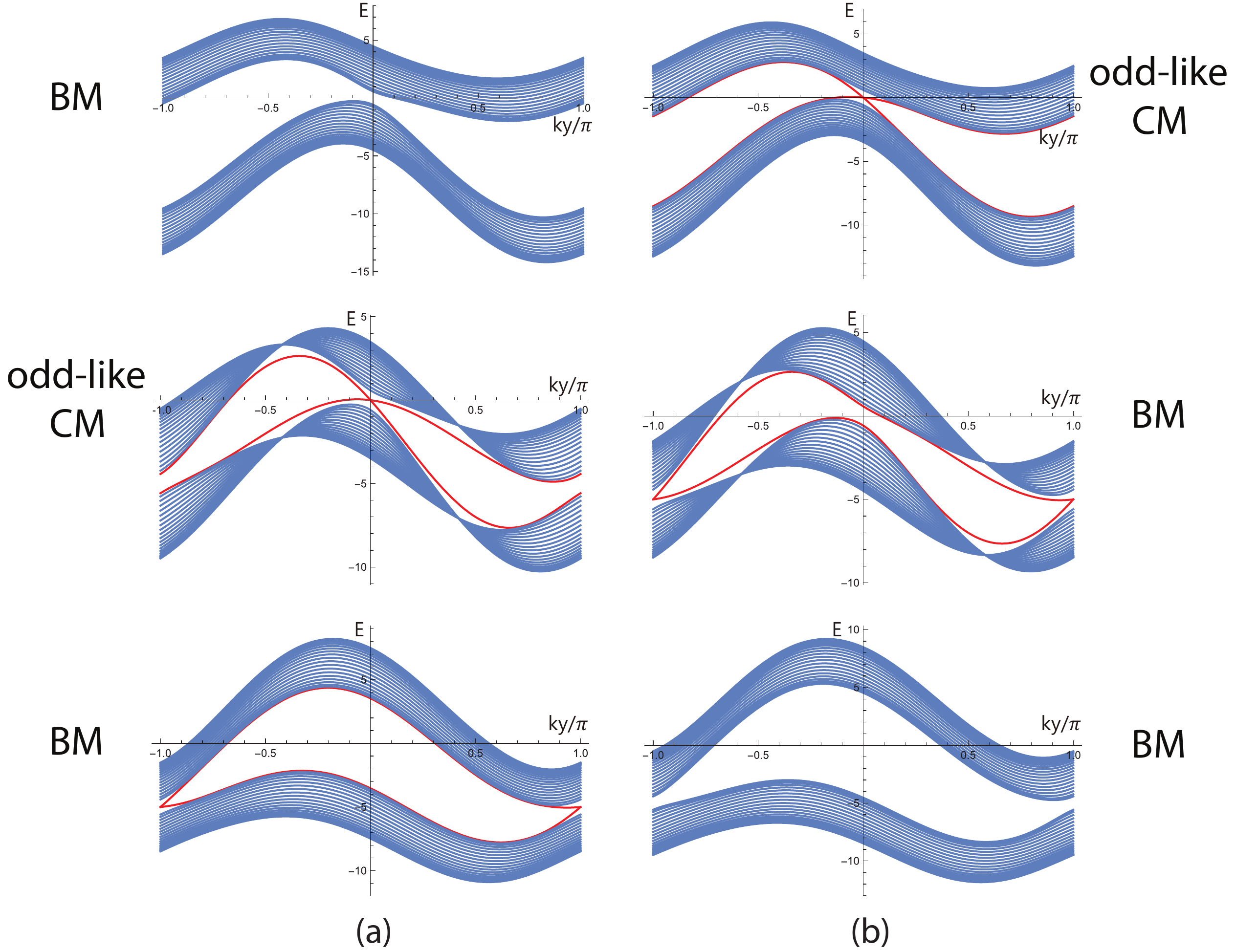}
    \caption{
    The longitudinal edge of the odd-like Chern metal in a stripe geometry at different $ h/t $ values.
    Other parameters are $t=1$, $t_s=1$, $t_b=1.25 >1 $.
    Top (a) $h/t=-4.5$, no edge state, a bulk FS, a BM. (b) $h/t=-3.5$, an edge state at Fermi energy near $ k_y=0 $,
    a bulk FS, an odd-like Chern metal. Due to the useful exotic edge mode in (b),
    there is a universal non-integer Universal Hall conductance jump across the TPT from the BM to the odd-like CM.
    Middle (a) $h/t=-0.5$, an edge state at Fermi energy near $ k_y=0 $, a bulk FS, an odd-like Chern metal.
    (b) $h/t=+0.5$,  there is an edge state near $ k_y= \pi $, but it is well below the Fermi energy.
    despite an edge mode floating around the
    vast majority of bulk modes. Due to the useful exotic edge mode in (a), there is a universal non-integer Universal Hall conductance jump across the TPT from (a) to (b).
    Bottom (a) $h/t=+3.5$, there is an edge state near $ k_y= \pi $, but it is well below the Fermi energy, a BM. (b) $h/t=+4.5$, no edge state, a BM .
    There is no Hall conductance jump.
    Inside the odd-like CM, the exotic edge modes at the Fermi energy near $ k_y=0 $ satisfy $v_Lv_R> 0$,
    thus the Universal Hall conductance jump is a universal non-integer.   }
\label{oddlikeedge}
\end{figure}

\subsection{ The even-like Chern metal }

When $t_b\leq t_s$, the situation is similar to the even case II presented in the appendix B,
thus there is no Hall conductance jump at $h/t=4$, but a unit Hall conductance jump at $h/t=-4$ and $0$.
It is shown in Table III which is identical as the Table V for the even Chern metal, so the metallic phase in  $ -4t < h < 0 $
can be called even-like Chern metal.  
The change of Chern number $ \Delta Ch_{-} $ can be read from Fig.1, which is independent of $ \sigma_0 $ term. 

\begin{table}[!tbhp]
    \setlength{\tabcolsep}{10pt}
        \caption{ The  Universal Hall conductance jump from the even-like Chern metal to BM  extracted from 
        Fig.\ref{evenlikeberry}. It is defined as that of BM subtracts that of the even CM. }
        \begin{tabular}{c|c|c|c|c}
        \hline\hline
        $h$             &$\Delta\sigma_-$   &$\Delta\sigma_+$   &$\Delta\sigma$ &  $ \Delta Ch_{-} $   \\
        \hline
        $-4|t|$    &   +1              &  0                & 1   &     1     \\
        $0$        &   +2              & -1                & 1   &     2     \\
        $+4|t|$    &   -1              & +1                & 0   &    -1       \\ 
        \hline\hline
        \end{tabular}
\end{table}

\subsection{ The odd-like Chern metal }

When $t_b> t_s$, the situation is similar to the main text ``odd'' case.
thus there is no Hall conductance jump at $h/t=4$
and non-integer Hall conductance jump at $h/t=-4$ and $0$.
The value of non-integer is nothing but $t_s/t_b$.

\begin{table}[!tbhp]
    \setlength{\tabcolsep}{10pt}
        \caption{ The Universal Hall conductance jump of the  odd-like Chern metal to BM. 
        The change of Chern number $ \Delta Ch_{-} $ is the same as that in Table-III.}
        \begin{tabular}{c|c|c|c}
        \hline\hline
        $h$             &$\Delta\sigma_-$   &$\Delta\sigma_+$   &$\Delta\sigma$\\
        \hline
        $-4|t|$         &   $t_s/(2t_b)$    & $t_s/(2t_b)$      & $t_s/t_b$\\
        $0$             &   $+1+t_s/(2t_b)$ & $-1+t_s/(2t_b)$   & $t_s/t_b$\\
        $+4|t|$         &   -1              & +1                & 0\\
        \hline\hline
        \end{tabular}
\end{table}
   which is identical to the odd-Chern metal presented in Secs.II-VI, so the metallic phase below  $ -4t < h < 0 $
   can be called odd-like Chern metal. Note that due to the C- symmetry,
   the jump from one odd Chern metal to another odd Chern metal near $ h=0 $ is ``twice'' that from the odd CM to BM ( Fig.24 ).
   Here, due to the absence of the C- symmetry, there is only TPT from an odd CM to a BM, the TPT from one odd CM to another odd CM is absent,
   so does the ``doubling''.

\section{ "Topological invariants" and bulk-edge correspondence in gapless fermionic systems with extended Fermi surface  }

    Table I shows that the Chern number ( or Chern number jump )  may not be an effective way to distinguish even/odd Chern metal
    from the band metal. Because as shown in the appendix A and B, the band metal could also have non-vanishing Chern number.
    There could also be Chern number jump across two band metals which are essentially the same quantum phase.
    The band metal and even/odd CM themselves all have non-vanishing Hall conductances
    ( Only in the fine tuning case addressed in Sec.II and Sec.III, the band metal itself has zero Hall conductance ).
    It is the Universal Hall conductance jump which plays the role of the "topological invariant " to distinguish phases with different topological properties.
    However, they may not be an integer anymore: it remains an integer at the TPT from ECM to band metal.
    However, it becomes any number at the TPT from OCM to band metal.
    Chern number or its jump are not experimentally measurable, but the Hall conductance jumps are.
    There are also corresponding TPT in the L-edge mode. However, as shown in Table 1, these TPTs are not QPT.
    So far, they present the very first example of TPT which is not a QPT.
    These salient features describe completely the unique properties of "topological invariants"
    and bulk/edge correspondence in gapless topological phases with extended FS.
    Of course, the topological gapless phases studied in this manuscript have no in-direct gaps, but still direct gaps which give
    the room for bulk "topological invariants" and the associated L-edge modes.

\subsection{ Chern number (jump) versus the Hall conductance (jump) }

It was known that  the  topological invariant can not be experimentally measured directly.
What can be  experimentally measured directly is the Hall conductance $ \sigma_H $.
As shown in the introduction, there are intricate relations between the two \cite{z2TI}: in the
gapped insulating phases such as the even Chern insulator, odd Chern insulator and band insulator, they are the same,
but in the gapless metal phases such as Band metal, even CM and odd CM,
they maybe different: the Chern number $ Ch_{-} $ is evaluated over a given band,
independent of the fillings of the band, so can only be an integer \cite{bosonicQAH}. But the Hall conductance $ \sigma_H $
depends sensitively on the filling of the band, so need not to be quantized and can be any number even at $ T=0 $.
One can also evaluate $ \sigma_H(T) $ at a finite $ T $ by incorporating the Fermi distribution functions
on the fillings of the band at a finite $ T $. Most importantly, there are enriched bulk-edge correspondences
across the bulk TPT from the even Chern insulators to odd Chern Insulator, then to Odd Chern metals or
an alternative path from the even CI to even CM, then to Odd CM shown in Fig.\ref{sixfolds}.
The differences between $ Ch_{-} $ and $ \sigma_H $, especially their jumps in the Odd/Even Chern metal phases can be fully appreciated
in both longitudinal edge and transverse edge.

\subsection{ A first example of TPT without a QPT }

It is interesting to dig further from Sec.II-D on the universal unit conductance jump from the even CM to the BM listed in Table III
and the universal non-integer conductance jump from the odd CM to the BM listed in Table IV.
Just from the change in the ground state energy which reflects the quantum fluctuations, it is not a QPT.
As demonstrated in the last section, both odd and even CM have a non-vanishing Chern number and
associated edge modes near the Fermi energy inside the gap.
But some BM also has non-vanishing Chern number and some floating L-edge modes near the Fermi energy only.

However, from the change in the bulk Berry curvature, especially near the Dirac points, which reflects the global topological features,
it does show the universal Hall conductance jump. So it is clearly a TPT.

From the change in the edge modes which also reflects the global topological features,
it does show the change in the edge mode: the even CM has edge modes at the Fermi energy
near the projection of the Dirac points on the edge satisfying $ v_L v_R < 0 $.
the odd CM has edge modes at the Fermi energy near the projection of the Dirac points on the edge satisfying $ v_L v_R > 0 $.
These Dirac points are also the bulk singularities in the bulk Berry curvature.
The even CM and odd CM have the same bulk, but different edge with an edge reconstruction dynamic exponent $ z_L=2 $.
While the BM does not have such edge modes except some floating edge modes away from these projection points.

 It is also instructive to compare this universal Hall conductance jump to the universal jump in the superfluid density across the KT transition
 $ \Delta \rho_s/k_B T = \pi/2 $. It is a topological phase transition driven by vortex binding-unbinding, no order parameters, no symmetry breaking,
 has an essential singularity in free energy and also an infinite order transition.
 Numerically, we are not able to tell if there is also such a essential singularity in the ground
 state energy across the even/odd CM to BM transition.

In summary, this maybe the very first example of a TPT with the corresponding edge reconstruction, but not a QPT.
So far, all the previous known examples are TPT is always accompanied by a QPT \cite{weyl}.
Namely, any changes in topology and topological invariants is also accompanied by the change in the ground state energy.
Of course, a QPT need not necessarily a TPT. This very first example maybe contrasted to the bosonic 2d KT which is
in its own class in 2d classical and 1+1 d quantum critical phenomena.

\section{ Experimental detections by injecting current or in a moving sample  }

 This work consists of two parts. In the first part, we explore the QAH in a NN injecting current.
 In the second part, we examine the QAH subject to a NN gauge invariant current plus a NNN current.
 As shown in \cite{moving} and reviewed in the appendix E, the second part may also apply to the QAH in
 cold atom systems where the SOC is generated artificially, therefore non-relativistic.
 In this section, we discuss the experimental detections in the following three different and complementary set-ups respectively.

\subsubsection{In a NN injecting current}

 The first experiment to do is to directly inject current into the QAH systems shown in Fig.\ref{frames}a,b.
 In fact, the injecting current is nothing but the  $ d_0(k) $ term in Eq.\ref{dvector} in the momentum space
 which breaks the Parity.
 Of course, as alerted at the very beginning,
 a parity-even component $ \epsilon_0(k)=-2 t_e(2-\cos k_x -\cos k_y) $ may also exist in any real material
 in the first place. Its effects can be addressed similarly as in Sec.VII-VIII.
 Injecting currents maybe just the most effective way to tune this P- breaking term to drive the TPTs
 in Fig.\ref{fig:phaseLattice} or Fig.\ref{evenoddphase}.
 As shown in Fig.\ref{fig:phaseLattice}, it needs $ t_b/t=1 $
 to reach the odd CM directly from the even CI.
 As shown in Fig.\ref{evenoddphase}. Near $ h/t \sim 0^{-} $,  it is easy to reach the even-like CM
 from the even CI due to $ t_b/t \to 0^{-} $, but it still needs $ t_b/t=1 $ to reach the odd-like CM.

 All the measurements can be routinely performed in a static sample.
 The whole global phase diagram Fig.\ref{fig:phaseLattice} and Fig.\ref{evenoddphase}  can be explored by just
 tuning the strength of the injecting current $ t_b $. For example, various light \cite{lightatom1,lightatom2},
 atom, X-ray ( or ARPES ) and neutron scattering  can be used to detect the bulk FS geometry,
 the bulk excitation spectrum  and the associated edge reconstructions   in all the phases
 in Fig.\ref{fig:phaseLattice2} and Fig.\ref{evenoddphase}.
 The conventional 2-terminal or 4- terminal  transport measurements can be applied to measure the Hall conductance $ \sigma_H $
 and its jump $ \Delta \sigma_H $ from even/odd CM to the band metal.
 The specific heat $ C_v $, the compressibility $ \kappa_u $ and the Wilson ratio $ R_w $
 can be separately measured by various established thermodynamic measurements in both materials and cold atom systems.

\subsubsection{In a NN gauge invariant current plus NNN current}

 The injecting current term  could also be the gauge-invariant current in Eq.\ref{boostLattice}.
 We assume the NN dimensionless ratio $ \kappa_{b1} $ and the NNN hopping $ t_{b2} $  in Eq.\ref{boostLattice} can be controlled independently.
 Then the full phase diagram  Fig.\ref{fig:phaseLattice2}, especially the odd CI phase and its associated edge modes can be explored.

 From Eq.\ref{tts},  one can see that the band  insulator with $ Ch_{-}= 0 $ in the absence of the current near the
 phase boundary $ h/t \sim \pm 4 $ can be transformed to the Chern insulator with $ Ch_{-}= \pm 1 $ by injecting the current Eq.\ref{boostLattice}.
 This is because the gauge-invariant current effectively increases
 the hopping $ t $ and the SOC strength $ t_s $ according to Eq.\ref{tts}, therefore reduces the critical value of $ h/t $.
 This qualitative change of ground state can be detected by measuring the  Hall conductance $ \sigma_H $ and also detecting the edge modes by ARPES.
 $ t_b $ term Eq.\ref{tbform} leads to the Doppler effect in the electronic spectrum which changes sign when $ \kappa_{b1} = 1 $.
 This sign change can be captured by the various light scattering detections.
 By keeping $ \kappa_{b1} \ll 1 $, increasing the NNN $ t_{b2} $ current
 will drive the even CI to odd CI, then the odd CI to odd CM in Fig.\ref{fig:phaseLattice2} .
 It needs $ t_b/t=1 $ to reach the odd CI, then a even larger  $ t_b/t > 1 $ to reach the odd CM.
 The evolution of edge states, the edge reconstruction, the enriched bulk-edge correspondence, especially in the odd CI and odd/even CM
 can be mapped out by light scattering detections
 in the longitudinal or transverse moving sample in Fig.\ref{fig:Edge_Lattice_n2} and \ref{fig:Edge_Lattice2_n2}.

\subsubsection{In a moving sample  for the artificially generated QAH}

 As discussed in \cite{moving},
 if the SOC is artificially created in cold atom system \cite{haldaneexp,2dsocbec,QAHboson},
 it is not a relativistic effect. The spin in the SOC is just a pseudo-spin consisting of two hyperfine states.
 Then the gauge invariant current plus a NNN current in Eq.\ref{boostLattice} can be generated by a Galileo transformation (GT)
 as analyzed in the appendix E, except the two terms are not independent anymore.
 So some of the results achieved in Sec.V and VI can be adopted to this case also with some stringent restrictions.

\begin{figure}[tbhp]
\centering
\includegraphics[width=0.7 \linewidth]{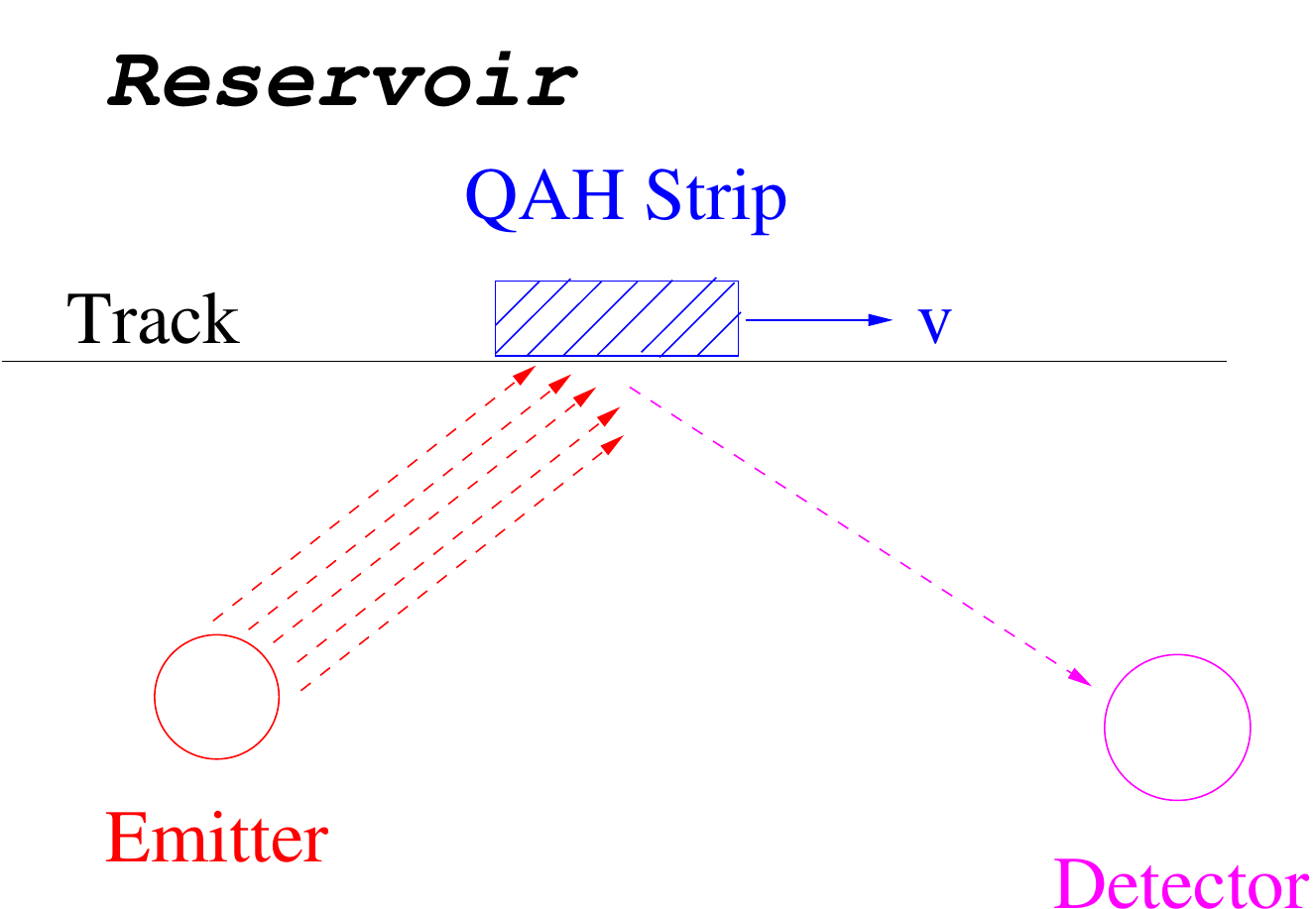}
\caption{ Light ( ARPES), atom or neutron scattering on a moving QAH sample with a strip geometry and a velocity $ \vec{v} $.
The longitudinal moving means the sample edge is along the velocity $ \vec{v} $.
The transverse moving means the sample edge is normal to the velocity $ \vec{v} $.
As shown in Sec.V, VI and appendix E, the edge states show quite different behaviours in the longitudinal or the transverse moving case
which can be detected by the scattering cross sections.
The multiple irradiating lines from the emitter declinate the irradiation regime where the sample enter and leave, then the scattered beams
can be detected by the receiver.  In a grand canonical ensemble,  the Reservoir shows exchange particles with the sample through the chemical potential $ \mu $.  }
\label{detector}
\end{figure}



 From Eq.\ref{ttbv},  one can see that the band  insulator with $ Ch_{-}= 0 $ in the static frame $ v=0 $ near the
 phase boundary $ h/t \sim \pm 4 $ can be transformed to the Chern insulator with $ Ch_{-}= \pm 1 $, but not the other way around.
 This counter-intuitive phenomenon can be best read from Eq.\ref{ttbv}: the Galileo boost effectively increases
 the hopping $ t $ and the SOC strength $ t_s $, therefore reduces the critical value of $ h/t $.
 However, due to $ t_b/t \ll 1 $, one is not able to even get close to the phase boundaries from the CI to odd CI,
 let alone the odd CI to odd CM in Fig.\ref{fig:phaseLattice2} which all happen when $ t_b/t_s > 1 $.
 So the NNN $ t_b $ term can only introduce a small Doppler effect in the electronic spectrum.

 Even so, we still have the three interesting results accessible to the detection in Fig.\ref{detector}:
 (1) The band  insulator near the 3rd order TPT from the Chern insulator to the band  insulator  will turn into
     a Chern insulator
 (2) There is a Doppler shift term in the fermionic spectrum in the moving sample which changes sign when $ v=v_c=t_0/t_{b1} $.
     For the cold atom systems in an optical lattice, the characteristic velocity can be estimated as $ v_c \sim 1 cm $
     which is easily within experimental reach \cite{moving}.
 (3) The shifted edge dispersions Eq.\ref{pmedge} in the longitudinal edge and Eq.\ref{squarerootedge}  in the transverse edge.
  These three predictions can be detected in the moving sample Fig.\ref{detector} by taking advantages of
  both bulk and longitudinal or transverse edge properties.

 First thing to try is  to set the sample static in the lab frame $ S $, but observed in
 the moving frame $ S^{\prime} $. In practice, the sample is very small, but the detecting device is
 usually heavy. Because exchanging the role of the lab and moving frame does not change the results,
 because both are related by Galileo transformation (GT) anyway. So,
 in a practical scattering detection experiment shown in Fig.\ref{detector},
 it is more convenient to set the emitter and the receiver static and make the sample moving with a constant velocity $ \vec{v}  $.
 Due to the small size of the sample, it is not easy to focus the beam on the sample when it is moving.
 To overcome this difficulty, one may just continuously shine the emitting beam, only when the sample move into its
 shadow, it will be scattered and collected by the detector. When it moves out of the shadow, there is no scattered beam anymore.



 However, when setting the sample moving with a constant velocity $ v $ in Fig.\ref{detector}, some of the measurements
 such as transports may become hard to implement. The Chern number $ Ch_{-} $ is not experimentally measurable in any inertial frames anyway.
 Fortunately, all kinds of scattering experiments mentioned in {\sl 1} and  sketched in  Fig.\ref{detector} remain
 applicable.
 One simply perform all the measurements in the lab frame $ S $, so for the light scattering,
 there is no need to consider the relativistic Doppler shift of the photon.
 Unfortunately, it seems difficult to measure the free energy, therefore the specific heat in the moving sample.
 Even more difficult things to measure is the Hall conductance $ \sigma_H $ which
 can be used to distinguish the  Chern insulator from the band  insulator, also the Odd Chern metal from the band metal.
 Fortunately, they can still be detected by the edge states either in longitudinal or transverse moving shown in Fig.\ref{detector}.


\section{Conclusions and perspectives }

The global phase diagram Fig.\ref{fig:phaseLattice} due to the injecting current and Fig.\ref{fig:phaseLattice2}
due to the gauge invariant current or a moving sample
contain several novel TPT/QPT.
Several new concepts also emerge from these phase diagrams: enriched bulk/L-edge/T-edge correspondences
which lead to intervening new topological phases such as OCI and ECM,
the L-edge dynamic exponent such as $ z_L=2,3 $, the T-edge critical behaviour such as $ \sim \sqrt{v^2-c^2} $,
the Universal Hall conductance jump replace the Chern number as "topological invariants",
especially a TPT which is not a QPT such as the OCM to BM and ECM to BM transition, etc.

It is instructive to compare the non-interacting spin-1/2 SOC fermionic TPTs in Fig.\ref{fig:phaseLattice} with that of the
interacting spin-0 bosonic Mott to the superfluid (SF)
QPT with $ z=1 $ in \cite{moving}. One can make the following loose analogy between the two systems:
$ h/t \sim U/t $, band  insulator $ \sim $ Mott phase, Chern insulator $ \sim $ SF phase,
Odd Chern metal/band metal phase $ \sim $ Boosted SF  (BSF) phase.
Here it has no symmetry breaking, the continuous $ U(1) $ symmetry and the discrete
C-symmetry are never broken, so no order parameters.
The change of FS topology and geometry across TPTs can be characterized by
the topological invariants such as
the Universal Hall conductance jump $ \Delta \sigma_H $.
There are also corresponding edge-states through the enriched bulk-edge correspondences.
However, in the interacting bosonic case, the continuous $ U(1) $ symmetry and the discrete
C- symmetry can be broken and characterized by various order parameters:
the $ U(1) $ is broken in the SF, both $ U(1) $ and C- are broken in the BSF phase.
There are no non-trivial edge modes, let alone any edge reconstructions in such topologically trivial interacting
bosonic systems. The boost favors the SF phase in the boson case.
Here it favors the Chern insulator.
For the interacting bosons, moving away from the C- symmetric point corresponds to $ z=2 $.
In Sec.II-VI, we focus on the half filling $ \mu=0 $ case which respects the C-symmetry.
In appendix A, we study the $ \mu \neq 0 $ case by doping the QAH materials.
We also investigate the much more interesting cases of breaking the C-symmetry in the appendix B and Sec.VII-VIII.

The Universal Hall conductance jump $ \Delta \sigma_H $ replace the Chern number jump to become
the new bulk topological invariant involved in the even/odd CI and the even/odd CM discovered in this work ( Fig.\ref{sixfolds} ).
There is also corresponding  new enriched bulk/L-edge/T-edge correspondence. It is these combined features which lead to a
likely complete classification of quantized, especially un-quantized AHE metals.
It is constructive to compare to the 3d Weyl metals where there is no new topological invariant involved, but a new structure of edge modes called Fermi-arc.
It also contributes to a un-quantized 3d  AHE $ \sigma_{xy}= e^2/h \times ( 2 k_0 ) $
where $ 2 k_0 $ is the momentum distance along $ \hat{z} $ between the $ + $ and $ - $ Weyl point.
As mentioned in the introduction, our approach to the classification of gapless AHE metallic phase is complementary to the
traditional one using SPT or SET to classify gapped or gapless topological phases:
We start from the known parent Hamiltonian whose topological phases are known, then add various symmetry breaking perturbations
to drive the known topological phases to new topological phases through new TPTs. Here, we
add the P-breaking currents leading to the odd Chern Insulator or odd Chern-metal. We also add
the C- breaking energy dispersion leading to possible even Chern metals. This specific Hamiltonian based classification scheme has the advantages
to discover new topological phases through possible novel TPTs, so can be used to classify topological phases and TPTs at the same time.
Furthermore it also automatically leads to various deformed Hamiltonians hosting these phases,
so may be directly connected to experimental realizations.
The disadvantages is that it is hard to prove it is a complete and exhaustive classifications.
It remains to show its completeness from a symmetry based approach such as K-theory on the Wannier basis \cite{Read}. However, it remains very challenging to
clarify completely and exhaustively  topological gapless phases with extended FS with $ z=2 $ from such an abstract algebraic/topological K-theory approach,
and also classify the bulk TPTs and the edge reconstructions among these phases.

  Recently, there are also new advances in the classification of 1d gapless interacting topological phases
  \cite{gapless2,wengapless,holgraphicorder} where 2d CFT
  and boundary CFT can be a useful tool. Due to the lack of local CFT, its extension to higher dimensions is more challenging.
  Here, we attempt a classification of 2d gapped Chern insulators and 2d gapless AHE metals with extended FS and $ z=2 $.
  Obviously, it has no conformal invariance in the first place,
  so 2d CFT and boundary CFT may not apply.
  It may also tempting to study the effects of the Hubbard interaction $ U $ in
  both the weak $ U/t \ll 1 $ and strong coupling $ U/t \gg 1 $ limit in Eq.\ref{QAH0} under an injecting current or in a moving sample.
  In the strong coupling limit and at integer fillings, both the bosonic and fermionic Hubbard model leads to various quantum spin models.
  Drawing the insights here achieved on the non-interacting topological phases,
  we expect that Similar methods may also be applied to study the strongly interacting SPT and SET,
  an injecting current will also drive SPT/SET to new SPT/SET through some novel TPTs.
  SPT/SET near a TPT may also depend on sensitively the observer in an inertial frame.

{\bf Acknowledgements}

We thank Prof. Congjun Wu  for helpful discussions and W. M . Liu for consistent encouragements.
We also thank Prof. Congjun Wu and Prof. Gang Tian for the hospitality during our visit at
the Institute for Theoretical Sciences.

\appendix

\section{ " Even Chern metal " Case I: $\epsilon_0(k)=-t_b=-\mu $}

As defined in Sec. VII, a even Chern metal breaks the $\mathcal{C}$-symmetry, but keeps $\mathcal{P}$-symmetry.
The simplest even case is to choose $\epsilon_0(k)=-t_b=-\mu$ in Eq.\ref{QAHevenodd}.
However, in this appendix, we show that simply doping the Chern insulator only leads to a band metal (BM).
The naively thought even Chern metal is nothing but the same phase as the BM.
As stressed in Sec.IX, this is also the simplest example to show the Chern number and its jump may not be effective anymore in
characterizing gapless fermionic topological phases with extended FS.

  This corresponds to the doped case away from the charge neutral point $ \mu=0 $. The total Hamiltonian Eq.\ref{QAHevenodd} takes the form
\begin{align}
    H(k)=H_\text{QAH}(k)-\mu
\label{eq:H-Even-1}
\end{align}
   where the chemical potential $ \mu $ is tuned by the reservoir in Fig.\ref{detector}.

At the QPT from the even Chern metal phase to the band metal,
there is no singularity in the ground-state energy density
and no Hall conductance jump.
The two energy bands are 
\begin{align}
E_\pm(k)=\pm E_\text{QAH}(k)-\mu
\end{align}
where $E_\text{QAH}(k)=\sqrt{[h+2t(\cos k_x+\cos k_y)]^2+4t_s^2(\sin^2k_x+\sin^2k_y)}$.
To exam if there is a Universal Hall conductance jump, without loss of generality, we take $\mu>0$.
Near $h/t=-4$, the Berry curvature singularity is located at $K_0=(0,0)$,
$E_\pm(K_0)=-\mu<0$, both upper band and lower band Fermi Surface (FS) enclose the $K_0$.
So according to Eq.\ref{Delta_sigma}, the Universal Hall conductance jump eventually cancels as shown in the Table V.
Similar situation happens for $h/t=0$ and $+4$.


\begin{figure}[tbhp]
    \centering
    \includegraphics[width=\linewidth]{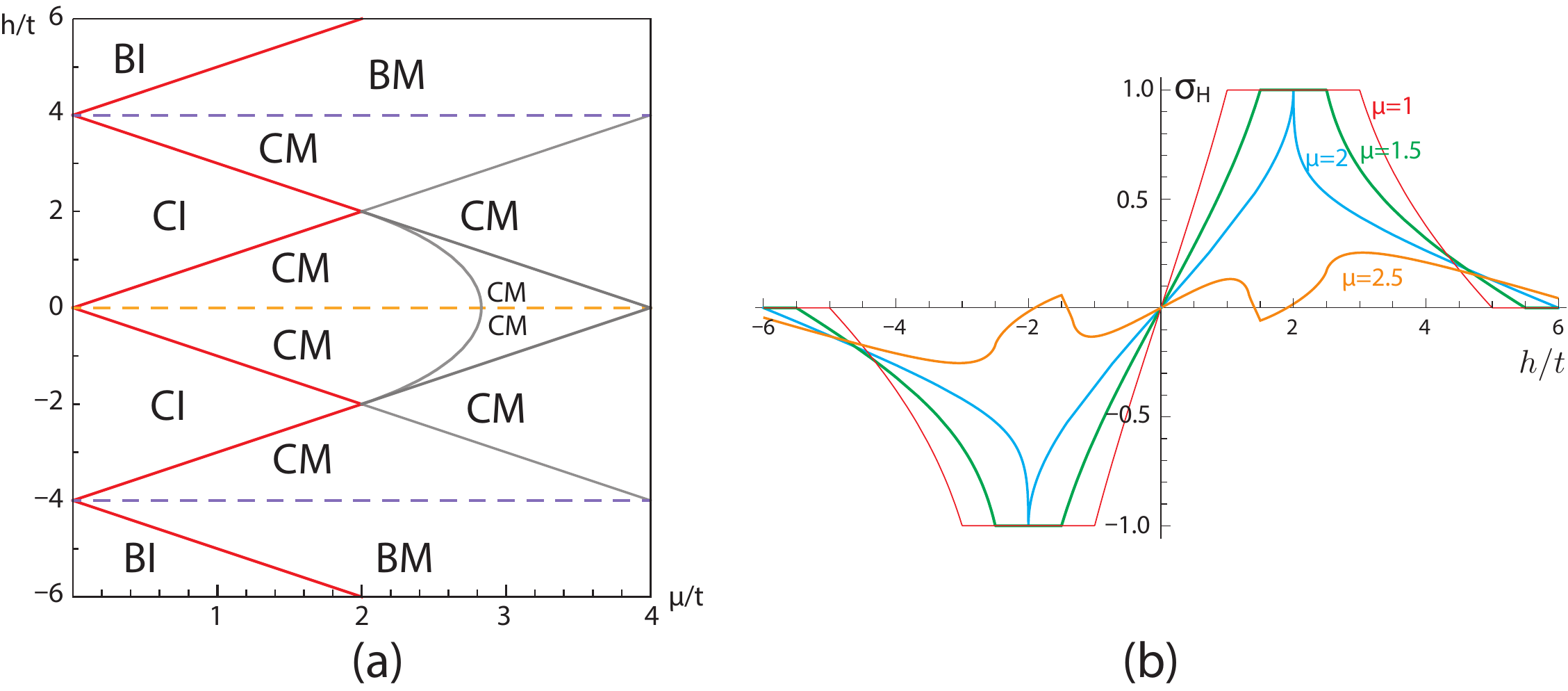}
    \caption{(a) The global phase diagram of the Lattice Hamiltonian \eqref{eq:H-Even-1} for the even Chern Metal I.
    (b) The Hall conductance as a function of $h/t$ for various fixed values of $\mu/t=1.0,1.5,2.0,2.5$.
     The Hall conductance shows no jump at $h/t=-4,0,+4$. As shown in this appendix,
     the `` even '' Chern metal phase is essentially the same as the band metal phase, except it may just has a larger AHE than the BM.
     One may simply call all the metallic phase just BM.
     For a similar analog in the interacting bosonic
     quantum anomalous Hall system, see \cite{QAHboson}. }
\label{even1phase}
\end{figure}

\begin{figure}[tbhp]
    \centering
    \includegraphics[width=0.8\linewidth]{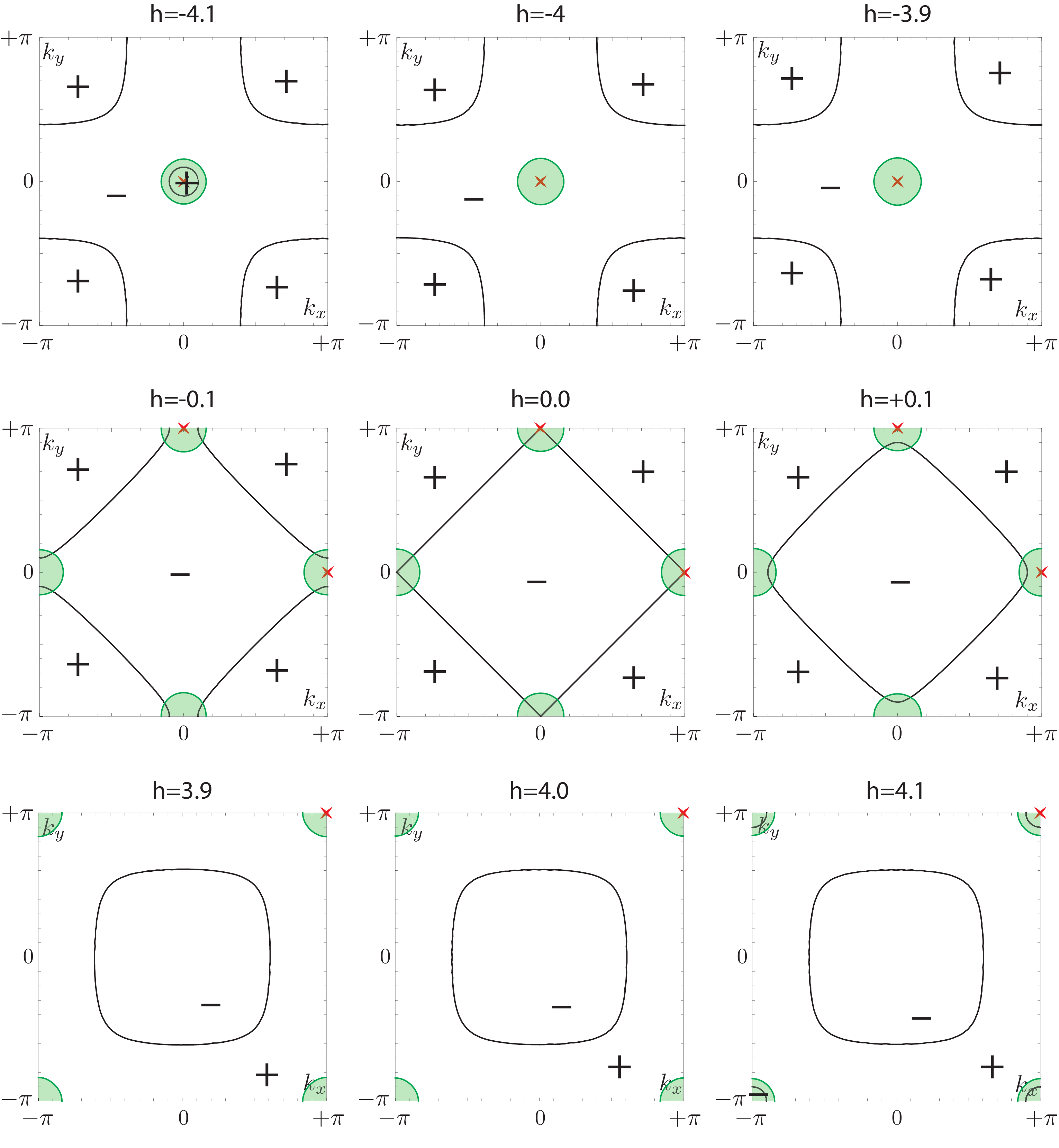}
    \caption{The Berry curvature $\Omega_{+}(\mathbf{k})$ and the Fermi surfaces (FS) of the even Chern metal case I:
    Other parameters are $t=1$, $t_s=1$, $\mu=1$.
    The black curve is the contour of $\Omega_+(\mathbf{k})=0$,
    which separates the positive part denoted by ``+'' from the negative part denoted by ``-''.
    The red $\times$ denotes the singular part of $\Omega_{+}(\mathbf{k})$ listed in Eq.\ref{Delta_sigmaexpress}.
    The green line denotes the electron FS.
    Top, near $h/t=-4$,  Middle, near $h/t=0$,  Bottom, near $h/t=+4$.
    The upper band FS always encloses the singularity of $\Omega_{+}(\mathbf{k})$.
    They lead to $ \Delta \sigma_{+} $ in the Table V.
    The lower band is always occupied, so always encloses the singularity of $\Omega_{-}(\mathbf{k})$,
    it leads to $ \Delta \sigma_{-} $ in the Table V.
    Thus there is no Hall conductance jump $ \Delta \sigma= \Delta \sigma_{+}+ \Delta \sigma_{-} $ as shown in the Table V.
    One may simply call all the metallic phase just as BM.}
\label{even1berry}
\end{figure}

\begin{table}[!tbhp]
    \setlength{\tabcolsep}{10pt}
        \caption{ The Table of the Universal Hall conductance jump of the even Chern metal case I. 
        The change of Chern number $ \Delta Ch_{-} $ is the same as that in Table-III.}
        \begin{tabular}{c|c|c|c}
        \hline\hline
        $h$             &$\Delta\sigma_-$   &$\Delta\sigma_+$   &$\Delta\sigma$\\
        \hline
        $-4|t|$   &   +1              & -1                & 0\\
        $0$       &   +2              & -2                & 0\\
        $+4|t|$   &   -1              & +1                & 0\\
        \hline\hline
        \end{tabular}
\end{table}
   which shows there is no difference between the ``even'' Chern metal here and the BM. So
   One may simply call all the metallic phase as just BM.

   The bulk-edge correspondence in this simplest even-Chern metal case can be extracted from Fig.\ref{fig:Edge_Lattice}
   and Fig.\ref{fig:Edge_Lattice2} at $ t_b=0 $ by cutting the edge modes at the chemical potential $ \mu > 0 $ instead of at $ \mu=0 $.
   Because $ t_b=0 $, there is no difference between longitudinal and transverse.
   If the cut remains inside the bulk gap, it is still a Chern insulator phase.
   When it also cuts some bulk states, it moves into the ``even'' Chen metal phase.
   Then it never cuts any edge mode near $ k_y=0 $ when $ h <0 $ or $ k_y=\pi $ when $ h > 0 $, so there is no
   jump in the Hall conductance, as indicated in Sec.VII-C.
   One may still see one edge mode floating above the vast majority of bulk modes,
   the edge mode does not play any significant role anymore.
   In fact, this floating edge mode can be simply absorbed into the vast majority of bulk modes.

   In summary, there is no ground state energy singularity, no Hall conductance jump from the tentative ``even'' Chern metal to the band metal.
   Despite to so called ``even'' Chern metal has a non-vanishing Chern number in its whole band structure,
   there is really no any physical measurable bulk or edge quantities to distinguish the two phases,
   they are really the same phase which is nothing but the conventional BM phase. So there are only three phases in Fig.\ref{even1phase}:
   CI, BI and BM. In fact, similar phenomenon also happens in the interacting bosonic  quantum anomalous Hall (QAH) system, see \cite{QAHboson}.

\section{Even Chern metal Case II: $\epsilon_0(k)=-2t_b(1-\cos k_y)$. }

 In the last appendix, we show that simply doping the Chern insulator may not lead to a real even Chern metal.
 Here we discuss a second example by choosing $\epsilon_0(k)=-2t_b(1-\cos k_y)$ which does lead to a real even Chern metal
 The total Bloch Hamiltonian takes the form
\begin{align}
    H(k)=-2t_b(1-\cos k_y)+H_\text{QAH}(k)
\label{eq:H-Even-2}
\end{align}

At the QPT from the even Chern metal phase to the band metal,
there is no Hall conductance jump at $h/t=4$
but a unit Hall conductance jump at $h/t=-4$ and $0$.
The two energy bands are 
\begin{align}
E_\pm(k)=\pm E_\text{QAH}(k)-2t_b(1-\cos k_y)
\end{align}

To exam the Hall conductance jump, we just examine the FS  according to Eq.\ref{Delta_sigma}.
At $h/t=-4$, the Berry curvature singularity is located at $K_0=(0,0)$ and $E_\pm(K_0)=0$.
For $h/t\approx -4$, $E_+(K_0)>0$, $E_-(K_0)<0$,
the upper band FS does not enclose the $K_0$,
but the lower band FS encloses the $K_0$.
Thus, there is a unit Hall conductance jump.
At $h/t=0$, the Berry curvature singularity is located at $K_1=(\pi,0)$ and $K_2=(0,\pi)$,
thus $E_\pm(K_1)=0$ and $E_\pm(K_2)\neq 0$.
For $h/t\approx 0$, $E_+(K_1)>0$, $E_-(K_1)<0$,
the upper band FS does not enclose the $K_1$,
but the lower band FS encloses the $K_1$.
Meanwhile, $E_\pm(K_2)$ always greater than 0 or smaller than 0,
the upper band and the lower band FS both enclose or exclude the $K_2$.
Thus, there is a unit Hall conductance jump contribution from $K_1$.
At $h/t=+4$, the Berry curvature singularity is located at $K_\pi=(\pi,\pi)$ and $E_\pm(K_\pi)\neq 0$.
For $h/t\approx +4$, $E_+(K_\pi)$ always greater than 0 or smaller than 0,
the upper band and the lower band FS both enclose or exclude  $K_\pi$.
Thus, there is no Hall conductance jump. In short,  only the metallic phase
below $ h=0 $ can be named as even Chern metal, all the other metallic phases are just BM.

\begin{figure}[tbhp]
    \centering
    \includegraphics[width=\linewidth]{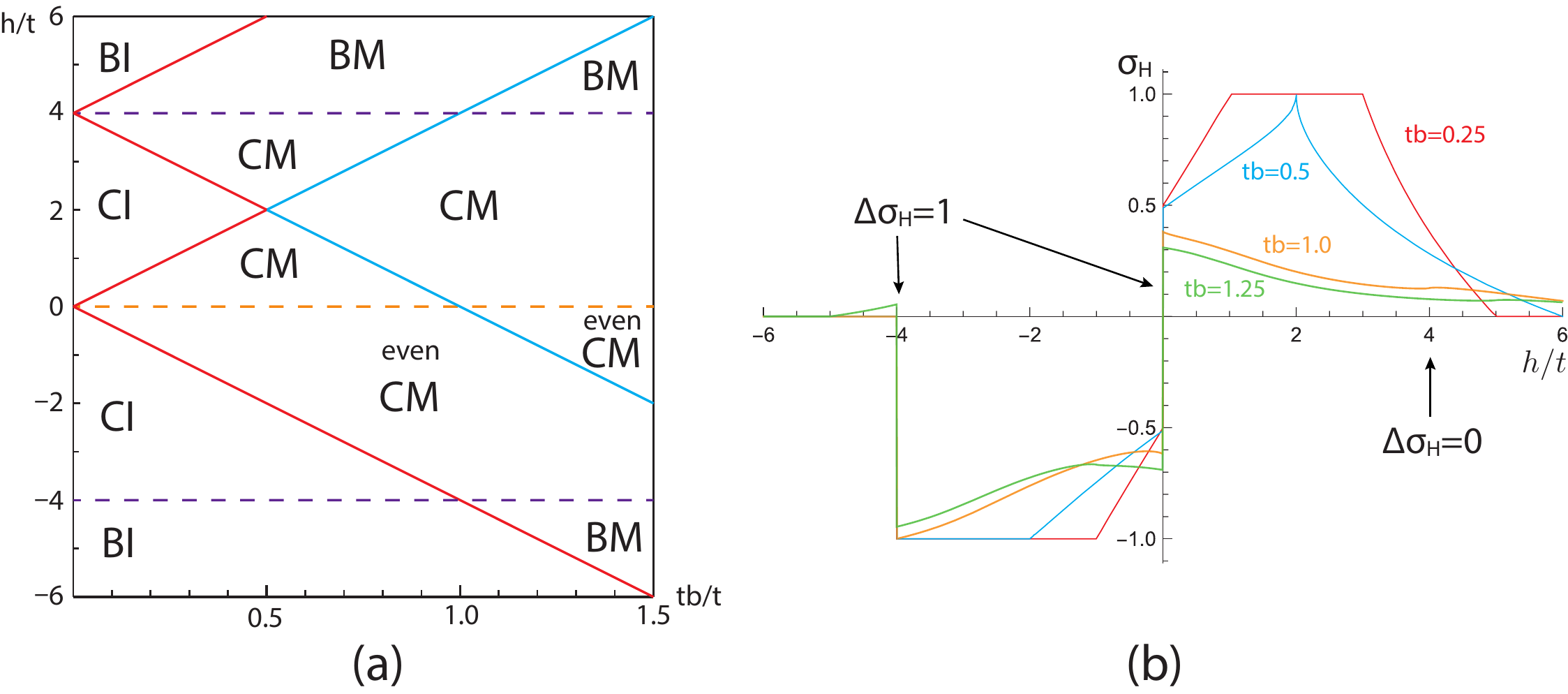}
    \caption{ In both (a) and (b), we also fixed $t=t_s=1$.
    (a) The global phase diagram of the Lattice Hamiltonian \eqref{eq:H-Even-2} for the even Chern metal case II.
    The even CM exist when $ -4 < h/t < 0 $.
    While all the CM at $ 0 < h/t < 4 $ are essentially the same
    as the BM despite superficially their bands have a non-vanishing Chern number.
    Due to this superficial difference and its relatively larger AHE, we still keep the symbol CM in the figure. See also Fig.\ref{evenoddphase}
    which also hosts odd-like CM.
    (b) The Hall conductance as a function of $h/t$ for various fixed values of $t_b/t=0.25,0.50,1.00,1.25$.
    The Hall conductance only shows a unit jump at $h/t=-4,0$.
    Especially, the $t_b/t=0.25$ curve shows a unit jump near $h/t=0$  from the even CM to a BM ,
    and also a unit jump near $h/t=-4$  from the CI to BI;
    the $t_b/t=1.25$ curve shows  the even CM to a BM with a  unit jump near $h/t=0$ and $h/t=-4$ ( see Table VI and Fig.\ref{even2edge} ).
    As shown in Table VI and Fig.\ref{even2edge}, only the metallic phase
    $ -4t < h < 0 $ can be named as even Chern metal, all the other metallic phases are just BM.
    Near $ h/t \sim 0^{-} $, it is easy to reach the even CM  from the CI due to $ t_b/t \to 0^{-} $. }
\label{even2phase}
\end{figure}

\begin{table}[!tbhp]
    \setlength{\tabcolsep}{10pt}
        \caption{The Universal Hall conductance jump of the even Chern metal case II to BM in the bulk.
        It has the same $ \Delta \sigma_{-} $ as those in Table V. But $ \Delta \sigma_{+} $ is different,
        so the total $\Delta\sigma$ is different. The change of Chern number $ \Delta Ch_{-} $ is the same as that in Table-III. }
        \begin{tabular}{c|c|c|c}
        \hline\hline
        $h$             &$\Delta\sigma_-$   &$\Delta\sigma_+$   &$\Delta\sigma$\\
        \hline
        $-4|t|$    &   +1              &  0                & 1\\
        $0$        &   +2              & -1                & 1\\
        $+4|t|$    &   -1              & +1                & 0\\
        \hline\hline
        \end{tabular}
\end{table}
  Due to the C- symmetry breaking, $ h \to -h $ is not related by the Time-reversal transformation anymore.

\begin{figure}[tbhp]
    \centering
    \includegraphics[width=0.8\linewidth]{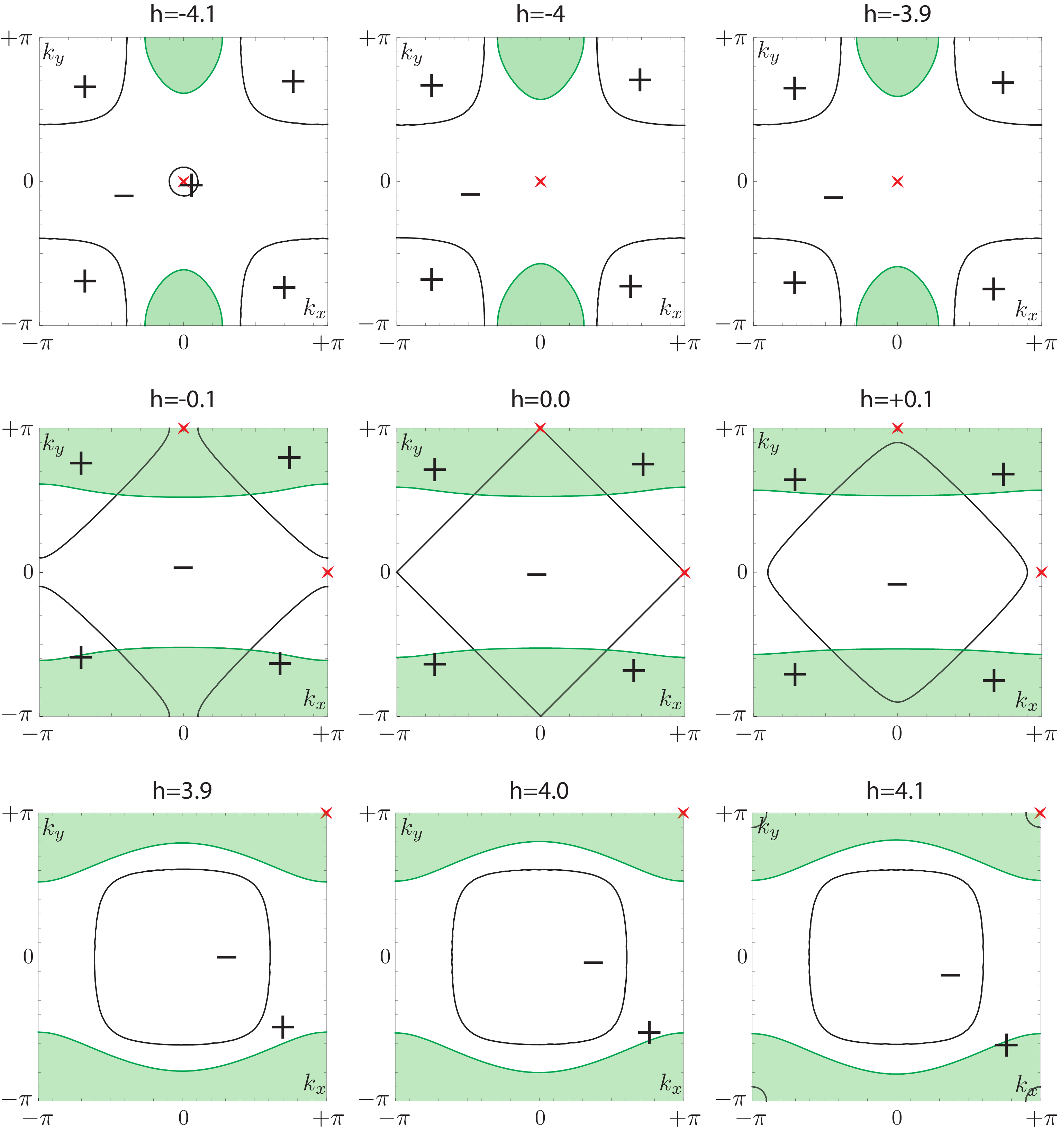}
    \caption{Berry curvature $\Omega_{+}(\mathbf{k})$ and the FS of the even Chern metal case II, it leads to $ \Delta \sigma_{+} $ in the Table VI.
    Other parameters are $t=1$, $t_s=1$, $t_b=1.25$.
    The black curve,  the red $\times$ and the green line denote the same information as Fig.\ref{even1berry}.
    Top: near $h/t=-4$:
    the upper band FS always excludes the singularity of $\Omega_{+}(\mathbf{k})$ leading to  $ \Delta \sigma_{+}=0 $,
    thus the Hall conductance jump is 1;
    Middle: near $h/t=0$:
    the upper band FS only encloses one of the two singularities of $\Omega_{+}(\mathbf{k})$ leading to $ \Delta \sigma_{+}=-1 $,
    thus the Hall conductance jump is also 1;
    Bottom: near $h/t=+4$:  the upper band FS always encloses the singularity of $\Omega_{+}(\mathbf{k})$  leading to $ \Delta \sigma_{+}=1 $,
    thus the Hall conductance jump is 0.
    The lower band is always occupied, so always encloses the singularity of $\Omega_{-}(\mathbf{k})$,
    it leads to $ \Delta \sigma_{-} $ in the Table VI.
    Only the metallic phase   at $  -4t < h < 0  $ can be named as even Chern metal, all the other metallic phases are just BM.}
\label{even2berry}
\end{figure}

\begin{figure}[tbhp]
    \centering
    \includegraphics[width=\linewidth]{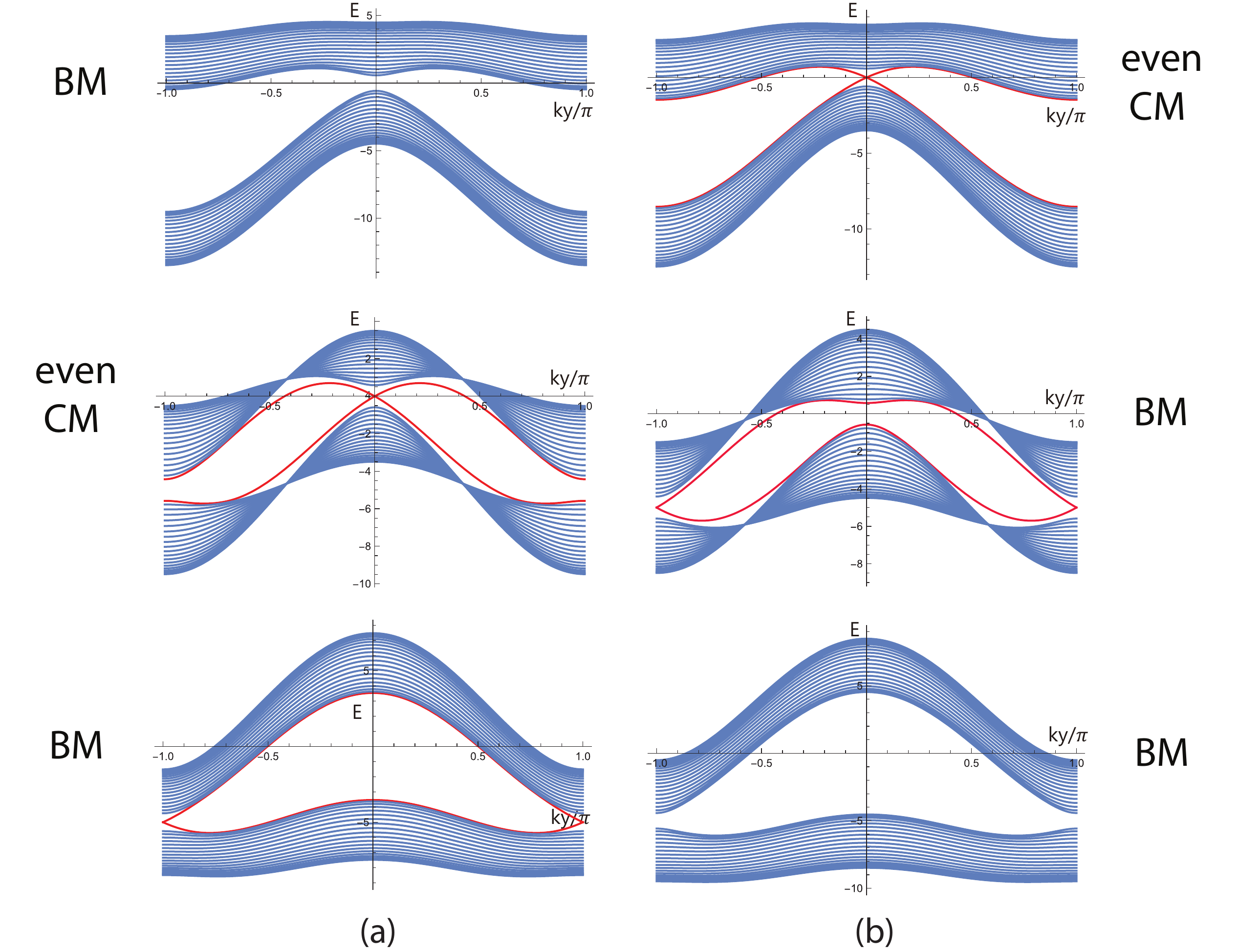}
    \caption{ The longitudinal edge structure in a stripe geometry of the even Chern metal case II at different $h/t$ values.
    Other parameters are $t=1$, $t_s=1$, $t_b=1.25$.
    Top (a) $h/t=-4.5$, no edge state, has a bulk FS, it is a BM.  (b) $h/t=-3.5$; an edge state at Fermi energy near $ k_y=0 $, also a bulk FS.
    It is a even Chern metal phase. Due to the useful edge state in (b),
    there is a unit Hall conductance jump across the TPT from the BM to the even Chern metal phase.
    Middle (a) $h/t=-0.5$, an edge state at Fermi energy near $ k_y=0 $, also a bulk FS.
    It is a even Chern metal phase.
    (b) $h/t=+0.5$;
    there is an edge state near $ k_y= \pi $, but it is well below the Fermi energy.
    Despite  there is an edge mode floating near the vast majority of
    bulk modes. It is still a BM. Due to the useful edge state in (a),
    there is a unit Hall conductance jump across the TPT from the even Chern metal  to the BM  phase.
    Bottom (a) $h/t=+3.5$, there is an edge state near $ k_y= \pi $, but it is well below the Fermi energy.
    it is a BM. (b) $h/t=+4.5$, no edge state. It is also a BM.
    As expected, there is no Hall conductance jump from the BM to the BM.
    Inside the even Chern metal, the edge modes at the Fermi energy near $ k_y=0 $  satisfy $v_Lv_R<0$.
    The Universal Hall conductance jump from the even Chern metal to the BM is an integer.  }
\label{even2edge}
\end{figure}

\section{The universal non-integer Hall conductance jump of the ``odd'' Chern metal }

We also have a graphics interpretation of the Hall conductance of ``odd'' Chern metal discussed in the main text.
The Hall conductance is an integral of Berry curvature over occupied states.
The zero temperature Hall conductance can be evaluated as $ \sigma_H=\sigma_-+\sigma_+ $:
\begin{align}
    \sigma_s=\frac{1}{2\pi}\int_{\mathbb{R}^2} \Omega_s(\mathbf{k} )\Theta(-\epsilon_s(\mathbf{k} ))d^2\mathbf{k}
\end{align}
where we have use the continuum theory to demonstrate the results.

Due to the $\mathcal{C}$-symmetry in the odd Chern metal,
the zero temperature Hall conductance can be also rewritten as
\begin{align}
    \sigma_H=\mathrm{Ch}_-+2\sigma_+
\label{upperonly}
\end{align}
 Note that due to the role played by the $\mathcal{C}$-symmetry,
 Eq.\ref{upperonly} does not hold in the even Chern metal presented in the appendix A and B or
 odd-even mixed case studied in Sec.VIII. So it only applies to the odd CM here.
 of course, the evaluations in the appendix A and B and Sec.VIII also applies here.

Below we will consider the $\alpha_x\alpha_y>0$ case
and $\alpha_x\alpha_y<0$ case separately.

\subsection{  $\alpha_x\alpha_y>0$ case: from the odd CM to the BM }

Without loss of generality, we consider $h/t\sim -4$ cases,
which belongs to the $\alpha_x\alpha_y>0$ case in Eq.\ref{eq:Dirac0},
\begin{align}
    H_0(\mathbf{k})=[\Delta+\alpha (k_x^2+k_y^2)]\sigma_z
        +vk_x\sigma_x+vk_y\sigma_y-ck_y\sigma_0
\label{eq:Dirac0copy}
\end{align}
where $\Delta=-(h+4t)$, $\alpha=t$, $v=2t_s$, $c=2t_b$.

\begin{figure}[tbhp]
    \centering
    \includegraphics[width=\linewidth]{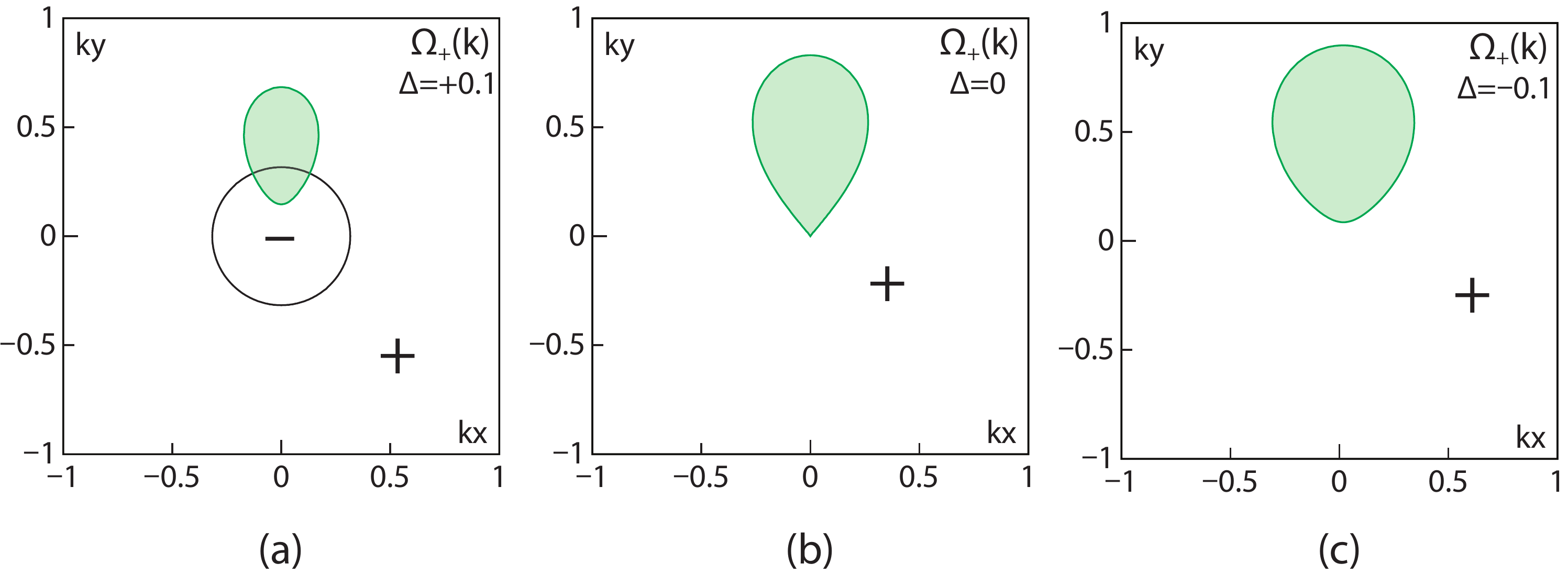}
    \caption{ According to Eq.\ref{upperonly}, these pictures explain the universal non-integer  Hall conductance jump around $h/t=\pm 4$.
    The distribution of the Berry curvature $\Omega_{+}(\mathbf{k})$
    and the electron FS at fixed $\alpha_x=\alpha_y=1$, $v=1$, $c=1.2$
    and varying (a) $\Delta=+0.1$ in the BM (b) $\Delta=0$ (c) $\Delta=-0.1$ in the Odd CM.
    The black curve is the contour of $\Omega_+(\mathbf{k})=0$,
    which separates the positive part denoted by ``+'' from the negative part denoted by ``-''.}
\label{fig:CM-1}
\end{figure}

The distribution of the Berry curvature $\Omega_+(\mathbf{k})$
and the electron Fermi surface is shown in Fig. \ref{fig:CM-1}.
Without loss of generality, we have fixed $\alpha_x=\alpha_y=1$, $v=1$, $c=1.3$.
When $\Delta>0$ in the BM, $\Omega_+(\mathbf{k})$ over FS contains positive part and negative part,
thus the integral $\sigma_+$ is almost get cancelled,
(exactly cancelled when $n=1$),
which suggests a negligible Hall conductance contribution from the upper band.
Since the $\mathrm{Ch}_-=0$, thus the total Hall conductance is negligible.
When $\Delta<0$ in the odd CM, $\Omega_+(\mathbf{k})$ over the FS only contains positive part,
thus the integral $\sigma_+$ is also a positive number.
Due to the $\Omega_+(\mathbf{k})$ over entire $\mathbf{k}$ gives $+1$,
if $c/v$ is not too large, we also know that $\sigma_+\ll 1$.
Since the $\mathrm{Ch}_-=-1$, thus $\sigma_H=\mathrm{Ch}_-+2\sigma_+=-1+2\sigma_+$ is still a negative number,
and the total Hall conductance takes a non-negligible negative value.
As varying $\Delta$ from positive to negative values,
the Hall conductance has to show a jump in order to connecting a non-negligible value to a negligible value.
This is the case for the Hall conductance jump around $h/t=\pm 4$.

\subsection{ $\alpha_x\alpha_y<0$ case: from odd CM to its the time-reversal partner odd CM. }

When $h\sim 0$ and near the two valleys $K_1$ and $K_2$, we have
\begin{align}
    H_1&=[\Delta-\alpha (k_x^2-k_y^2)]\sigma_z-vk_x\sigma_x+vk_y\sigma_y-ck_y\sigma_0
                                           \nonumber  \\
    H_2&=[\Delta+\alpha (k_x^2-k_y^2)]\sigma_z+vk_x\sigma_x-vk_y\sigma_y+ck_y\sigma_0
\label{eq:Dirac12copy}
\end{align}
where $\Delta=-h$ and other parameters are the same as the $h\sim 4t$ case discussed in Sec.III-A.
Note the opposite sign of the velocities $v$ between $k_x$ and $k_y$
and opposite sign of $\alpha$ between $k_x^2$ and $k_y^2$ indicating $\alpha_x\alpha_y<0$.

One may also identify the relation between $H_1$ and $H_2$,
$\Omega_{1,+}(\mathbf{k};\Delta)=-\Omega_{2,+}(\mathbf{k};-\Delta)$,
and $\sigma_{1,+}(\Delta)=-\sigma_{2,+}(-\Delta)$
Thus the total Hall conductance is
\begin{align}
    \sigma_H&=\mathrm{Ch}_-+2\sigma_+,\quad
    \mathrm{Ch}_-=\mathrm{Ch}_{1,-}+\mathrm{Ch}_{2,-},\quad
 \nonumber  \\
    \sigma_+&=\sigma_{1,+}+\sigma_{2,+}
    =\sigma_{2,+}(\Delta)-\sigma_{2,+}(-\Delta)
\label{upperonly12}
\end{align}

\begin{figure}[tbhp]
    \centering
    \includegraphics[width=\linewidth]{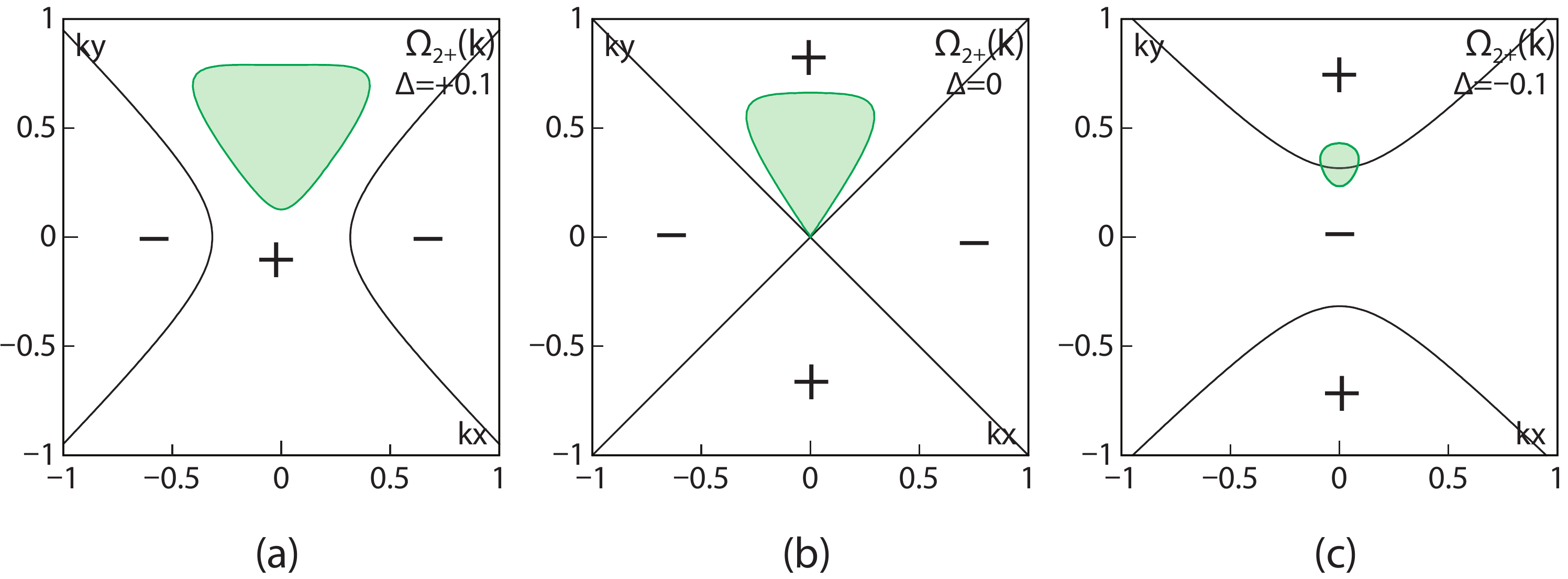}
\caption{ According Eq.\ref{upperonly12}, these pictures explain the universal non-integer Hall conductance jump around $h/t=0$.
    The distribution of the Berry curvature $\Omega_{2,+}(\mathbf{k})$
    and the electron FS at fixed $\alpha_x=-\alpha_y=1$, $v_x=v_y=1$, $c=1.3$
    and varying (a) $\Delta=+0.1$, odd CM (b) $\Delta=0$ (c) $\Delta=-0.1$, the time-reversal partner of the odd CM. }
\label{fig:CM-2}
\end{figure}

The distribution of the Berry curvature $\Omega_{2,+}(\mathbf{k})$
and the electron Fermi surface is shown in Fig. \ref{fig:CM-2}.
Without loss of generality, we have fixed $\alpha_x=-\alpha_y=1$, $v=1$, $c=1.2$.
When $\Delta>0$ in the odd CM, $\Omega_{2,+}(\mathbf{k})$ over the FS only contains positive part,
thus the integral $\sigma_{2,+}$ is also a positive number.
When $c/v$ is not too large, we also know that $\sigma_{2,+}\ll 1$.
When $\Delta<0$ in the time-reversal partner of the odd CM, $\Omega_{2,+}(\mathbf{k})$ over the FS contains positive part and negative part,
thus the integral $\sigma_{2,+}$ is almost get cancelled, (exactly cancelled when $n=1$).
Combining $\Delta>0$ and $\Delta<0$ case, we conclude:
if $\Delta>0$, $\sigma_+(\Delta)=\sigma_{2,+}(\Delta)-\sigma_{2,+}(-\Delta)\approx \sigma_{2,+}(\Delta)$;
if $\Delta<0$, $\sigma_+(\Delta)=\sigma_{2,+}(\Delta)-\sigma_{2,+}(-\Delta)\approx -\sigma_{2,+}(-\Delta)$.
We also know ${\rm Ch}_-=-\sgn(\Delta)$,
so the total Hall conductance is:
\begin{align}
\sigma_H & \approx-\sgn(\Delta)+2\sgn(\Delta)\sigma_{2,+}(|\Delta|)
     \nonumber  \\
 & =-\sgn(\Delta)[1-2\sigma_{2,+}(|\Delta|)]
\label{double12}
\end{align}
When $c/v$ is not too large, $\sigma_{2,+}(|\Delta|)$ is a small quantity,
thus the Hall conductance has to show a jump in order to connecting a positive value to a negative value
from the odd CM to its the time-reversal partner.
This is the case for the Hall conductance jump around $h/t=0$ from OCM/OCM which is twice the value of OCM/BM discussed in the last subsection.

\section{The Berry phase ( or singularity )  of  the $ z=1 $ Dirac point}

Consider the  doped ( even ) case,
\begin{align}
    H(\mathbf{k})=(\Delta+\alpha_x k_x^2+\alpha_y k_y^2)\sigma_z
        +v_xk_x\sigma_x+v_yk_y\sigma_y-\mu\>,
\label{eq:Dirac_d}
\end{align}
For simplicity, we can use the simplified version for $ \alpha_x=\alpha_y=0 $:
\begin{align}
    H(\mathbf{k})=\Delta\sigma_z+vk_x\sigma_x+vk_y\sigma_y-\mu\>,
\label{eq:Dirac_d0}
\end{align}
 For $ \mu <|\Delta|$, the Berry curvature takes the form:
\begin{align}
    \Omega_\pm(k)=\mp\frac{\Delta}{2(\Delta^2+k^2)^{3/2}}
\end{align}
Under the limit $\Delta\to 0$, the $\Omega_\pm(k)$ becomes a $\delta$-function.
From the definition, we can check that
$\lim_{\Delta\to 0}\Omega_\pm(k)=\mp\pi\mathrm{sgn}(\Delta)\delta(\mathbf{k}-0)$
which is the well known $ \pi $ Berry phase for a Dirac point.
In order to see this, we only need to verify two conditions
1) $\lim_{\Delta\to 0}\Omega_\pm(k)=0$ when $k\neq 0$
and $\lim_{\Delta\to 0}\Omega_\pm(k)=\mp\mathrm{sgn}(\Delta)\infty$ when $k=0$;
2) $\int d^2k \Omega_\pm(k)=\mp\pi\mathrm{sgn}(\Delta)$.
The condition 1) is obvious, the condition 2) is due to
the well-known result $\frac{1}{2\pi}\int d^2k\Omega_\pm(k)=\mp \mathrm{sgn}(\Delta)/2$.
They were used to evaluate the bulk Universal Hall conductance jump in Sec. VII-B.

Note that if one introduces $\alpha\neq0$, then $\lim_{\Delta\to 0}\Omega_\pm(k)\neq 0$ when $k\neq 0$,
which means $\lim_{\Delta\to 0}\Omega_\pm(k)$ is not just a pure $\delta$-function.

For $\mu>|\Delta|$, we can solve $k_F=\sqrt{\mu^2-\Delta^2}$ in the upper band.
The Berry curvature from the lower band $\phi_-$ which is always occupied and the upper band $\phi_+$ which
has a FS are
\begin{align}
    \phi_+&=\int_{0}^{k_F} 2\pi kdk \Omega_+(k)
    =-\pi (1-|\Delta/\mu|)\sgn(\Delta)\\
    \phi_-&=\int_{0}^{\infty} 2\pi kdk \Omega_-(k)= +\pi\sgn(\Delta)
\end{align}
and then:
\begin{align}
   \phi= \phi_+ +\phi_-=\pi|\Delta/\mu|\sgn(\Delta)=\pi\Delta/|\mu|=\pi\Delta/\mu
\end{align}
 which leads to the Hall conductance $ \sigma_H= \frac{\phi}{2 \pi}= \Delta/2 \mu $.

\section{  Galileo transformation for the artificially generated non-relativistic QAH   }

 As discussed in \cite{moving}, if the SOC is artificially created in cold atom system \cite{haldaneexp,2dsocbec,QAHboson},
 it is not a relativistic effect.
 The spin in the SOC is just a pseudo-spin consisting of two hyperfine states.
 Then the gauge invariant current plus a NNN current in Eq.\ref{boostLattice} can be generated by a Galileo transformation (GT).
 However, the QAH in real materials \cite{QAHexp,QAHexp2,TBGAHE} comes from the relativistic effect at the order of $ (v/c)^2 $.
 So the conventional GT does not apply anymore, one need to apply a low velocity expansion of the Lorentz
 transformation (LT) upto the order of $ (v/c)^2 $ where the time contraction effects of the LT must be taken into account.
 So here we only focus on the QAH in artificial materials in a moving sample,
 leave how the QAH in real materials change in a moving sample to a future publication.

  To perform a GT on the lattice, the first thing to do is to do it directly one the QAH model Eq.\ref{QAH0}.
  The action corresponding to the QAH model Eq.\ref{QAH0} is:
\begin{align}
	\mathcal{S}_{QAH}=\int d\tau \sum_i c_i^\dagger \partial_\tau c_i
	+H_{QAH}[c_i^\dagger, c_i]
\end{align}
In the continuum theory, a Galilean boost with a constant velocity $\vec{v}_b$ will lead to
$\partial_\tau\to\partial_\tau+i\vec{v}_b\cdot\vec{\nabla}$.
In the lattice theory, one need replace the spatial derivative $\partial_{\hat{\mu}}$
by its discrete version (lattice derivative) $\Delta_{\hat{\mu}}$
via $\partial_{\hat{\mu}}\phi_i\to
\Delta_{\hat{\mu}}\phi_i=a^{-1}(\phi_{i+\hat{\mu}}-\phi_i)+ \cdots $
and $\Delta_{\hat{\mu}}^*\phi_i=a^{-1}(\phi_i-\phi_{i-\hat{\mu}}) + \cdots $,
where $a$ is the lattice constant where $ \cdots $ means that one should add infinite number of higher order terms
which still lead to the same contume limit

Thus, under a Galilean boost, the lattice action can be written as
\begin{align}
	\mathcal{L}_{QAH,b}&=\sum_i c_i^\dagger \partial_\tau c_i
	+i \sum^{\infty}_{n=1 } [ t_{bn,x} c_i^\dagger c_{i+nx}+t_{bn,y} c_i^\dagger c_{i+ny} ]
                                     \nonumber  \\
	&+  h.c. + H_{QAH}[c_i^\dagger,c_i]
\label{BoostedQAHform}
\end{align}
  which sets up the form of the Boosted QAH model.
  In principle, one need to include the infinite sum of terms. The simplest thing to do is to include only the $ n=1 $ NN term
  $ i (t_{b1,x} c_i^\dagger c_{i+x}+t_{b1,y} c_i^\dagger c_{i+y}-h.c.) $
  where $t_{b,x}=\hbar v_{b,x}/(2a)$ and $t_{b,y}=\hbar v_{b,y}/(2a)$  has the energy dimension.

  For the interacting bosonic system studied in \cite{moving}, this leading term is proportional to the $ U(1) $ conserved current term,
  so can be absorbed by a unitary transformation into the hopping term in
  $ H_{BH}[b_i^\dagger,b_i] $. So one need also consider also the subleading term $ n=2 $ NNN current term
  in the series. Unfortunately,  one can not determine the ratio of $ t_{b2}/t_{b1} $ just from the substitution.
  One need to repeat the derivation from the ionic model to the BH model to determine the whole series in Eq.\ref{BoostedQAHform}.
  This was achieved in \cite{moving}.

  For the present non-interacting fermionic QAH, the $ n=1 $ term is not proportional to any conserved current term,
  one can just take it as an injecting current as done in Sec.II  which leads to various new phases and TPTs in the main text.
  However, the Galilean boost in a lattice in the presence of SOC ( or Non-Abelian gauge field )
  should take the different form Eq.\ref{boostLatticev} where the $ n=1 $ term is indeed the NN gauge invariant current.
  Then it still can be absorbed into the SOC term in $ H_{QAH} $ by the unitary transformation Eq.\ref{tildebasis}.
  Then one must consider the $ n=2 $ NNN current term.
  So under a GT boost with the velocity $ v $ relative to the lattice along the $ \hat{x} $ direction, Eq.\ref{boostLattice}
  need to be replaced by:
\begin{equation}
  H_{bx}= - v [ \frac{t_{b1}}{t_0} \sum_{i} J_{ix} + i \sum^{\infty}_{n=2} (t_{bn}/n)  c^{\dagger}_i c_{i+nx}] + h.c.
\label{boostLatticev}
\end{equation}
  where $ J_{ix} $ is the NN gauge-invariant current and $ t_{b1}, t_{b2} $ are completely determined by
  the Wannier functions $  \phi( |\vec{x}| ) $ of the lattice system:
\begin{align}
  t_{bn}  = \hbar \int d^2 x \phi( |\vec{x}| ) \frac{ \partial}{\partial x } \phi( |\vec{x} + na \hat{x} | ),~~~n=1,2
\label{tttt}
\end{align}
  where $ t_{bn} v $ carry the dimension of the hopping ( for notational simplicity, we still use the same symbols as
  in Eq.\ref{BoostedQAHform}). So $ t_{b1}, t_{b2} $ are not dependent anymore with $ t_{b1} \ll t_{b2} $ in the tight-binding limit.

  One can adopt the derivation in \cite{moving} straightforwardly by identifying:
\begin{align}
  t   & =t_s= t_{b1} \sqrt{v^2 + v^2_c},~~~v_c=\frac{ t_{0} }{t_{b1} }
               \nonumber   \\
  t_b & = \frac{  v^2_c-v^2 }{ v^2_c + v^2  } v t_{b2}
\label{ttbv}
\end{align}
  where $ v $ is the velocity of the moving frame, $ v_c $ is the characteristic velocity where $ t_b $ changes sign.
  It was used in the main text Sec.IX-3.
  As shown in Sec.V, $ t_b $ cause the same Doppler shift to the 4 nodes as expected for a moving observer.

\onecolumngrid
\pagebreak
\widetext
\begin{center}
	\textbf{\large Supplemental Materials}
\end{center}
\setcounter{equation}{0}
\setcounter{figure}{0}
\setcounter{table}{0}
\setcounter{page}{1}
\makeatletter
\renewcommand{\theequation}{S\arabic{equation}}
\renewcommand{\thefigure}{S\arabic{figure}}
\renewcommand{\bibnumfmt}[1]{[S#1]}
\renewcommand{\citenumfont}[1]{S#1}
\makeatother
\vspace{4mm}

In the supplemental materials,
we provide the enlarged picture for the edge modes in Fig.20, 21 and Fig.27, 28 respectively.

\begin{figure}[tbhp]
    \centering
    \includegraphics[width=\linewidth]{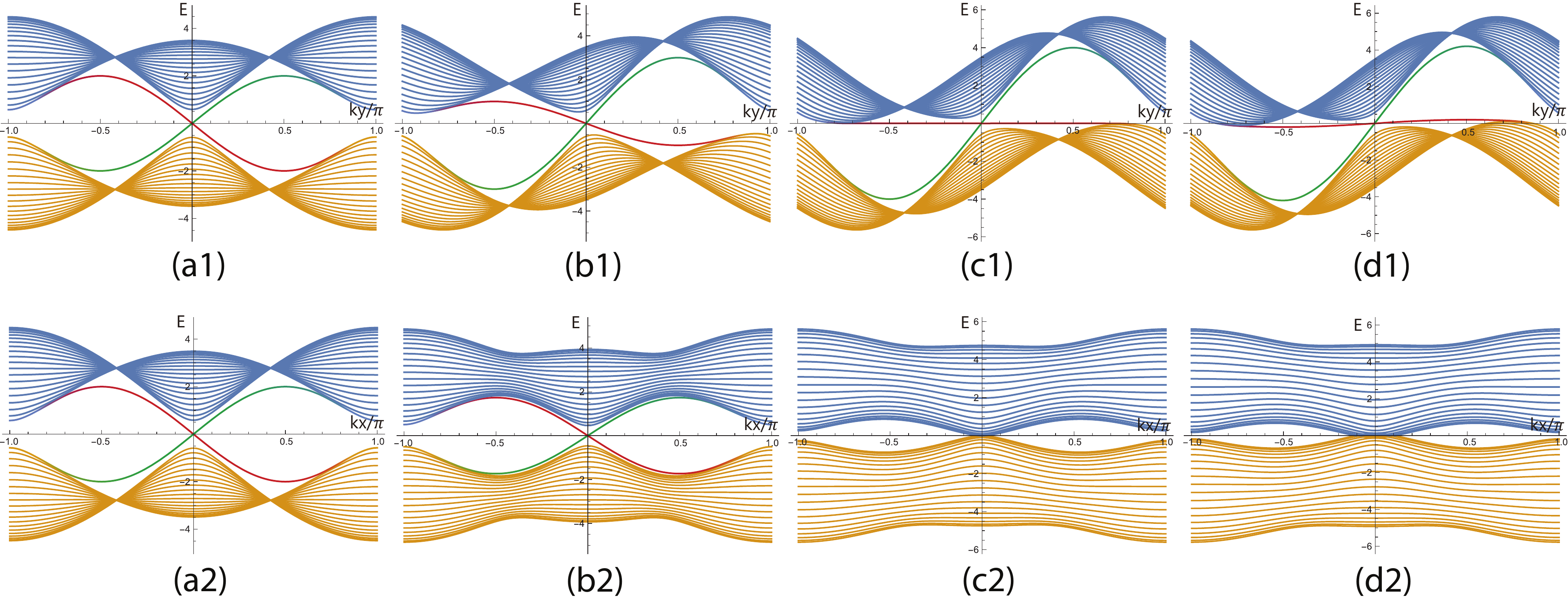}
    \caption{The edge states of the lattice Hamiltonian in a strip geometry. We fixed $h=-0.5$.
    From left to right, the parameter $t_b/t_s$ is $0,0.5,1.0,1.1$, respectively.	
    (Top) Longitudinal injection: With the periodic boundary condition in the $y$-direction and the open
    boundary condition in the $x$-direction. The edge modes always exist, but undergoes the edge reconstruction at $ t_s/t =1 $.
    The two edges move along the opposite directions when $ t_s/t < 1 $ in (a1) and (b1) in the CI,
    then  one edge becomes flat at $ t_s/t= 1 $ in (c1),
    then two edges move along the {\em same } direction when  $ t_s/t > 1 $ in (d1) in the odd CM,
	(Bottom) Transverse injection:
    Exchanging the role of $ x $ and $ y $ direction. The edge modes exist only when $ t_s/t < 1 $ in (a2) and (b2),
    but squeezed out at $ t_s/t= 1 $ in (c2) where the direct bulk gap closes, completely disappear when  $ t_s/t > 1 $ in (d2) in the odd CM.
    Although the edge modes show quite different behaviours in the line 1 and the line 2,
    there seems a one to one Longitudinal/Transvese edge-edge correspondence between them.
    In both figures, one can shift $k\to k+\pi$ to reach $h=+0.5$ results.
    See also Fig.\ref{fig:xEdge} and Fig.\ref{fig:yEdge} for the continuum calculations. }
\label{figS:Edge_Lattice}
\end{figure}

\begin{figure}[tbhp]
    \centering
    \includegraphics[width=\linewidth]{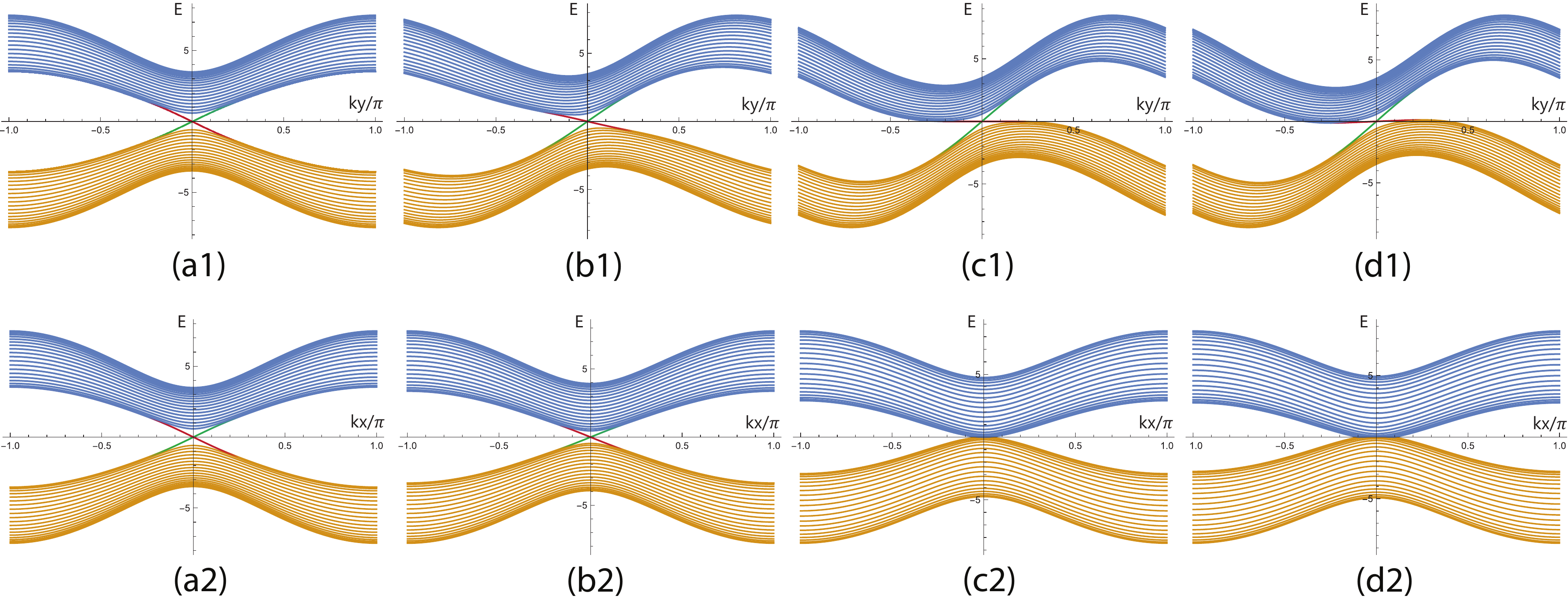}
    \caption{ The same situation as in Fig.\ref{fig:Edge_Lattice} except
	 at a fixed $h=-3.5$. In both figures, one can shift $k\to k+\pi$ to reach $h=+3.5$ results.
     It shows qualitatively the same edge TPTs as those in Fig.\ref{fig:Edge_Lattice}.}
\label{figS:Edge_Lattice2}
\end{figure}

\begin{figure}[tbhp]
    \centering
    \includegraphics[width=\linewidth]{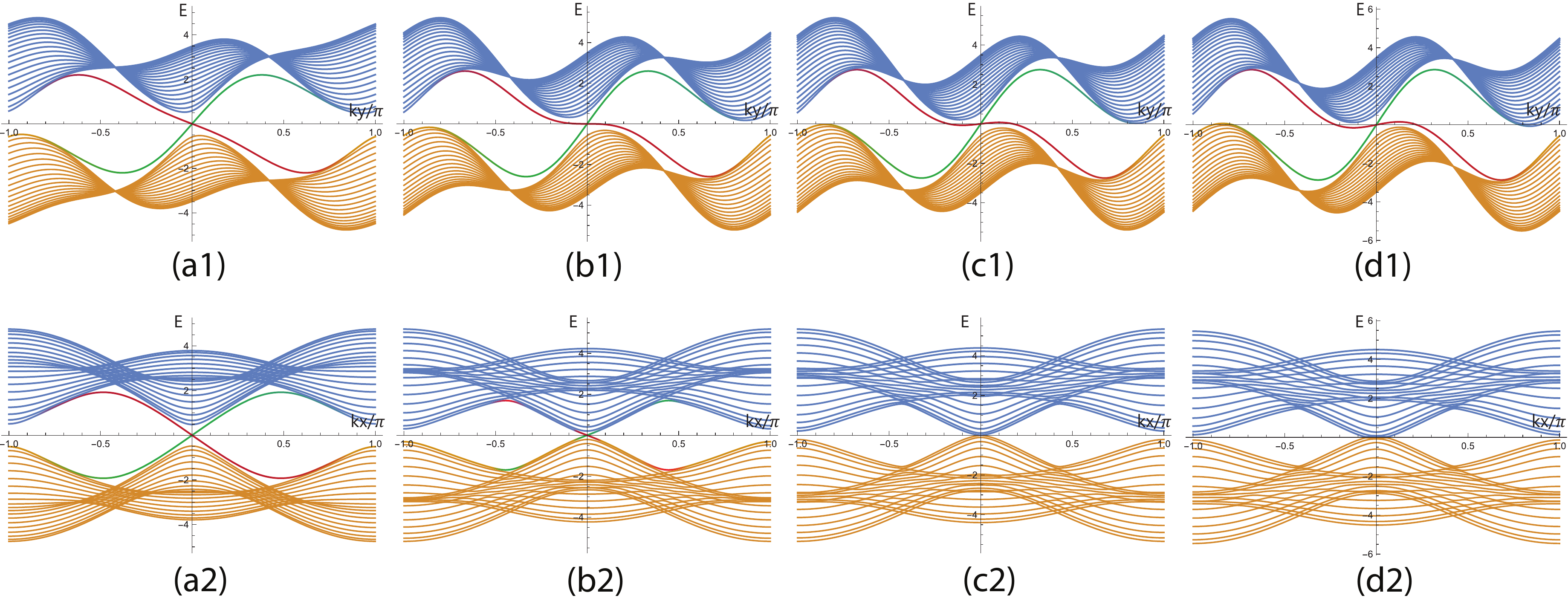}
    \caption{The edge state of the lattice Hamiltonian Eq. \eqref{eq:Boosted_Hk_n}.
    From (a) to (d), the parameter $t_b/t_s$ is $0.5,1.0,1.17,1.3$, respectively. We fixed $h=-0.5$.
    (Top) Longitudinal boost: With periodic boundary condition in the $y$-direction and open
    boundary condition in the $x$-direction. The edge modes always exist in this case.
    The two edge mods move in the opposite direction near $ k_y=0 $ in (a1) Chern Insulator where $ t_b/t_s < 1 $,
    then one edge mode's slope vanishes in (b1) where $ t_b/t_s = 1 $
    with the edge dispersion $ \omega \sim k^3_y $, namely the longitudinal edge dynamic exponent $ z_L=3 $.
    then the two edge modes move along the same direction near $ k_y=0 $ in (c1) odd Chern insulator where $ t_b/t_s = 1.17 > 1 $.
    At the same time, the system's (in-direct) gap vanishes which corresponds to the $ z=2 $ bulk TPT from the
    $ C=-1 $ odd Chern insulator to A2 Odd Chern metal in Fig.\ref{fig:phaseLattice2}.
    It gets to the Odd Chern metal phase in (d1) where $ t_b/t_s = 1.3 > 1.17 $, the two edge modes still move along the same direction.
	(Bottom) Transverse boost:
    Exchanging the role of $ x-$ and $y-$ direction.
    The edge mode exists upto (c2) where $ t_b/t_s = 1.17 > 1 $. So the odd CI between (b2)  and (c2) still has the transverse edge mode.
    At (c2), the system's direct gap vanishes which corresponds to the TPT from the
    $ C=-1 $ odd Chern insulator to A2 Odd Chern metal in the bulk in Fig.\ref{fig:phaseLattice2}.
    It is in the Odd Chern metal phase in (d2) where $ t_b/t_s = 1.3 > 1.17 $, no edge mode.
    The T-edge disappears at the same time as the bulk TPT with its velocity still vanishing  $ \sqrt{v^2-c^2} $ as in Fig.\ref{fig:yEdge}.
    One can shift $k\to k+\pi$ to reach $h=+0.5$ results.	}
\label{figS:Edge_Lattice_n2}
\end{figure}

\begin{figure}[tbhp]
    \centering
    \includegraphics[width=\linewidth]{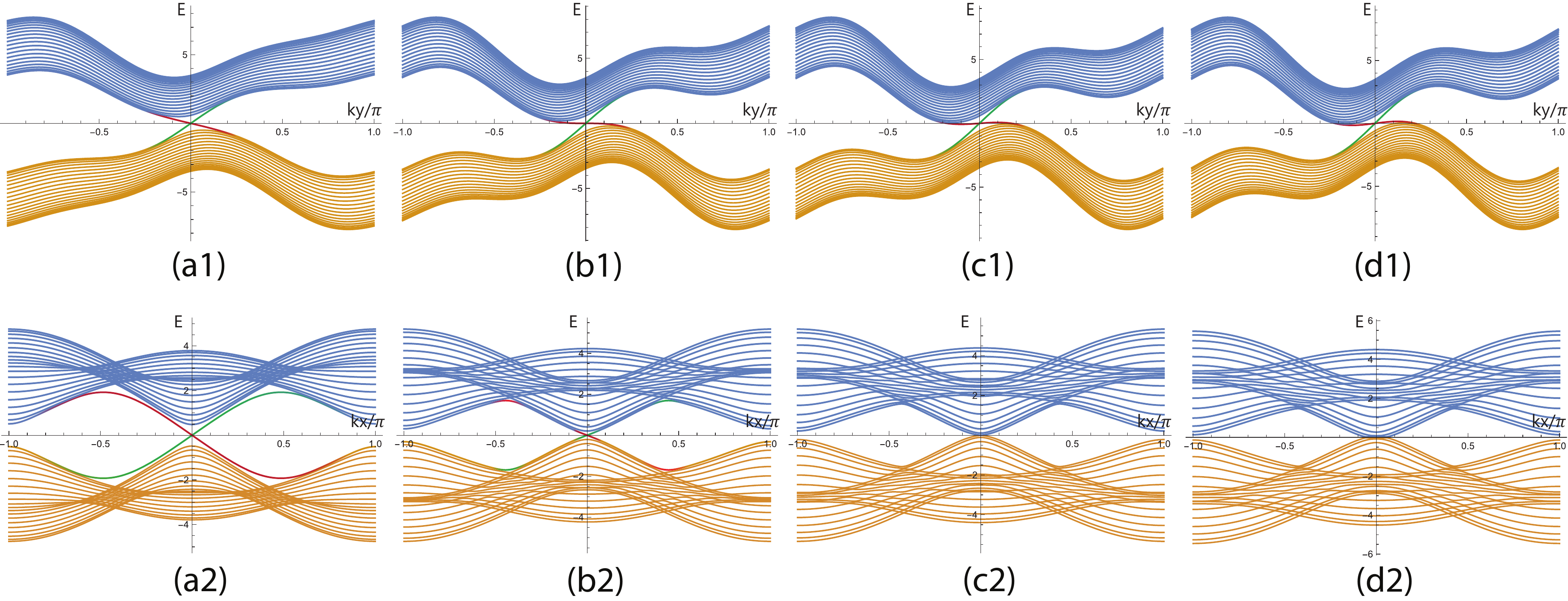}
    \caption{ The same situation as Fig.\ref{fig:Edge_Lattice_n2} except $ h=-3.5 $ which is Mirror reflected image of $ h=-0.5 $.
     As alerted in  Fig.\ref{fig:phaseLattice2},
     despite the bulk phase boundary in Fig.\ref{fig:phaseLattice2} has such a Mirror symmetry at $ t=t_s $,
     it is not persevered in the presence of the strip boundaries.
     One can shift $k\to k+\pi$ to reach $h=+3.5$ results  which is Mirror reflected image of $ h=0.5 $.
     It shows qualitatively the same edge TPTs, odd CI and odd CM as those in Fig.\ref{fig:Edge_Lattice_n2}. }
\label{figS:Edge_Lattice2_n2}
\end{figure}


\begin{thebibliography}{999}
\bibitem{footnote00} Note that Eq.\eqref{eq:C} does not holds for the $\alpha_x\alpha_y=0$ case,
a more general form would be
${\rm Ch}_-=\frac{1}{2}\sgn(v_xv_y)
[\sgn(\Delta)-(1-\frac{1}{2}\sgn(\alpha_x\alpha_y))(\sgn(\alpha_x)+\sgn(\alpha_y))]$. But this case does not happen here.
\bibitem{SSS} On the intermediate ( intervening ) phase, it is instructive to make some analogy to interacting boson systems,
there are two possibilities on the transition between two phases breaking different symmetries
the SF phase breaking the $ U(1) $ symmetry, the solid breaking the translational symmetry.
The first one is a direct first order transition.
the second is through the intervening phase called supersolid which breaks both the $ U(1) $ symmetry and the translational symmetry,
so the putative direct first order transition splits into 2 second order ones.
\bibitem{Nooverlap} Note that in the 2d BZ, B1 phase in Fig.2, there is no overlap between any of the P- or H- Fermi pockets
near $K_1=(\pi,0)$ and near $K_2= (0, \pi)$. It looks like so in Fig.\ref{fig:Dirac12_FS} is only because one is expanding
around the two different minima, but putting them on the same $ k_y $ axis.
\bibitem{class3}
Because the collisions between two FFs takes a saddle point (cone) geometry near a Van Hove singularity (VHS),
the dispersion Eq.3 in \cite{weyl} seems also suggest one may define the dynamic exponent $ z=2 $.
However, due to its saddle point nature, we feel it may not be suitable to introduce $z$ in this class.
\bibitem{extendtolarge}
The following evaluations can be extended to $|\alpha\Delta/v^2| > 1/2$ with more complicated manipulations without changing the results qulitatively.
\bibitem{footnote3}
Here $T\ll |\Delta|$ should be interpreted as $T\ll |\Delta|$ only when $\Delta\neq 0$.
If $ \Delta = 0$, then setting $ \Delta = 0$, one $ T $ remains the lowest energy scale compared to other non-vanishing energy scale.
\bibitem{bosonicQAH} As an analogy, for interacting bosonic QAH,
because the interacting bosons stay near the minima of a band, so the Chern number defined for a Bogliubov quasi-particle band
at the quadratic level only has mathematical sense \cite{QAHboson}. At first sight, it may not practically very useful, but it may still be
experimentally measured by some non-equilibrium pumping detection.
\end{thebibliography}
\end{document}